\documentclass[usenatbib,useAMS]{mn2e}

\newcommand{\dust}{850$\,$\micron }
\newcommand{\vunits}{km$\,$s$^{-1}$}
\usepackage{aas_macros} 
\usepackage[pdftex]{graphicx}

\newcommand{\Msun}{M$_{\odot}$}
\newcommand{\Tstara}{T$_A^*$}
\newcommand{\twelco}{\ensuremath{\mathrm{^{12}CO}}}
\newcommand{\ceighto}{\ensuremath{\mathrm{C^{18}O}}}
\newcommand{\thirtco}{\ensuremath{\mathrm{^{13}CO}}}
\newcommand{\htwo}{\ensuremath{\mathrm{H_2}}}
\newcommand{\jtt}{$J=\!3\!\rightarrow\!2$}
\newcommand{\kms}{km\,s$^{-1}$}
\usepackage{mathptmx}
\usepackage{hyperref}

\title[GBS: First look at Serpens] {The JCMT Legacy
  Survey of the Gould Belt: a first look at Serpens with HARP}

\author[The GBS Serpens Team]{S.F. Graves$^{1,2}$\thanks{E-mail s.graves@mrao.cam.ac.uk},
J.S. Richer$^{1,2}$,
J.V. Buckle$^{1,2}$,
A. Duarte-Cabral$^{3}$, 
G.A. Fuller$^{3}$,\newauthor
M.R. Hogerheijde$^{4}$,
J.E. Owen$^{5}$, 
C. Brunt$^{6}$, 
H.M. Butner$^{7,8}$,
B. Cavanagh$^{8}$, \newauthor
A. Chrysostomou$^{8,9}$,
E.I. Curtis$^{1,2}$,
C.J. Davis$^{8}$,
M. Etxaluze$^{10,11}$, 
J. Di Francesco$^{12,13}$,\newauthor 
P. Friberg$^{8}$,
R.K. Friesen$^{14}$,
J.S. Greaves$^{15}$, 
J. Hatchell$^{6}$,  
D. Johnstone$^{12,13}$, 
B. Matthews$^{12}$,\newauthor
H. Matthews$^{16}$, 
C.D. Matzner$^{17}$, 
D. Nutter$^{18}$, 
J.M.C. Rawlings$^{19}$, 
J.F. Roberts$^{20}$, 
S. Sadavoy$^{12,13}$,\newauthor 
R.J. Simpson$^{18}$, 
N.F.H. Tothill$^{6,21}$,
Y.G. Tsamis$^{22}$,
S. Viti$^{19}$, 
D. Ward-Thompson$^{18}$,\newauthor 
G.J. White$^{10,11}$,
J.G.A. Wouterloot$^{8}$,
J. Yates$^{19}$  \\
$^{1}$Astrophysics Group, Cavendish Laboratory, J J Thomson Avenue, Cambridge, CB3 0HE, UK\\
$^{2}$Kavli Institute for Cosmology, c/o Institute of Astronomy, University of Cambridge, Madingley Road, Cambridge, CB3 0HA, UK \\
$^{3}$Jodrell Bank Centre for Astrophysics, School of Physics and Astronomy, The University of Manchester, Oxford Road, Manchester M13 9PL, UK\\
$^{4}$Leiden Observatory, Leiden University, PO Box 9513, 2300 RA, Leiden, The
Netherlands\\
$^{5}$Institute of Astronomy, University of Cambridge, Madingley
Road, Cambridge, CB3 0HE, UK\\
$^{6}$School of Physics, University of Exeter, Stocker Road, Exeter, UK\\
$^{7}$Department of Physics and Astronomy, James Madison University, 901
Carrier Drive, Harrisonburg, VA 22807, USA\\
$^{8}$Joint Astronomy Centre, 660 N. A'Ohoku Place, University Park, Hilo,
Hawaii 96720, USA\\
$^{9}$School of Physics, Astronomy and Mathematics, University of
Hertfordshire, College Lane, Hatfield, UK\\
$^{10}$Science and Technology Facilities Council, Rutherford Appleton
Laboratory, Chilton, Didcot, UK\\
$^{11}$Department of Physics and Astronomy, Open University, Walton Hall, Milton
Keynes, UK\\
$^{12}$National Research Council Canada, Herzberg Institute of Astrophysics, 5071 West Saanich Rd, Victoria, BC, V9E 2E7\\
$^{13}$Department of Physics \& Astronomy, University of Victoria,  3800 Finnerty Rd., Victoria, BC, Canada\\
$^{14}$North American ALMA Science Center, National Radio Astronomy Observatory, 520 Edgemont Rd, Charlottesville VA 22903, USA\\
$^{15}$Scottish Universities Physics Alliance, Physics \& Astronomy, University
of St Andrews, North Haugh, St Andrews, Fife, UK\\
$^{16}$National Research Council of Canada, Dominion Radio Astrophysical
Observatory, 717 White Lake Rd., Penticton, BC, Canada\\
$^{17}$Department of Astronomy \& Astrophysics, University of Toronto, 50 St George Street, Toronto, Canada\\
$^{18}$School of Physics \& Astronomy, Cardiff University, 5 The Parade,
Cardiff, UK\\ 
$^{19}$Dept of Physics \& Astronomy, University College London, Gower Street,
London, UK\\
$^{20}$Centro de Astrobiolog\'{i}a (CSIC/INTA), Ctra de Torrej\'{o}n a Ajalvir km 4, E-28850 Torrej\'{o}n de Ardoz, Madrid, Spain\\
$^{21}$School of Physics, University of New South Wales, Sydney, NSW 2052, Australia\\
$^{22}$Instituto de Astrof\'isica de Andaluc\'ia (CSIC), Camino Bajo
de Hu\'tor 50, 18008 Granada, Spain\\
}

\begin{document}
\date{Accepted June 2nd 2010}
\pagerange{\pageref{firstpage}--\pageref{lastpage}} \pubyear{2010}
\maketitle
\label{firstpage}


\begin{abstract}

  The Gould Belt Legacy Survey (GBS) on the JCMT has observed a region
  of 260 square arcminutes in \twelco\ \jtt\ emission, and a 190 square
  arcminute subset of this in \thirtco\ and \ceighto\ towards the Serpens
  molecular cloud. We examine the global velocity structure of
  the non-outflowing gas, and calculate excitation temperatures and
  opacities. The large scale mass and energetics of the
  region are evaluated, with special consideration for high velocity
  gas. We find the cloud to have a mass of 203\,\Msun, and to be 
  gravitationally bound, and that the kinetic energy of the outflowing
  gas is approximately seventy percent   of the turbulent kinetic energy of the
  cloud.  
  We identify compact outflows towards some of the submillimetre
  Class 0/I sources in the region.

\end{abstract}



\section{Introduction}

\subsection{The JCMT Gould Belt Survey} 

The James Clerk Maxwell Telescope's (JCMT) Gould Belt Legacy Survey
\citep{WardThompson2007} is a large scale programme to observe nearby
($<$500\,pc) regions of active star formation in both submillimetre
continuum emission and three CO species \jtt\
emission\footnote{http://www.jach.hawaii.edu/JCMT/surveys/gb/}. The
survey will observe molecular clouds with SCUBA-2 \citep{Holland2006}
to detect the continuum emission, and has been carrying out heterodyne
spectrometer observations of filamentary regions in a subset of these
clouds. These observations are currently being carried out with
HARP/ACSIS \citep{Buckle2009} in the \twelco, \thirtco\ and \ceighto\
isotopologues, in the \jtt\ transition.  POL-2 \citep{Bastien2005},
the SCUBA-2 polarimeter, will also be used to study some of the
SCUBA-2 observed regions. A first look at the data from the Orion-B
region has already been presented, by \citet{Buckle2010}, and from
the Taurus region by \citet{Davis2010}.

The survey seeks to perform some of the first large scale analyses of
very large, high resolution submillimetre maps and data cubes of the closest
molecular clouds and examine the star formation occurring within
them. In particular, the spectral part of this survey will examine the
properties of higher density gas than most large scale surveys, due to
the higher critical density of the \jtt\ transitions
($10^{4}-10^5$\,cm$^{-3}$). This is better matched to the densities
probed in submillimetre continuum emission than the lower $J$
transitions. The aims of the spectral survey are to search for and
categorise high velocity outflows in order to identify protostars; to
examine the column density in cores; to study the support mechanisms
of protostars and their evolution; and to examine the turbulence and
velocity structure of the parent molecular clouds
\citep{WardThompson2007}. SCUBA-2 continuum observations will also be
made, and will cover an extremely large area compared to the current data sets
available from e.g. SCUBA. As well as looking at the individual
regions, the GBS will carry out statistical studies and comparisons on
all the data sets and objects found.

\subsection{The Serpens molecular cloud}
The Serpens molecular cloud is a much-studied region of low mass star
formation, located $\sim$230\,pc from the Sun \citep[and references
therein, in particular \citealt{Straizys2003}]{Eiroa2008}.  It is
forming a compact and high density cluster of new stars, and its close
proximity allows us to map its gas properties at high spatial
resolution (a FWHM beam size of 3300\,AU at the distance of Serpens for
\twelco).  The cloud consists of two clusters of embedded YSOs of
which many possible Class 0/I/II sources have been detected by the
\textit{spitzer} c2d survey using MIPS \citep{Harvey2007a}; and using
both MIPS and IRAC \citep{Harvey2007}. The region observed in this
study is focused on the Serpens main cloud core, which contains the
well known submillimetre Class 0-I objects SMM1-11
\citep{Casali1993,Davis1999}. \citet{Hogerheijde1999} presented high
resolution interferometric observations of some of these cores in a
variety of molecular tracers. Evidence for infall towards SMM 2, 3 and
4 has been found \citep[see e.g.~][]{Gregersen1997}, further
supporting the view that these are very young sources currently
undergoing star formation. \citet{DuarteCabral2010} find evidence that
the global structure of the cloud implies it consists of two colliding
sub clouds.

Much evidence of molecular outflows has been detected in the region in
previous JCMT observations \citep{White1995,Davis1999}.
\Citet{Hodapp1999} measured the proper motions of \htwo\ jets in the
north-west part of this region (in the vicinity of SMM 5, 9 and 10),
further connecting these jets to outflow sources.  The observations
were centred on the region known as the Serpens cloud core, containing
many protostars, dust sources, HH objects and potential
outflows. \citet{Eiroa2008} presented a recent review of the entire
Serpens region. This paper presents new observations and results of
outflow kinematics along with the general cloud properties.

\section{Observations}
\label{sec:obs}

\begin{figure}
\begin{center}
\includegraphics[width=80mm]{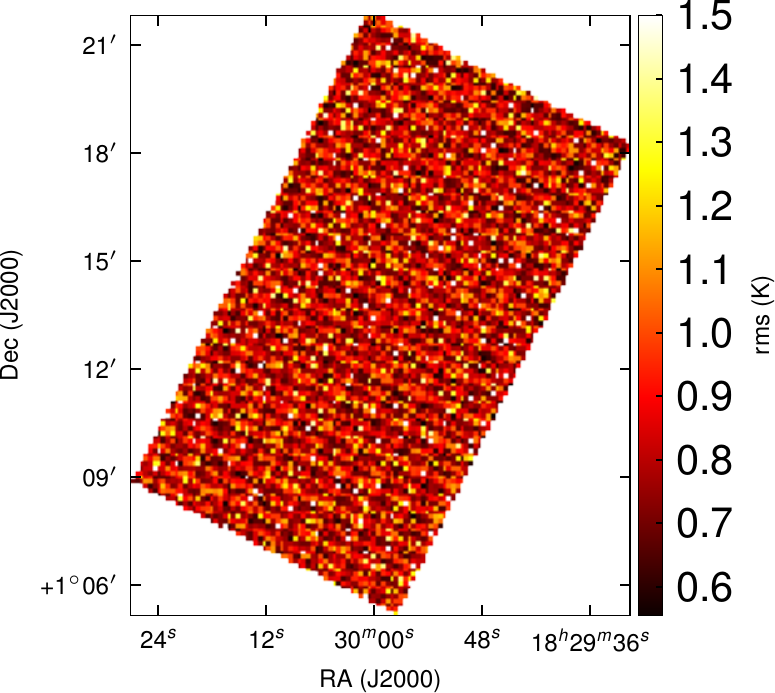}
\caption{The noise on the \thirtco\ map: the root mean square
  integration of a line free region from a nearest-neighbour gridded
  map with 7.27 arcsecond pixels and 1\,\kms\ channels.}
\label{fig:thirteenco_noise}
\end{center}
\end{figure}

Observations of the Serpens molecular cloud in the \jtt\ CO rotational
transition were obtained using HARP at the JCMT. \twelco\ observations
were obtained in April and July 2007, while \thirtco\ and \ceighto\
observations were obtained simultaneously in August 2007.

The telescope's main beam efficiency $\eta_{MB}$ is 0.63, taken from
\citet{Buckle2009}, where it was calculated from observations of
planets.  Each map was taken in `raster-scan' mode. The telescope
continually scans rows back and forth across the science region,
writing out the data every 7.27 arcseconds. The telescope regularly
observes a reference position to correct for the atmospheric
distortion. In order to minimise the variation of the noise across the
map half the observations scan along the longer axis of the
rectangular area being observed, and half along the shorter axis. The
JCMT beam size at 345.796$\,$GHz (\twelco\ \jtt) is 14.6\arcsec, at
330.588\,GHz (\thirtco) it is 15.2\arcsec, and at 329.331\,GHz
(\ceighto) it is 15.3\arcsec.

In total 3.5 hours was spent observing the \twelco\, and 6 hours
observing the \thirtco\ \& \ceighto.  The observations were taken with
a channel width of 0.42\, km\, s$^{-1}$ for \twelco\ and 0.06\, km\,
s$^{-1}$ for \thirtco\ and \ceighto. These were resampled onto
1\,\kms\ for the \twelco\ and 0.1\,\kms\ for the \thirtco\ and
\ceighto. The achieved noise levels on an un-smoothed, nearest
neighbour gridded map were 0.10\,K in 1\,\kms for \twelco\ and 0.93
and 0.88\,K in 0.1\,\kms\ channels for \ceighto\ and \thirtco\
respectively. The \twelco\ has significantly exceeded the survey target
rms level, which is 0.3\,K in 1.0\,\kms\ channels.  However, the
\thirtco\ and \ceighto\ noise levels are significantly higher than the
survey target rms (0.88 and 0.93\,K compared with a target of 0.3\,K
in 0.1\,\kms) therefore more observations are ongoing on this region,
to reduce the co-added noise to the required level. This will allow
more detailed follow up work on the core kinematics; however, the
current noise level is more than adequate to study the bulk gas
conditions and energetics across the entire region.

One of the requirements of the GBS is to achieve reasonable rms
uniformity across the scans. As a multi-detector array, HARP naturally
has some variation in system temperatures for each detector, and
additionally during the Survey science verification phase (in which
these data were taken) not all detectors in the 4x4 array could be
used, therefore completely even noise was not achieved. We illustrate
the uniformity of the noise achieved in
Fig.~\ref{fig:thirteenco_noise}, which shows the rms values across the
\thirtco\ map. As can be seen, the variation is low and although a
grid pattern can be seen, it is constant across the map.

After the data were taken, the scans were analysed and for each scan
the detectors showing anomalous baseline structure or excessive noise
were removed from the reduction.  The remaining data had a first order
baseline fitted to line free regions of the spectra. This baseline was
removed and the data were co-added, using nearest neighbour
gridding. The rest of the images shown here and used for analysis are
gridded onto 3 arcsecond pixels. The \twelco\ data from each detector
were re-scaled to correct for striping caused by varying detector
responses \citep[see][for technique]{Curtis2010}.  A
Gaussian kernel (using a FWHM of 9\arcsec) was used to grid these
cleaned raw files into position-position-velocity cubes for each
isotopologue (giving an effective beam size on the maps of 17\arcsec). 
The \ceighto\ and \thirtco\ data files were gridded into separate
files for each scan using a 10\arcsec\ FWHM Gaussian kernel, then the
resulting files were mosaiced together using an 8\arcsec\ FWHM
Gaussian kernel (giving an effective beamsize for the maps of 20\arcsec).  
\textrm{This
larger effective beamsize was used in order to increase the signal to
noise of the maps.}  In the \twelco\ reduced cube there was an
anomalous baseline feature present in channels from 35.8 to 44.8
\kms. The shape was similar to other known instrumental effects in
HARP/ACSIS data at the time the data were taken. The affected
velocity region was blanked out, and then these channels were
interpolated from surrounding data. Across most of the map there is no
emission at these affected velocities, with only a small region
containing the tail of a high velocity outflow having visible
emission. The structure of the high velocity tail affected in these
spectra appears to be well matched by the simple interpolation.

The resulting noise measurement for the Gaussian smoothed maps used in
the scientific analysis is 0.017\,K for the \twelco\ in 1\,\kms\ channels,
and 0.19\,K for both the \thirtco\ and \ceighto\ respectively, in 0.1\,\kms\
channels.

\subsection{\twelco\ map}
The \jtt\ transition of the CO isotopologues observed here have
critical densities of the order of 10$^{4-5}$\,cm$^{-3}$. Tracing
these high densities allow us to look at the dense gas more intimately
involved with the ongoing star formation than lower transitions,
tracing lower densities, allow. It also allows us to examine the higher
temperature emission from molecular outflows, as these transitions
occur at a correspondingly higher energy level -- 31-33\,K -- than the lower transitions. 

\begin{figure*}
  \begin{minipage}[c]{4.44in}
    \includegraphics{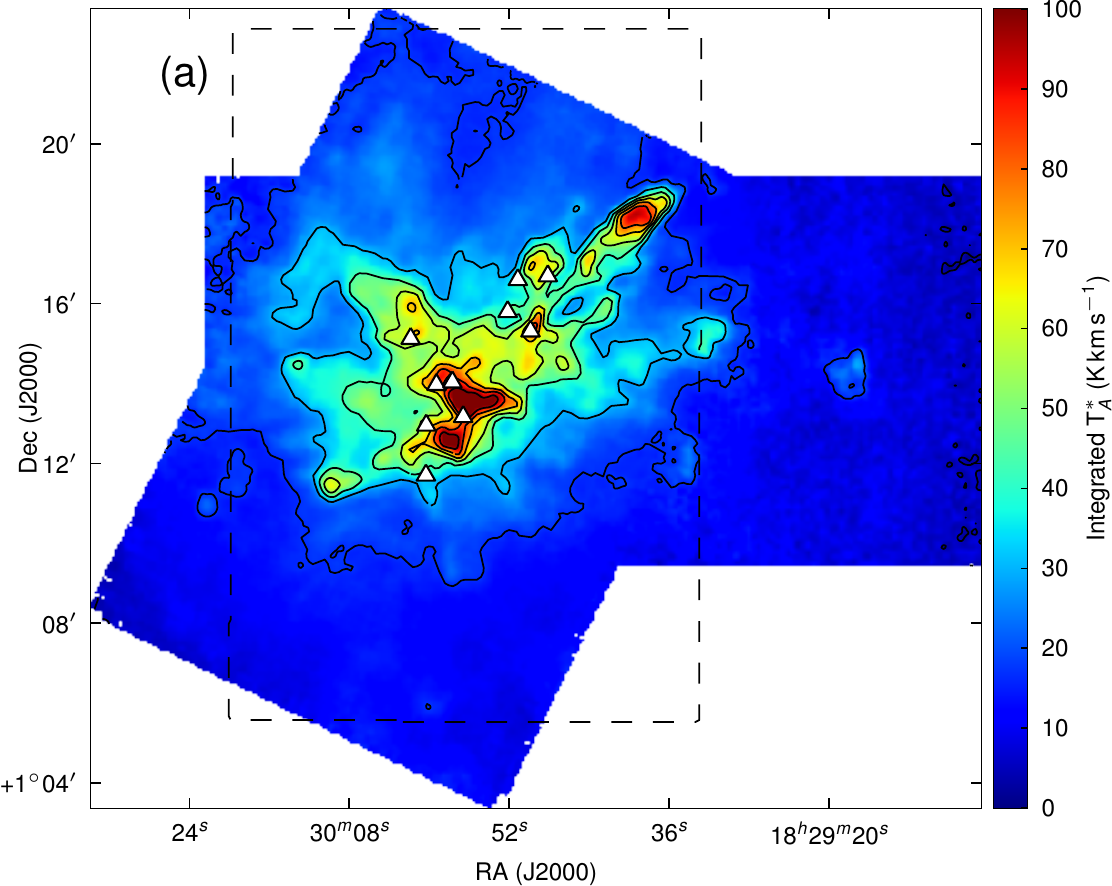}
  \end{minipage}
  \begin{minipage}[c]{2.50in}
     \includegraphics{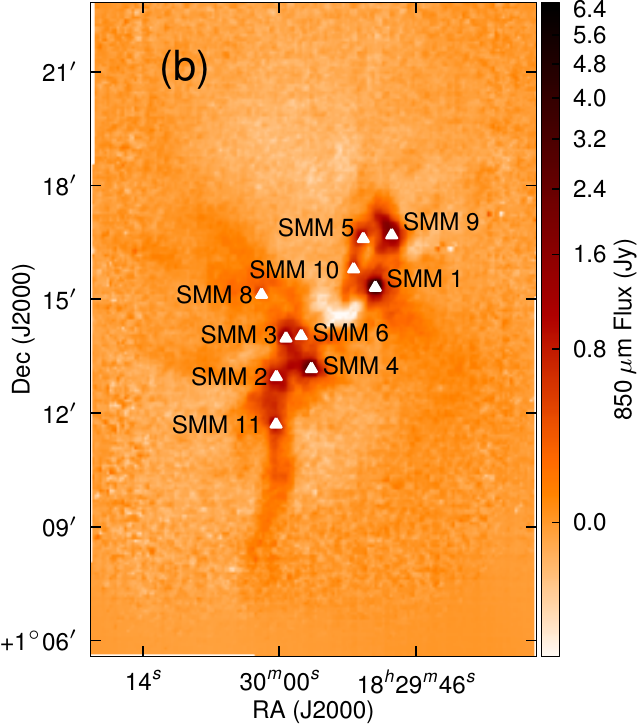}
  \end{minipage}
  \newline
  \vspace{0.2cm}
  \begin{minipage}[c]{1.0\textwidth}
    \includegraphics{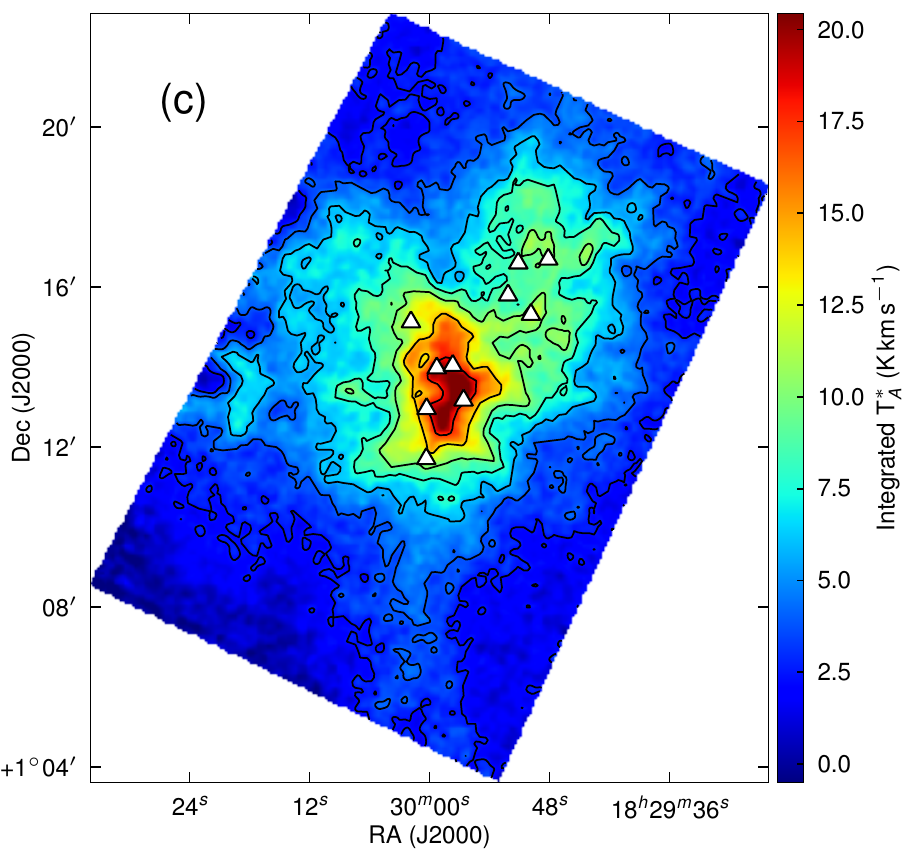} 
    \includegraphics{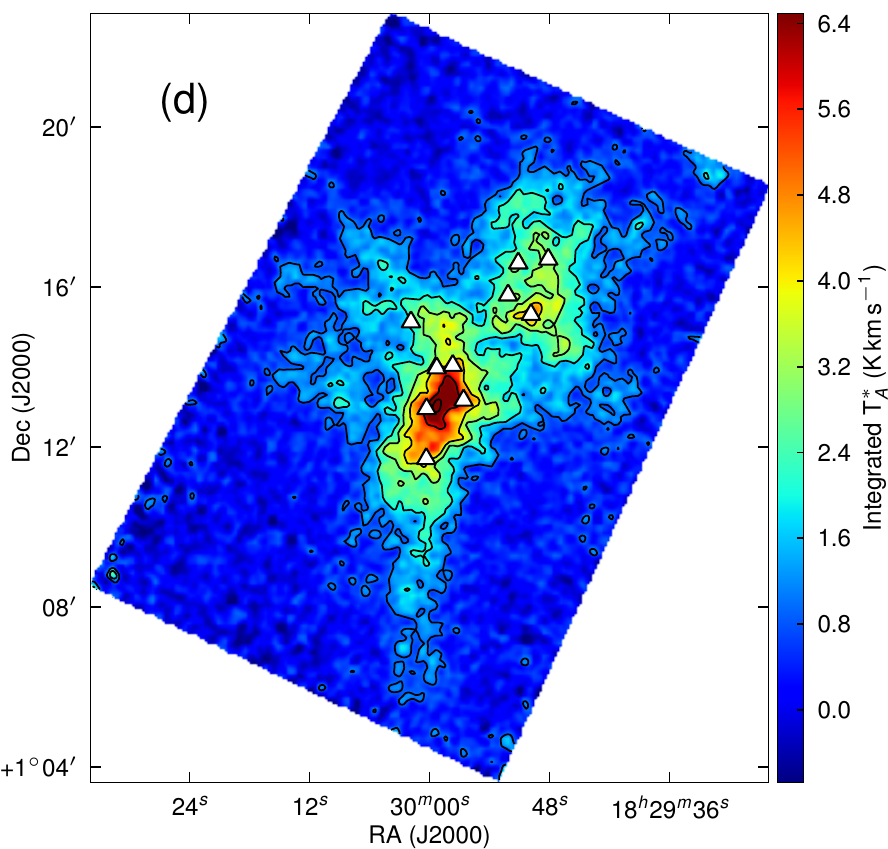}
  \end{minipage}
 
  \caption{Presentation of the CO integrated intensity images. For
    comparison the SCUBA data from \citet{Davis1999} is shown in
    (b). The \ceighto\ image qualitatively follows the SCUBA
    emission.\label{fig:data} (a) Integrated \twelco\ \Tstara\ emission
    in K\, \kms. Integrated from -1.12 to 15.88\,\kms. The region
    observed with SCUBA is outlined in grey. (b) SCUBA 850\,\micron\
    emission \citep{Davis1999}.  Positions of the known submillimetre
    cores are shown and labelled. (c) Integrated \thirtco\ \Tstara\
    emission in K\,\kms. Integrated from 4.54 to 10.44\,\kms. (d)
    Integrated \ceighto\ \Tstara\ emission in K\,\kms. Integrated from
    5.55 to 9.95\,\kms.}
\end{figure*}

\begin{figure*}
  \centering
  \includegraphics[width=4.0in]{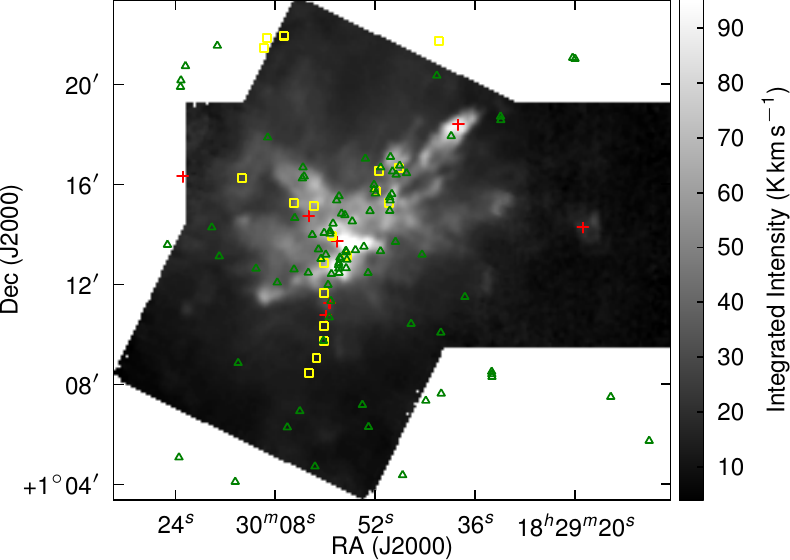}
  \caption{Integrated \twelco\ emission, with SCUBA cores
    \citep{DiFrancesco2008} shown as yellow squares, c2d Spitzer YSO
    candidates \citep{Evans2009} shown as green triangles, and HH
    objects shown as red crosses \citep{Davis1999}. The 850\micron
    SCUBA emission from \citet{Davis1999} is shown in contours.}
  \label{fig:12c-overlay}
\end{figure*}

Fig.~\ref{fig:data} (a) shows the integrated \twelco\ emission;
Fig.~\ref{fig:12c-overlay}  overlays IR
sources, submillimetre continuum sources and Herbig Haro (HH) objects. As is evident
from the integrated emission alone, the region is a complex, clustered
star-forming region. The `finger-like' extensions so common to CO
observations in many regions are clearly discernible; however, it should be noted
that these do not always correspond to obvious spatially
compact bipolar outflows.

\begin{figure}
\begin{center}
\includegraphics[angle=-90,width=1.6in]{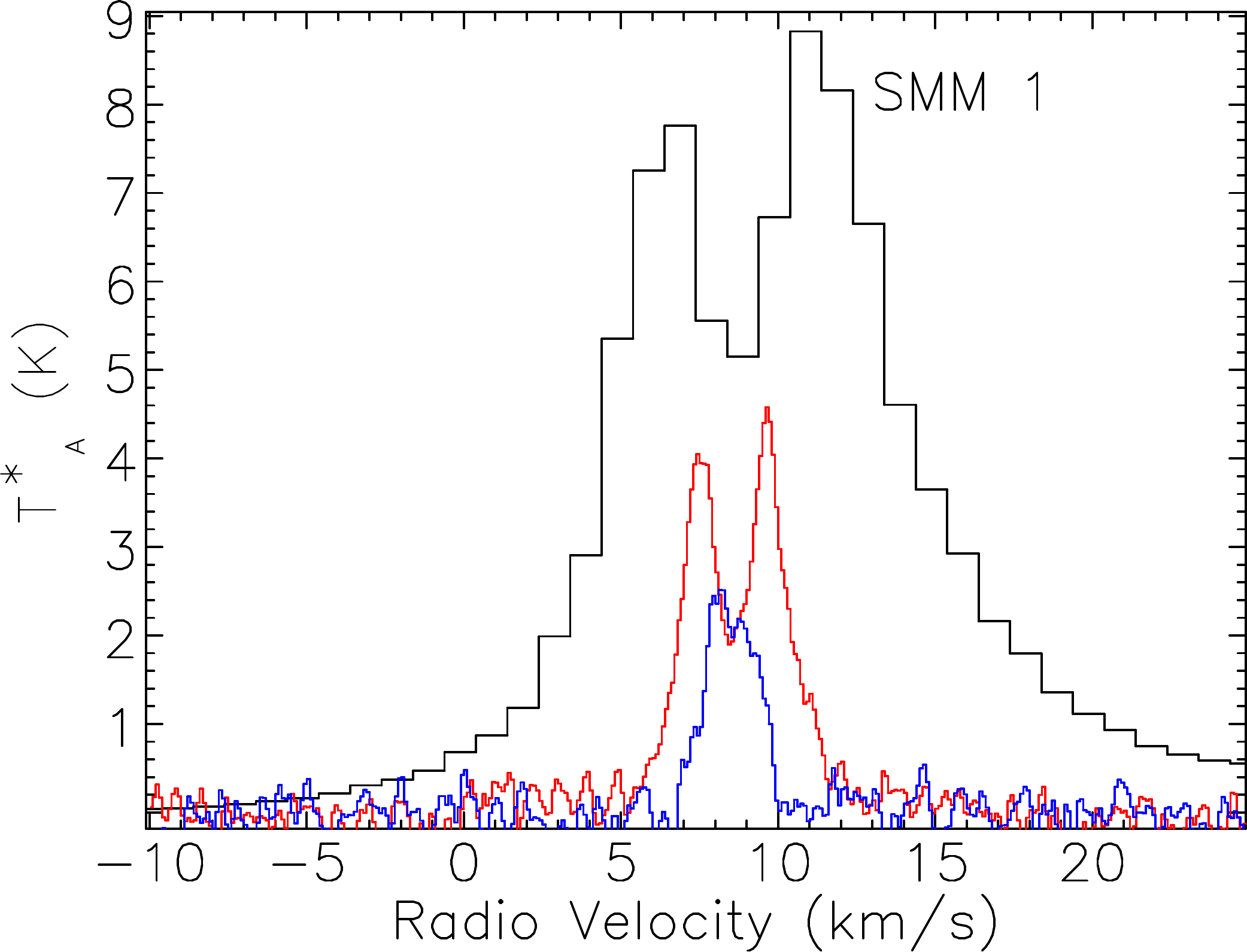}
\includegraphics[angle=-90,width=1.6in]{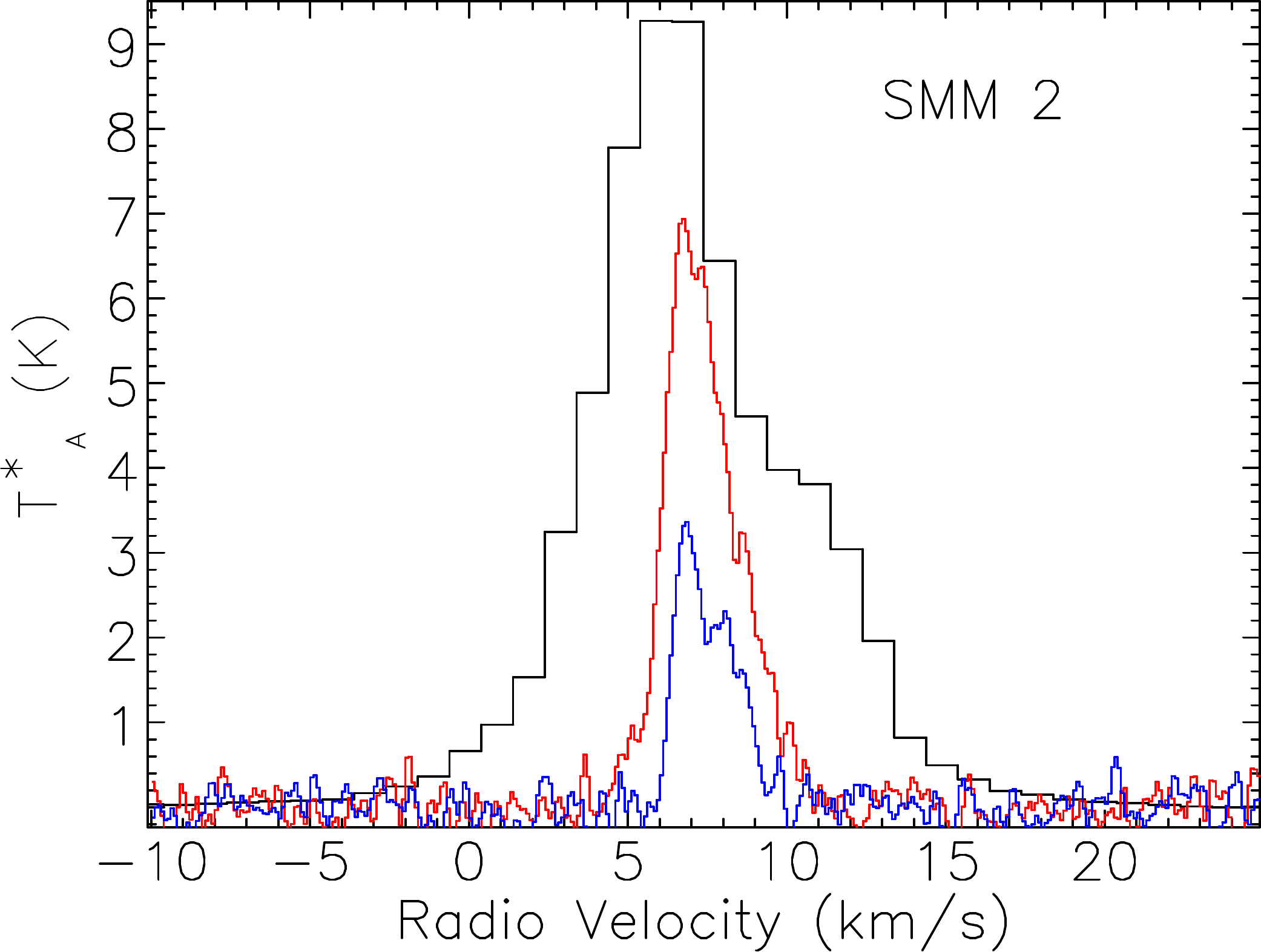}\\
\vspace{3mm}
\includegraphics[angle=-90,width=1.6in]{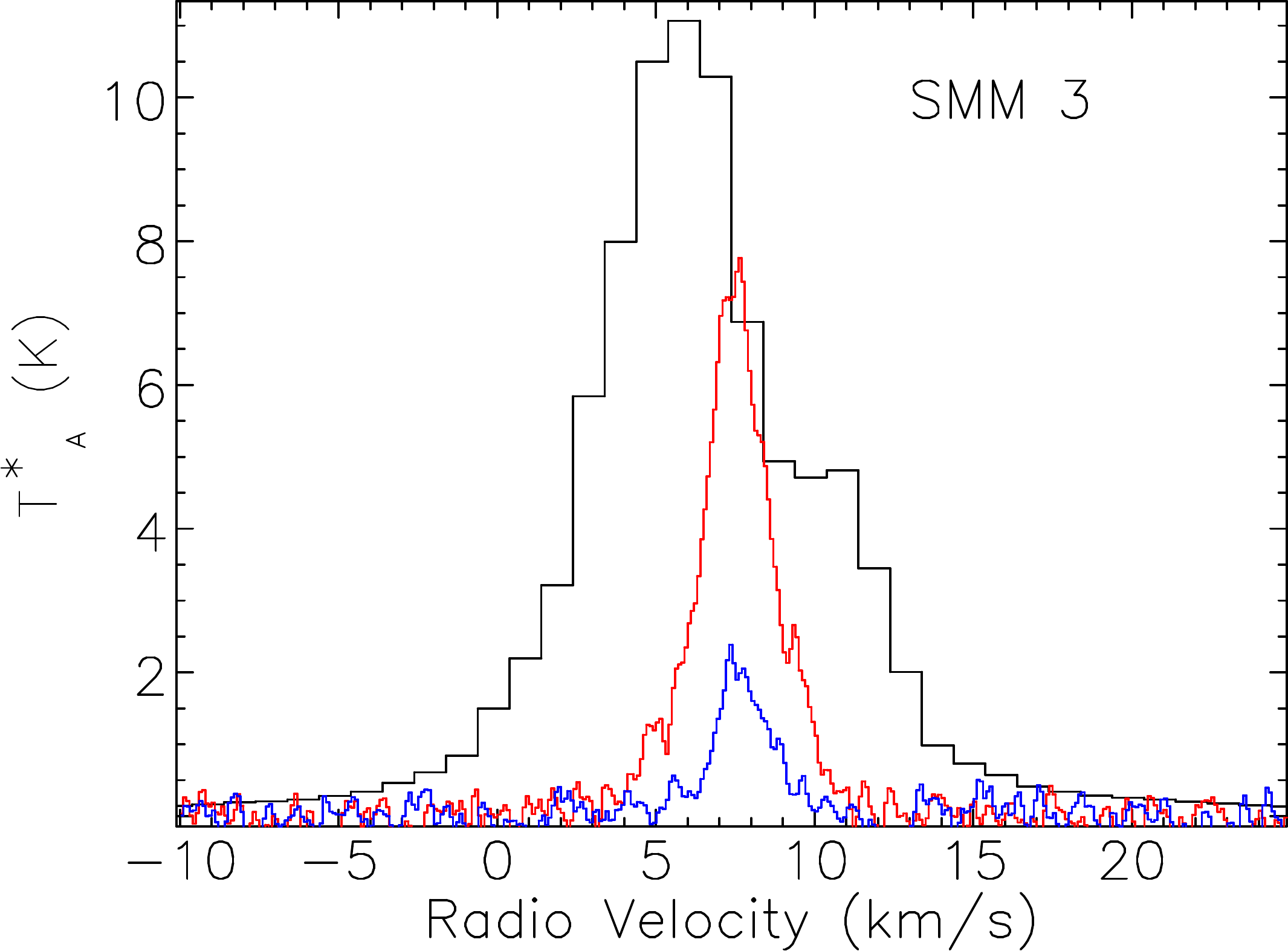}
\includegraphics[angle=-90,width=1.6in]{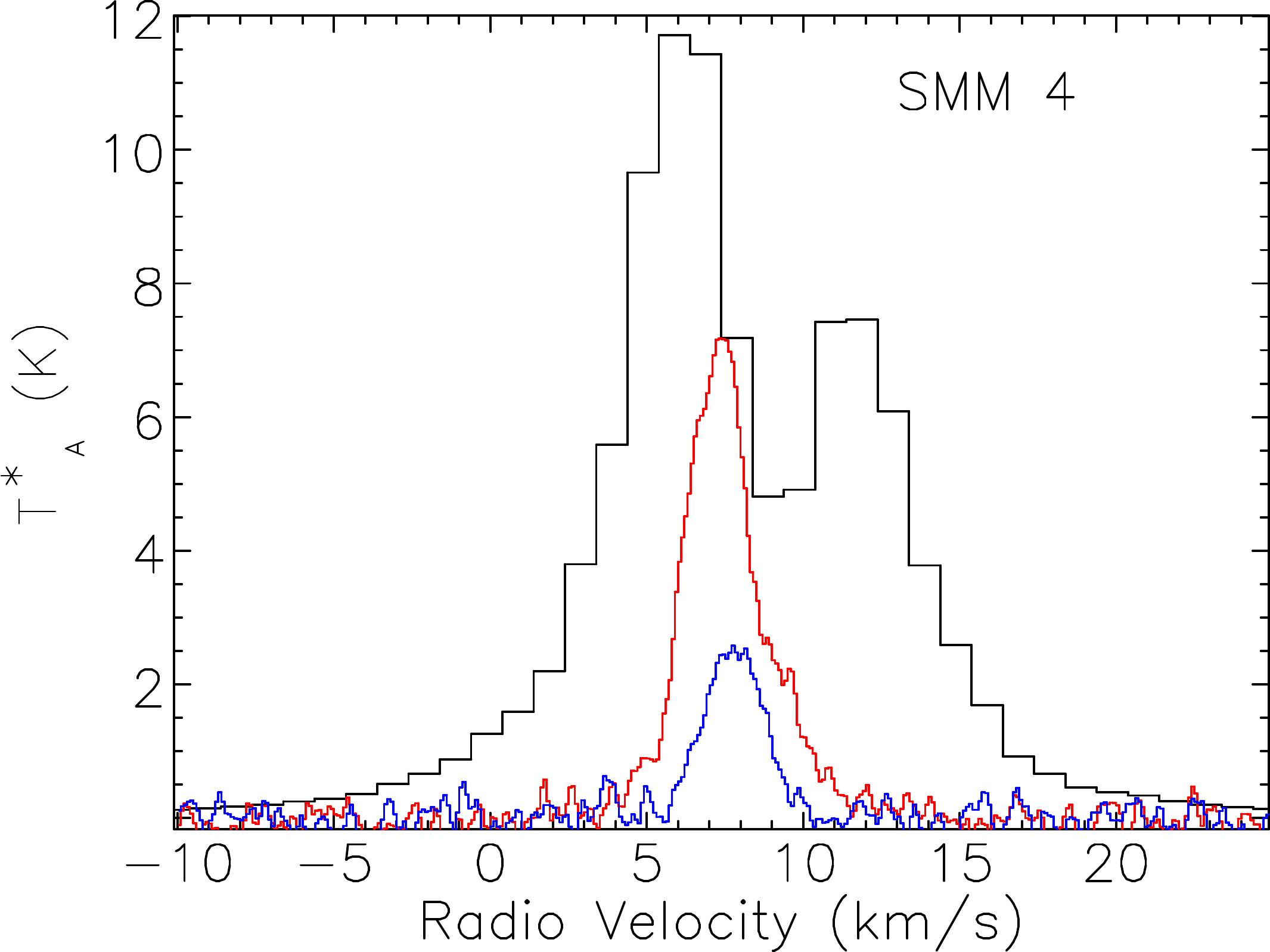}\\
\vspace{3mm}
\includegraphics[angle=-90,width=1.6in]{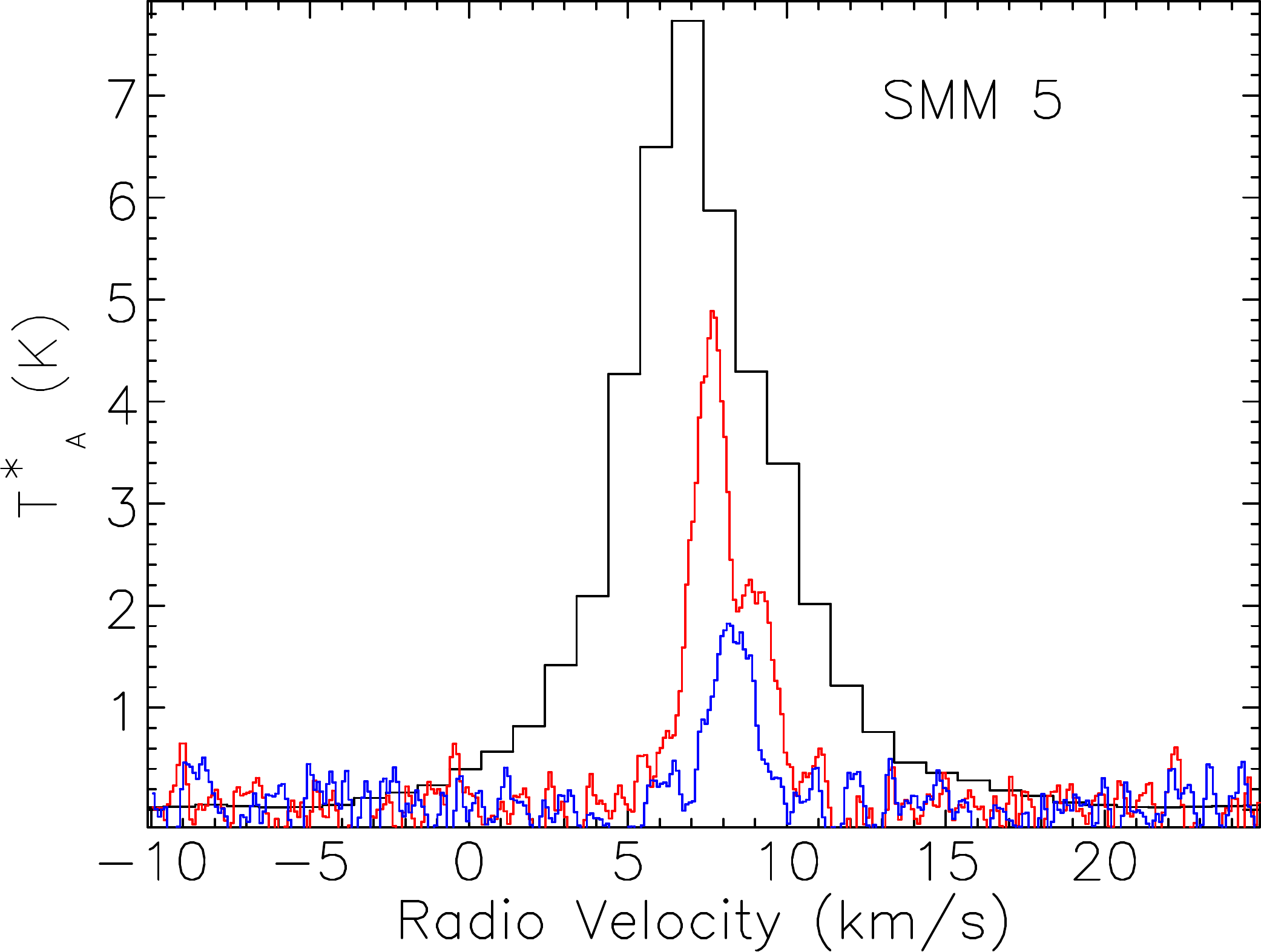}
\includegraphics[angle=-90,width=1.6in]{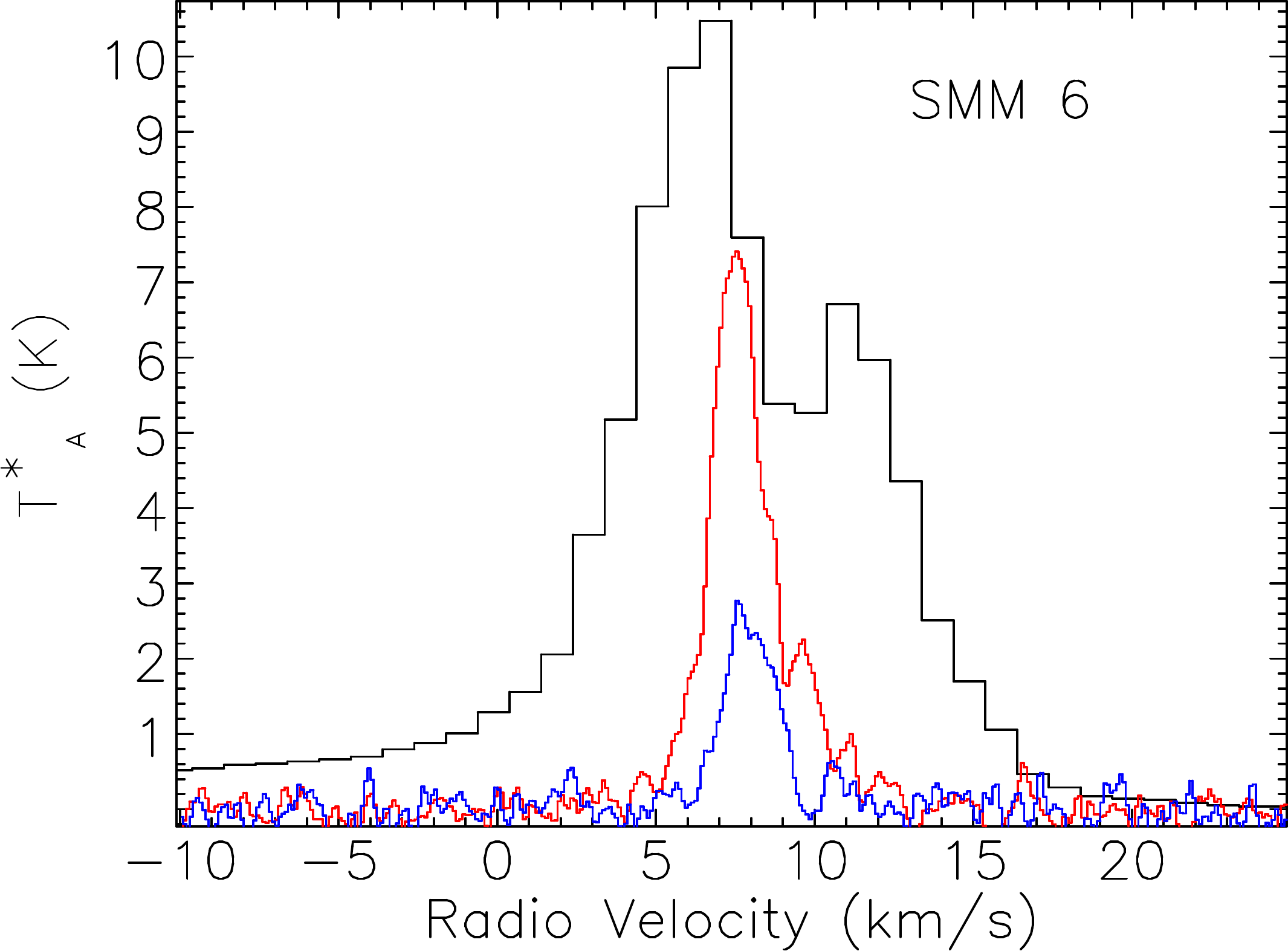}\\
\vspace{3mm}
\includegraphics[angle=-90,width=1.6in]{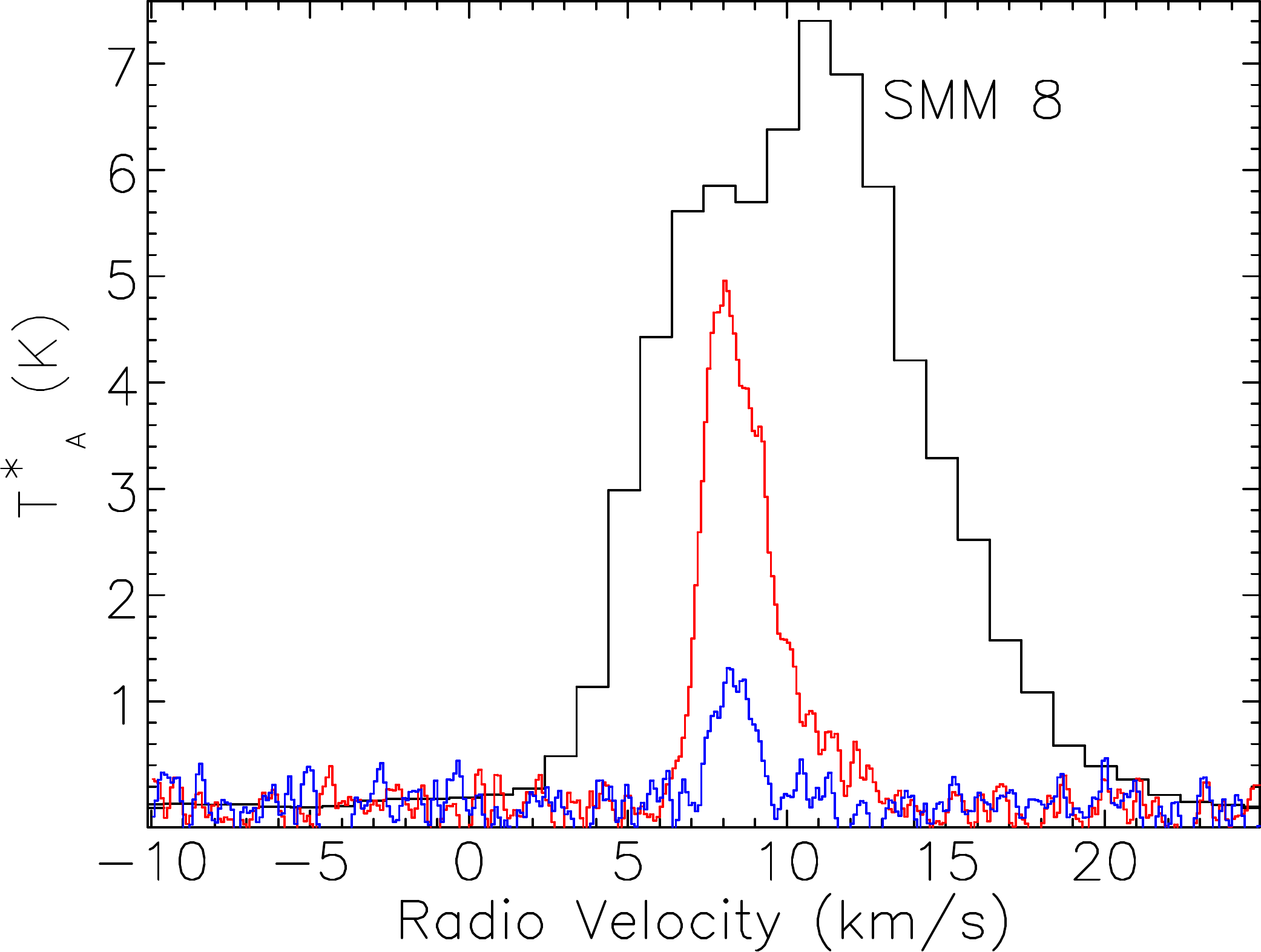}
\includegraphics[angle=-90,width=1.6in]{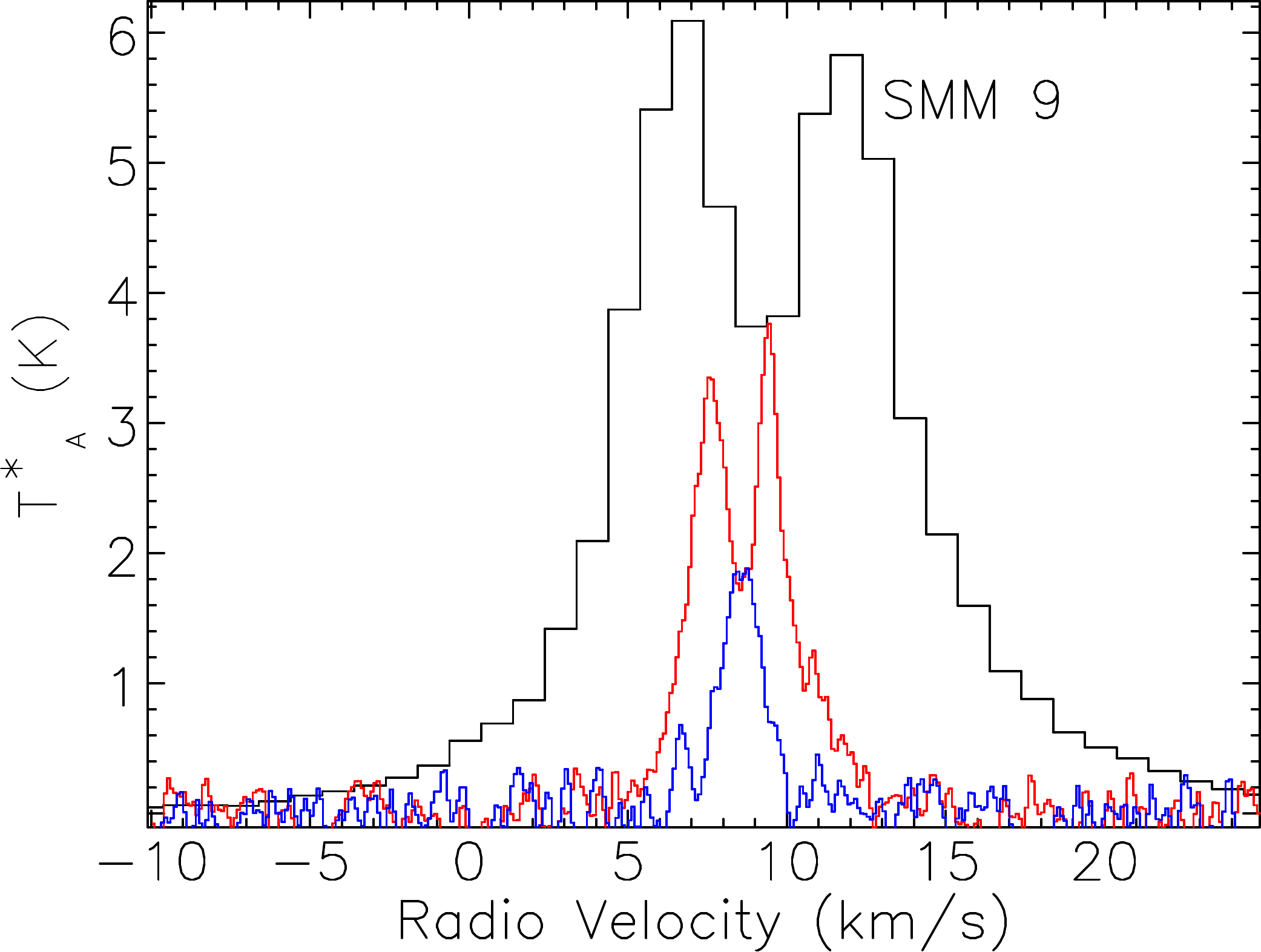}\\
\vspace{3mm}
\includegraphics[angle=-90,width=1.6in]{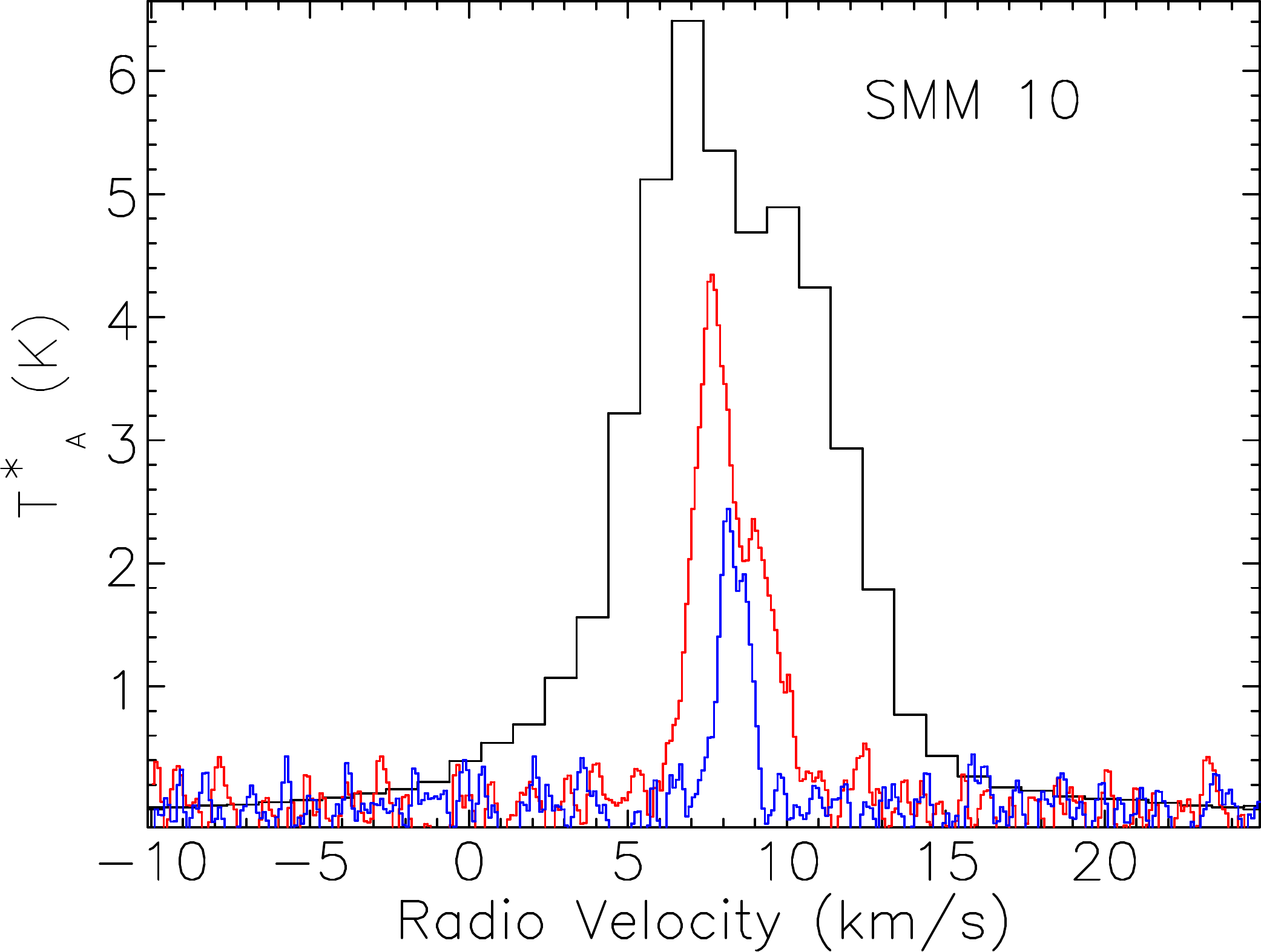}
\includegraphics[angle=-90,width=1.6in]{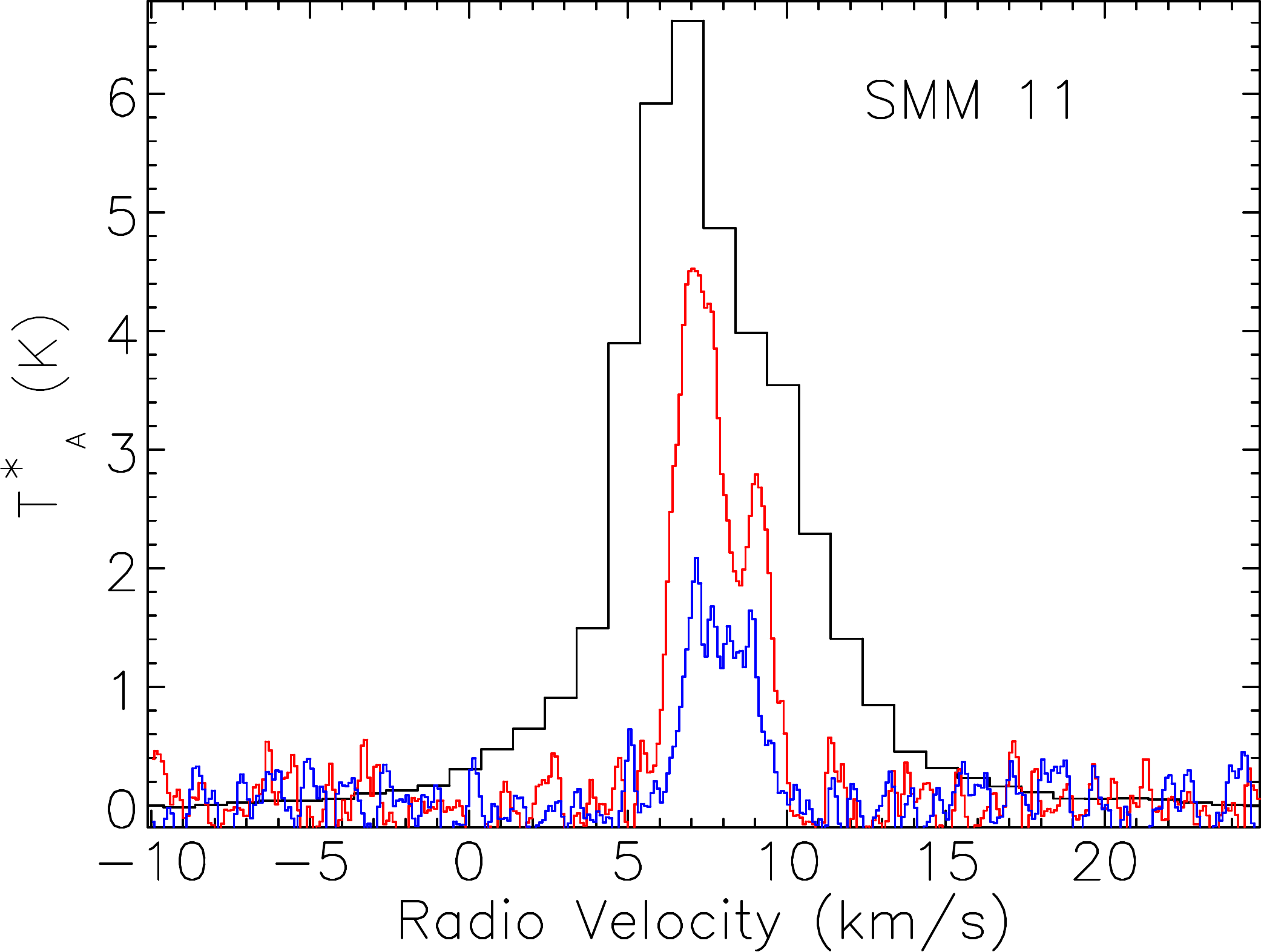}

\caption{Spectra from the 10 SMM positions shown in
  Fig.~\ref{fig:data}.  All three isotopologues are shown on the same
  axes. \twelco\ is the highest intensity spectra (black), and \ceighto\ is
  the weakest (blue). \thirtco\ is shown in red.  \label{co_spectra}}
\end{center}
\end{figure}

\begin{figure}
  \centering
  \includegraphics[angle=-90,width=1.6in]{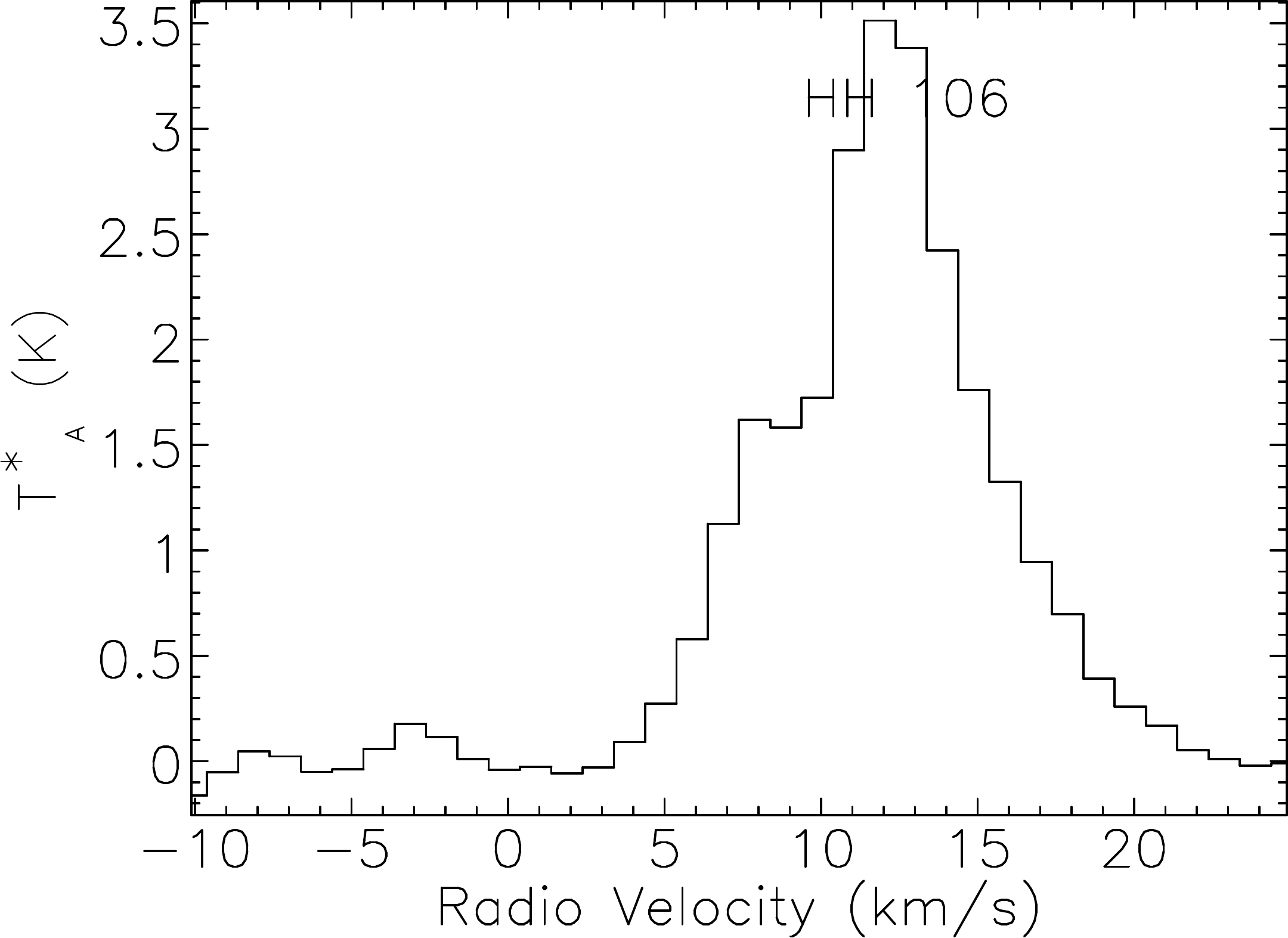}
  \includegraphics[angle=-90,width=1.6in]{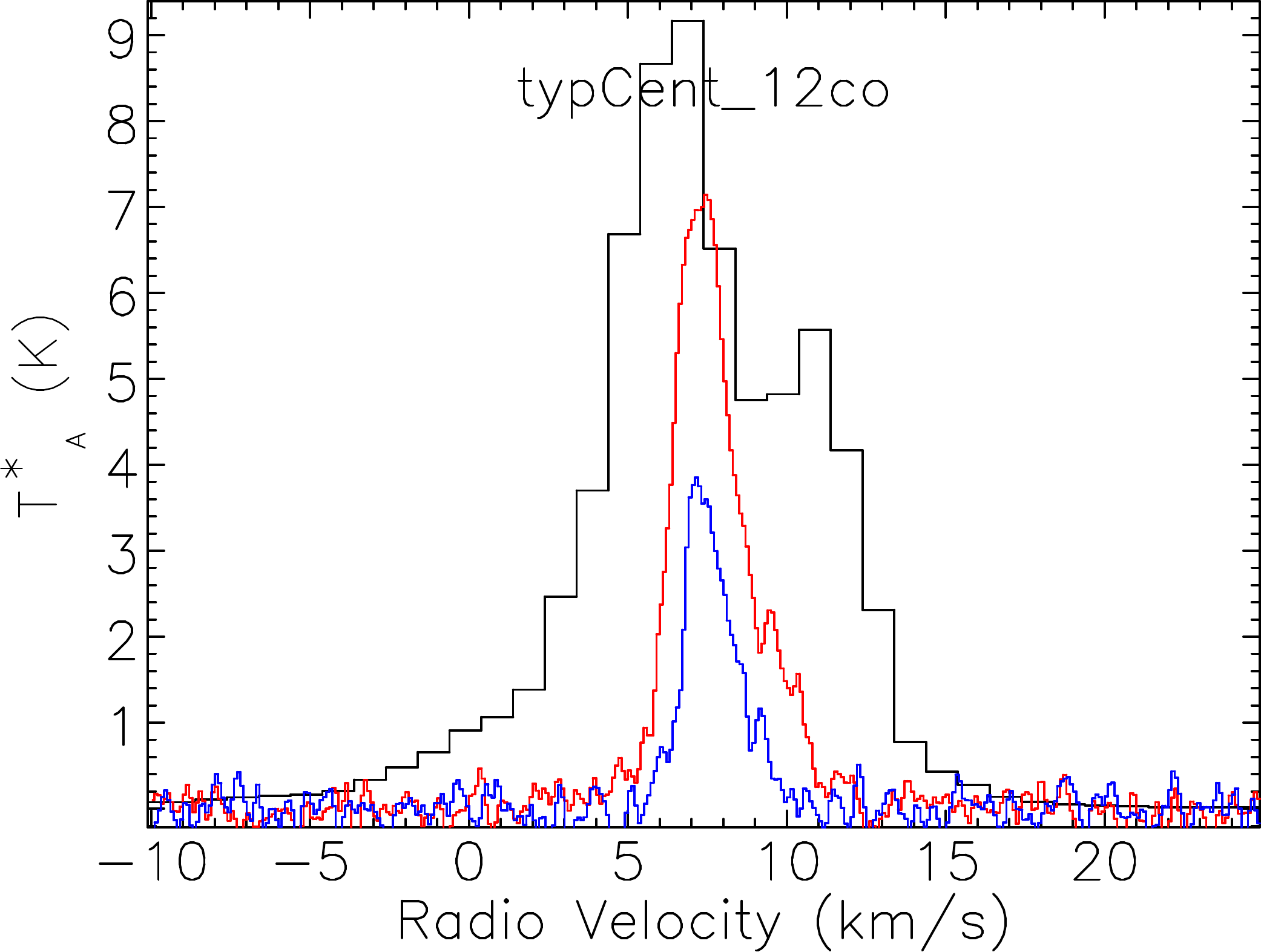}\\
  \vspace{3mm}
  \includegraphics[angle=-90,width=1.6in]{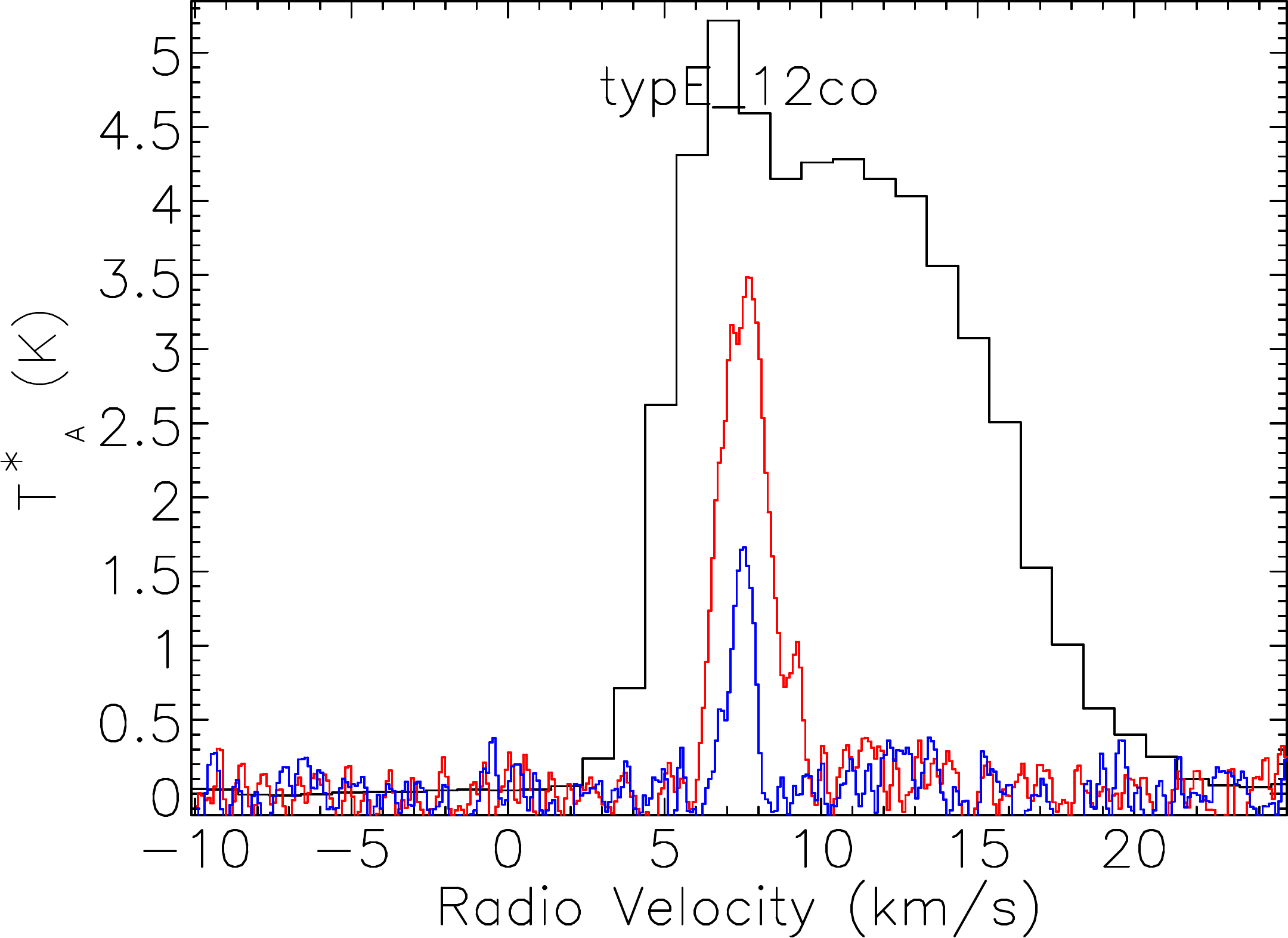}
  \includegraphics[angle=-90,width=1.6in]{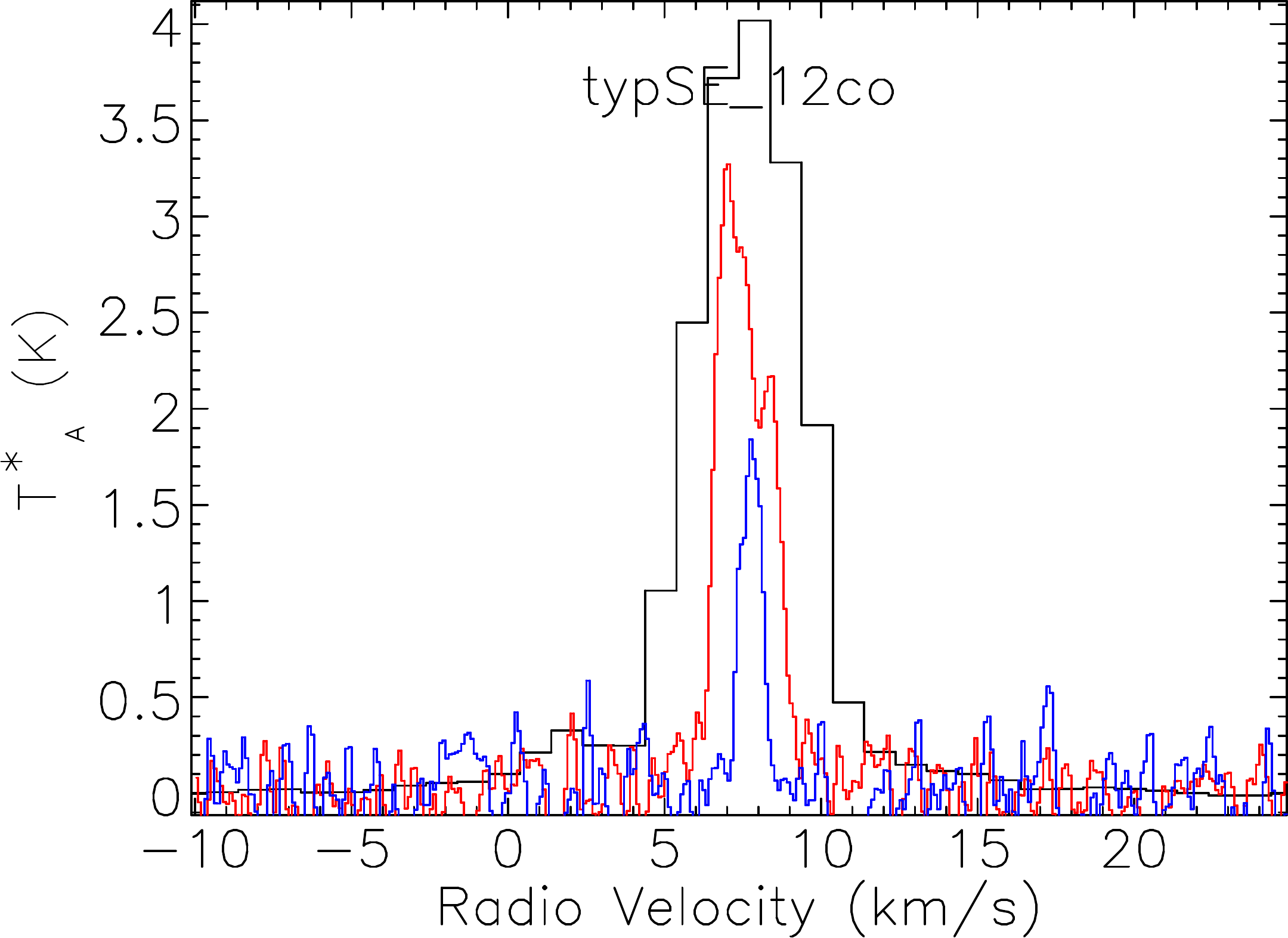}\\

  \caption{Spectra from HH 106 (\twelco\ only) and 3 typical non-core positions in the
    cloud. These were located at: 18:29:58.4 +1:13:21 (top right),
    18:30:05.4 +1:16:30 (bottom left) and 18:30:03.4 +1:10:51.0
    (bottom right). Colours as for Fig.~\ref{co_spectra}}
  \label{fig:extra_spec}
\end{figure}

Our \twelco\ \jtt\ map covers a region of 260 square arcminutes
centred on RA:$18^{h}29^{m}45^{s}$ Dec:$1\degr 13\arcmin 57\arcsec $
(J2000). The region was extended westwards to follow the strong
bow-shock that 
culminates  in 
the Herbig Haro object HH 106. This extended section has higher
noise and was not followed up in \thirtco\ and \ceighto.  The
integrated intensity map in Fig.~\ref{fig:12c-overlay} is not
dominated by emission at the systemic velocity of the cloud,
presumably at least in part due to the strong self absorption present
in the \twelco\ spectra. Fig.~\ref{co_spectra} presents the spectra
for all three isotopologues plotted on the same axes for each of the
SMM cores, and Fig.~\ref{fig:extra_spec} shows spectra from some
typical positions throughout the cloud. Many of the \twelco\ spectra exhibit either a double-peaked
profile or a distorted Gaussian shape. By comparing the peak intensity
and shape of the \ceighto\ spectra from the same position as the \twelco, it is clear
that the shapes of the \twelco\ line profiles appear to be caused by self absorption
rather than multiple velocity components.

The \twelco\ emission is, therefore, dominated by the less dense blue
and red-shifted high velocity gas of the cloud.  The absorption
maximum appears slightly red shifted with respect to the local rest
velocity--perhaps indicative of global infall in the cloud, as one
would expect in a star forming region. There are also prominent
velocity wings present in the \twelco\ spectra as can been seen in
Fig.~\ref{co_spectra}, also indicating the presence of molecular
outflows in the region, some with velocities up to
20$\,$km$\,$s$^{-1}$ with respect to the local rest velocity.  These
outflows have been observed before with molecular tracers
\citep{Olmi2002,McMullin2000,Davis1999,White1995} and with masers
\citep{Moscadelli2005}; however the data presented here improve on the
resolution of the previous single dish CO studies. Recent \textit{Spitzer}
observations \citep{Harvey2007a,Harvey2007} have identified several
deeply embedded sources within the Serpens molecular cloud--we
would expect these sources to be of a Class 0 or early Class I nature
and thus drive powerful molecular outflows.  Fig.~\ref{fig:12c-overlay} shows
the integrated \twelco\ emission, with the positions of YSO
candidates, SCUBA cores and HH objects also displayed. We can see that
the \textit{Spitzer} deep sources align with the sources in the \dust\ SCUBA
image, which are known in the literature as the 10 SMM sources, SMM
1-11, although note that there is no SMM 7. SMM 7 was first identified
by \citet{White1995} at RA: $18^h29^m 55.6^2$ Dec:$+01\degr 14\arcmin
52\arcsec$, but was not detected in the SCUBA map of
\citet{Davis1999}. There is not a clear \ceighto\ emission peak
present at its position in Fig.~\ref{fig:data}(d).

\begin{figure*}
\begin{center}
\includegraphics{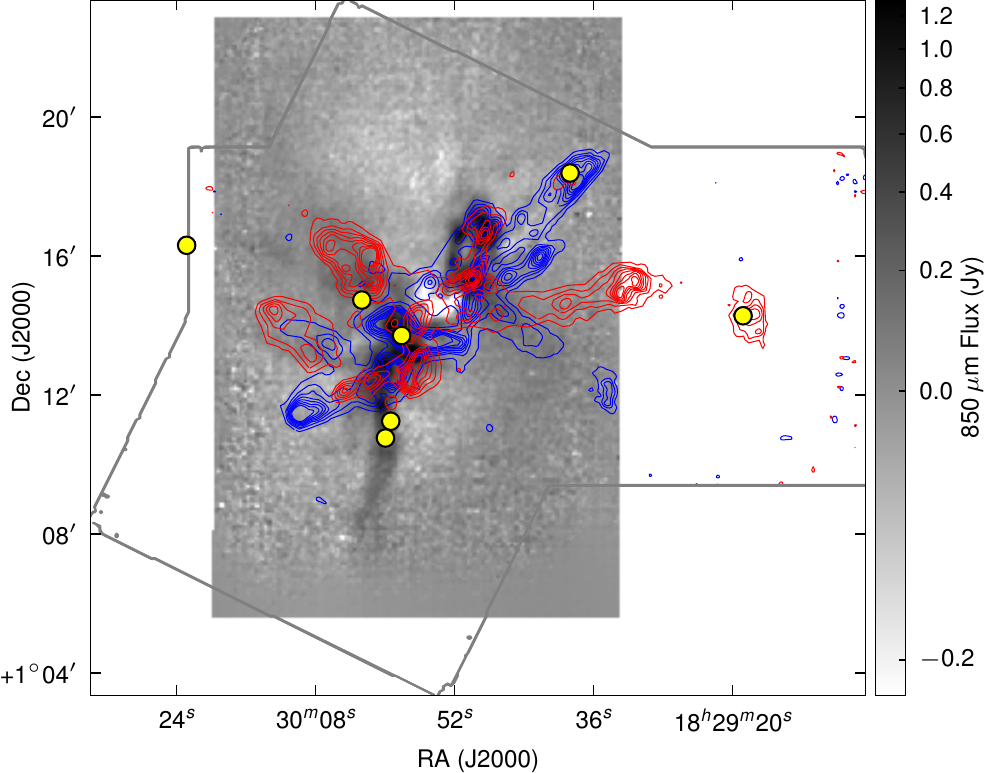}
\caption{Grey-scale image is the SCUBA 850\,\micron\ emission from
  \citet{Davis1999}. The blue contours represent blue shifted \twelco\
  gas and red contours represent red shifted \twelco\ gas. The contour
  levels go up in 2.5\,K\,\kms\ steps, starting from 2.5\,K\,\kms. The
  integration ranges are -4 to 4 km s$^{-1}$ for blue shifted gas and
  14 to 22 km s$^{-1}$ for red shifted gas. The positions of known HH
  objects (from \citet{Davis1999} are shown as circles, see
  Fig.\ref{fig-chanmaps} for labelling).\label{fig:co_outflows}}
\end{center}
\end{figure*}

\begin{figure}
  \centering
  \includegraphics[width=3.2in]{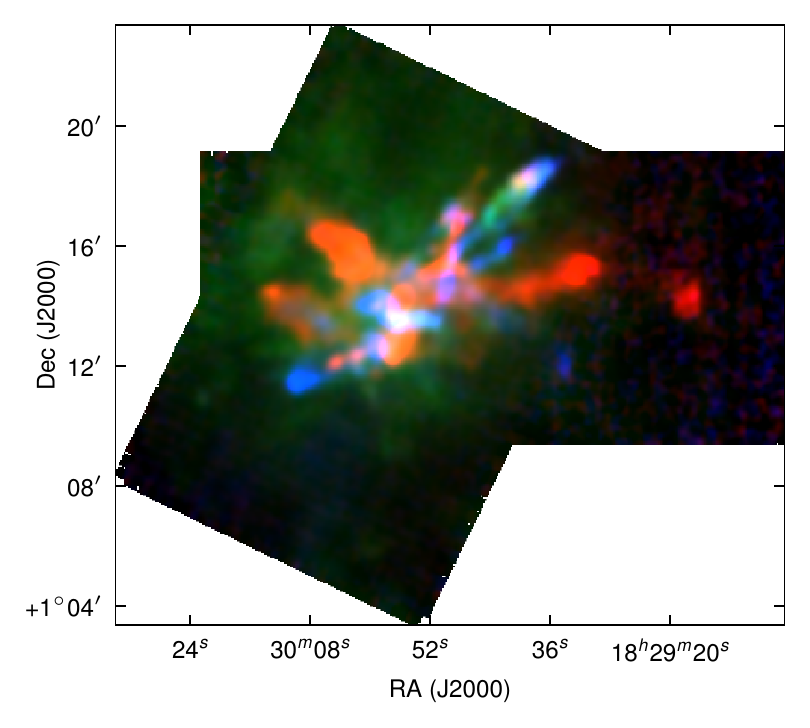}
  \caption{Red-green-blue image of integrated intensity structure in
    \twelco. Red: 11.8-57.8\,\kms. Green: 4.8-10.8\,\kms. Blue:
    -42.2-3.8\,\kms. }
  \label{fig-rgbvel}
\end{figure}

Fig.~\ref{fig:co_outflows} shows the SCUBA 850 \micron\ emission from
\citet{Davis1999} with contours of red and blue shifted gas in
\twelco\, demonstrating that there are a large number of molecular
outflows associated with the Serpens sources.  These maps are at a
higher resolution than previous \twelco\ studies, allowing more
accurate flow and source identification,
compared to the previous results of \citet{Davis1999}, where some of
the flows are unresolved. Fig.\ref{fig-rgbvel} shows another view of
the \twelco\ velocity structure of the cloud. In this three-colour
image showing the red-shifted, blue-shifted and ambient emission
(green) the strong red and blue lobes can be clearly seen.

\subsection{\thirtco\ and \ceighto\ maps}

\begin{figure}
  \includegraphics{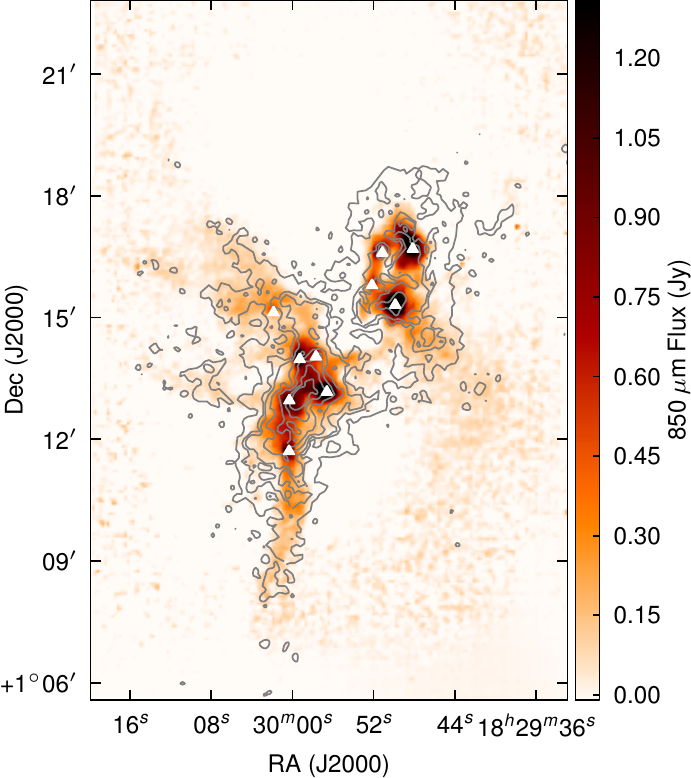}
  \caption{SCUBA 850\,\micron\ emission from \citet{Davis1999} with
    \ceighto\ integrated intensity contours overlaid. Contour levels
    are at 1-7 K\,\kms\, in 1\,K\,\kms\ steps, and the integration was
    from 5.3 to 10.6\,\kms.  \label{fig-scubac18o}}
\end{figure}

The isotopologues \thirtco\ and \ceighto\, are much less abundant than
\twelco, with the abundance ratio for the `local'\footnote{Measured in
  the Orion GMC} ISM being calculated as
$X^{{\twelco}}/X^{{\thirtco}}=77$ \citep{Wilson1994}. Being less
abundant, these species are only seen on the denser parts of the
sub-clusters, allowing us to probe the motions of the gas more closely
associated with the protostars rather than the outflowing gas.  Our
\thirtco\ and \ceighto\ maps (Fig.~\ref{fig:data}) cover a region of
17$\times$11 arcminutes centred on RA\,$18^{h}29^{m}58^{s}$, Dec\,
$1^{\degr}13^{\arcmin}02^{\arcsec}$(J2000). Fig.~\ref{fig-scubac18o}
shows the SCUBA 850 \micron\ emission and overlays contours of the
\ceighto\ integrated intensity emission. We can see the \ceighto\
emission traces the general shape of the dust and the filaments to the east and
south identified by \citet{Davis1999} very clearly, although
specific peaks do not align exactly with the position of the SCUBA
cores. The \ceighto\ spectra (in Fig.~\ref{co_spectra}) show some
evidence of multiple components along the line of sight. This
suggestion is strengthened by the analysis of \citet{DuarteCabral2010},
who suggest that there are multiple velocity components and that the
whole cloud can be modelled as two colliding flows or
sub-clouds. These multiple velocity components from the \ceighto\
emission should not be confused with the double-peaked structure of
much of the \twelco\ emission. This occurs at very different velocity
scales, hence our belief that it is due to self absorption of the line by nearer,
cooler gas.

The correlation between the \ceighto\ and the continuum emission
suggests that the \ceighto\ emission is not optically thick in the
bulk of the emission. This is also consistent with the CO spectra
shown in Fig.~\ref{co_spectra} where the \thirtco\ still shows some
self absorption from colder red shifted gas as seen in the \twelco\
spectra, whereas the \ceighto\ spectra do not show such a clear
pattern -- although the presence of multiple
components does make this less clear.
In Sec.~\ref{sec-optdepth} we investigate the opacities of the
isotopologues across the cloud.

It can be seen from these observations (see Fig.~\ref{fig:data}) that
the \twelco\ and \ceighto\ emission are tracing different
gas structures -- the \ceighto\ emission looks very similar to the
SCUBA emission, tracing a clumpy medium containing dense cores. The
\twelco\ emission on the other hand is bright throughout the 
cloud, but is dominated by lobe-like filaments extending out from
the cloud rather than clumpy emission in the centre of the cloud. The
\twelco\ does not appear to be tracing the SMM cores. This suggests
that the \twelco\ emission probably becomes optically thick in the gas
surrounding the cores, and does not see all the way to the central infrared
dense cores.

\begin{figure}
  \centering
  \includegraphics[width=3.2in]{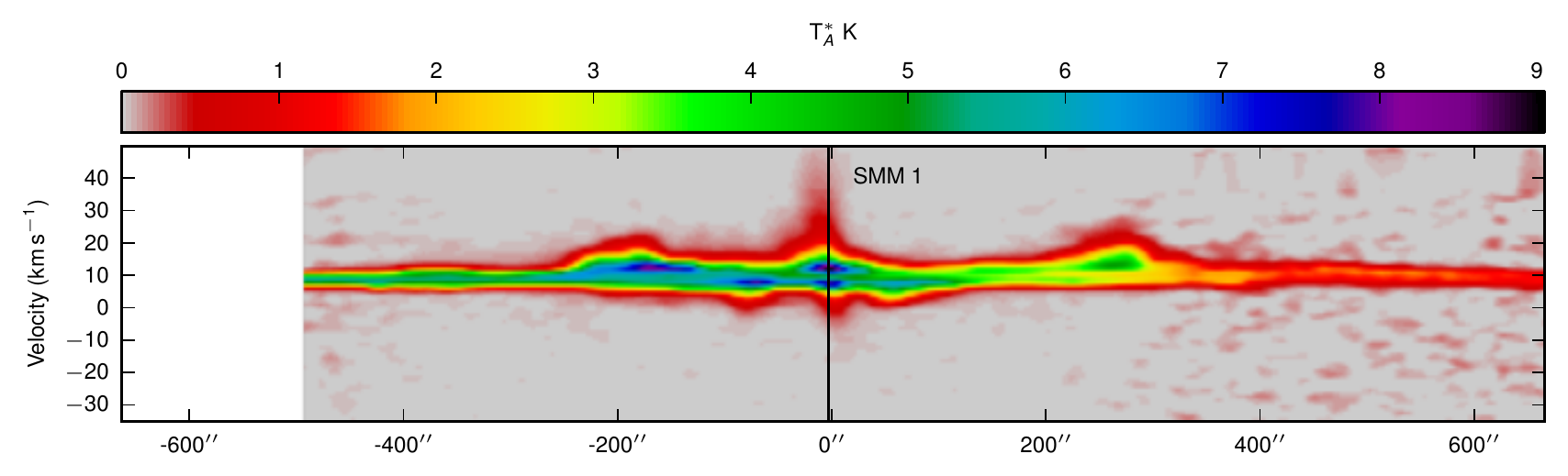}\\
  \includegraphics[width=1.6in]{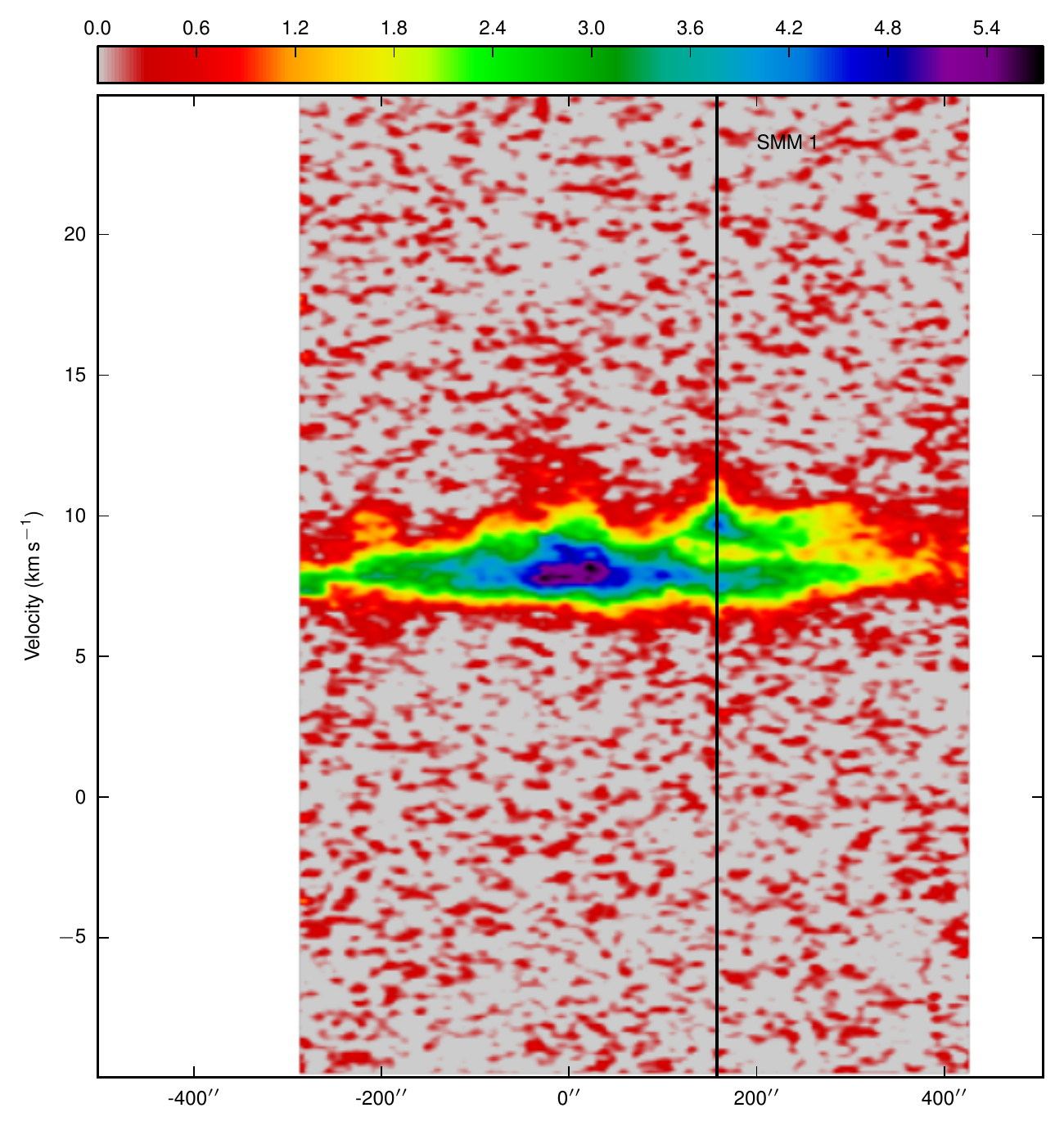}
  \includegraphics[width=1.6in]{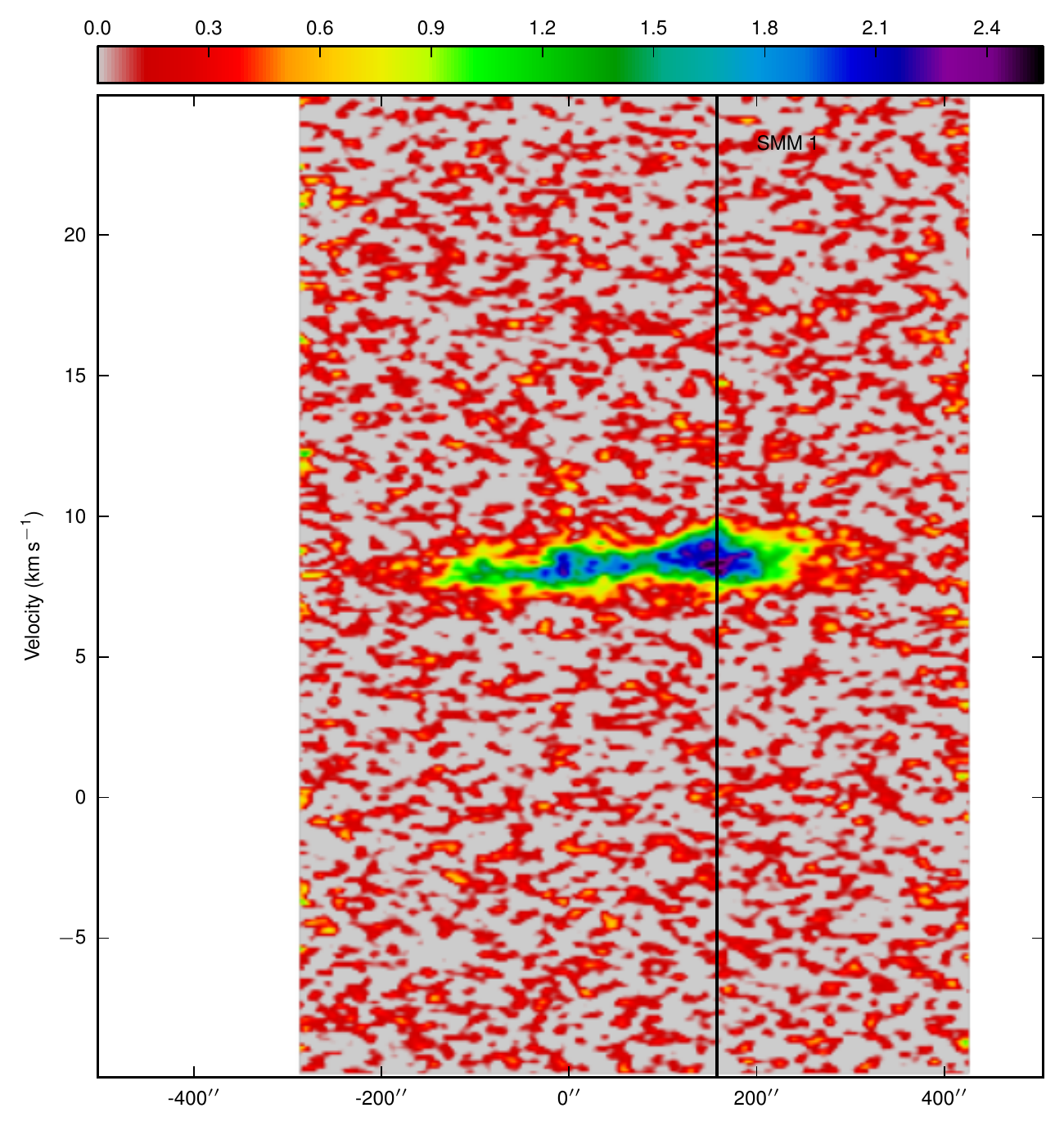}
  \caption{Position-velocity cuts through the position of SMM 1 along
    a line of constant declination, in all three isotopologues. The
    \twelco\ map is on the top, then the \thirtco\ is below on the
    left and the \ceighto\ on the right. The black lines indicate the
    position of SMM 1.\label{fig-smm1pv}}
  
\end{figure}

In order to look further at the different structures traced by the
three isotopologues, Fig.~\ref{fig-smm1pv} shows position-velocity
(PV) diagrams taken along a line of constant declination through the
position of SMM 1 in all three isotopologues. In this Figure the PV
diagrams of both \thirtco\ and \twelco\ show evidence of
self-absorption dips in the centre of the line, which are not seen in
the \ceighto. The \twelco\ emission also clearly shows strong evidence
for red- and blue-shifted emission in the line wings, which is not
clearly seen in the \thirtco\ and entirely absent in the \ceighto\
maps. Again, this emphasises that the isotopologues are truly tracing
different environments, and specifically that the \ceighto\ emission is
not influenced by the molecular outflows, at least at the sensitivities of these data.

\section{Cloud Conditions}

\subsection{Optical Depth}
\label{sec-optdepth}
The variation of optical depth across the cloud can be examined by
looking at the ratio of intensities of the different
isotopologues, $I_{\thirtco}/I_{\ceighto}$ and
$I_{\twelco}/I_{\thirtco}$, across the entire region being
studied. The relation between the line intensity ratio and the optical
depth \citep[e.g.~][]{Rohlfs2000} at a specific velocity is:

\begin{equation} \frac{T_{\twelco}}{T_{\thirtco}} = \frac{1 -
    \exp(-\tau_{12})}{1-\exp(-\tau_{13})}
\label{eqn:optdep}
\end{equation}

Where $T$ is the is the brightness temperature, and $\tau$ is the
optical depth.

Ratio maps were created, displaying the ratio of the peak intensities
of the isotopologues, thresholded at a five sigma peak value For
optically thin emission, the ratio of the intensities of two lines is
expected to be to tend towards the relative abundances of these
species as we look towards the edge of the map.

 For the \twelco/\thirtco\ ratio, we see maximum values of only 
 10 or so at most, compared with a measured abundance ratio of 77
 \citep{Wilson1994}, although this was from a different cloud. This
 implies that the \twelco\ is optically thick through the whole of
 the cloud. Due to the extremely strong self absorption in the
 \twelco\  we do
 not  attempt to calculate a valid optical depth from this
 ratio.

 The optical depth in the cloud for \ceighto\ and \thirtco\ can then
 be calculated by utilising Eqn.~\ref{eqn:optdep} above and the
 approximations that $\tau_{13} = X_{1318} \tau_{18}$ and
 $\tau_{12}=X_{1213}\tau_{13}$, where $X$ is the appropriate abundance
 ratio.  Often this is calculated at the velocity of the peak
 channel. However, as we can see in the spectra from these cubes (see
 Fig.~\ref{co_spectra}), there is visible self absorption in the
 \thirtco\ and the \twelco\, which would tend to limit the validity of
 this approach by underestimating the ratio by a varying amount across
 the map. Therefore, we follow the method in \citet{Ladd1998} and
 examine the ratio of the integrated intensities
 instead. \textrm{Although the self-absorption will still cause the
   integrated \thirtco\ intensity to be underestimated, this should be
   a lower fractional error than would occur if the intensity ratio at
   the velocity of the peak was used. This remaining systematic under-estimation
   will however boost the \ceighto\ optical
   depth, particularly in the central regions where the \thirtco\ self
   absorption is strongest. }

\begin{figure}
  \includegraphics[width=3in]{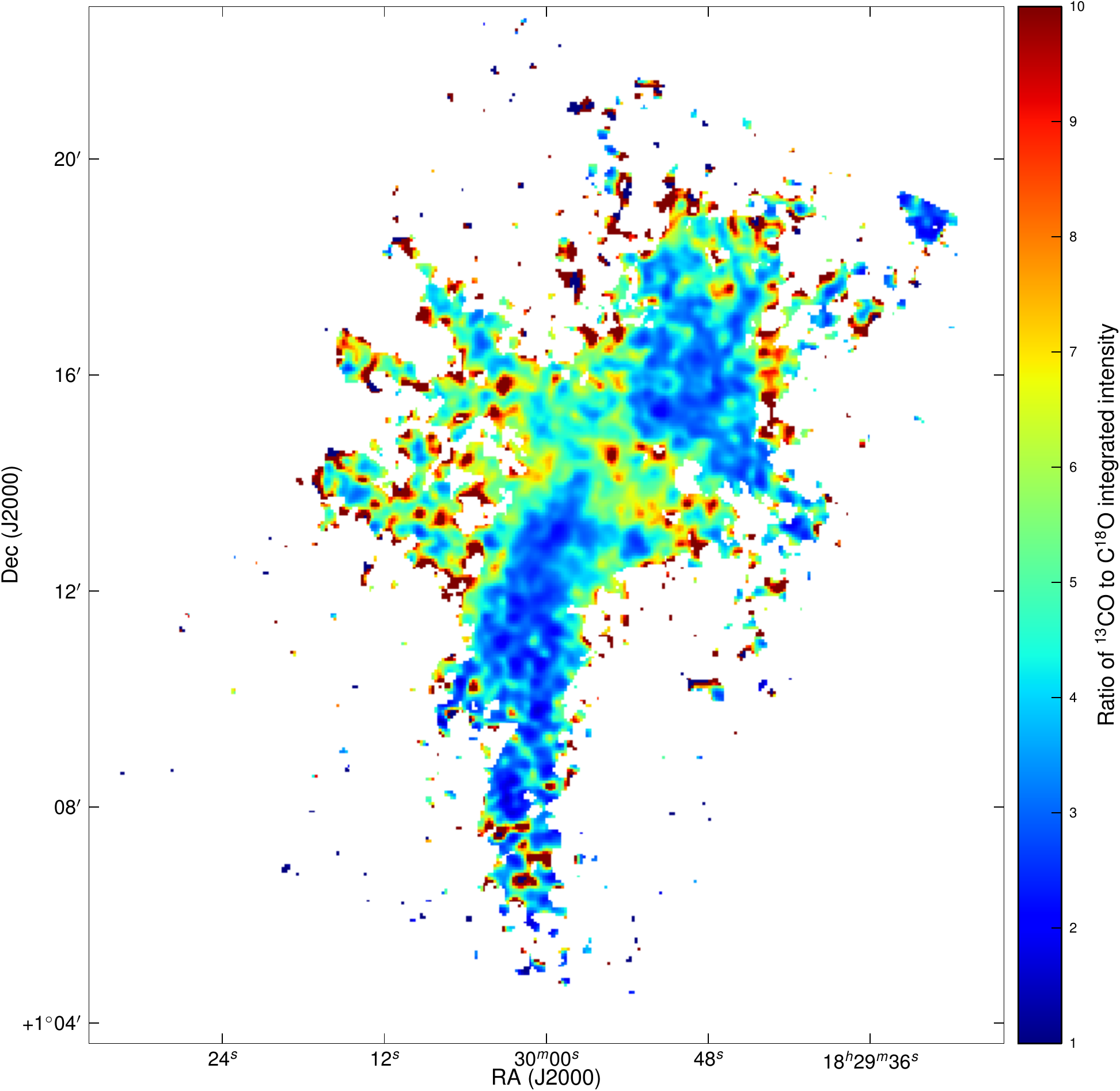}
  \caption{Ratio of integrated intensity of \thirtco\ over \ceighto\,
    both maps integrated from 0-15\,km\,s$^{-1}$.}
  \label{fig:1318integ}
\end{figure}

Fig.~\ref{fig:1318integ} shows
 the ratio map for the \thirtco/\ceighto\ integrated ratio. 

We then use:
\begin{equation}
\frac{\int^{\infty}_{-\infty} T_{13}(\nu)d\nu}{\int^{\infty}_{-\infty} T_{18}(\nu)d\nu}
=
\frac{\int^{\infty}_{-\infty} 1-\exp[ -\tau_{13}(\nu)]d\nu}{\int^{\infty}_{-\infty} 1-\exp[ -\tau_{13}(\nu)/f]d\nu}
\end{equation}

where $\tau_{13}(\nu) = \tau_{13}\exp[-\nu^2/2\sigma^2]$, and
$\tau_{18}(\nu) = \tau_{13}(\nu)/f$, to calculate the optical depth by
numerically minimising the following function:

\begin{equation}
\left|\frac{\int^{\infty}_{-\infty} 1-\exp[ -\tau_{13}\exp[-\nu^2/2\sigma^2]]d\nu}{\int^{\infty}_{-\infty} 1-\exp[ -\tau_{13}\exp[-\nu^2/2\sigma^2]/f]d\nu} - \frac{\int^{15 kms}_{0 kms} T_{13}(\nu)d\nu}{\int^{15 kms}_{0} T_{18}(\nu)d\nu}\right|
\end{equation}

This method is very stable to the value of $\sigma$ that is chosen,
{therefore we have simply used an average line-width instead of
  estimating the Gaussian line-width for each spectra} We used
$\sigma$ value corresponding to a line-width of 1.5\,\kms, the FWHM~of
an averaged \ceighto\ spectra. The abundance ratio can be calculated
for optically thin lines by looking at the intensity ratio towards the
edge of the cloud. We see values approaching the canonical value of 8
\citep{Frerking1982} therefore we feel confident both in using this as
the abundance ratio for \thirtco/\ceighto\, and that the \ceighto\
emission really is optically thin across the cloud.

\begin{figure}
   \centering
   \includegraphics[width=3.2in]{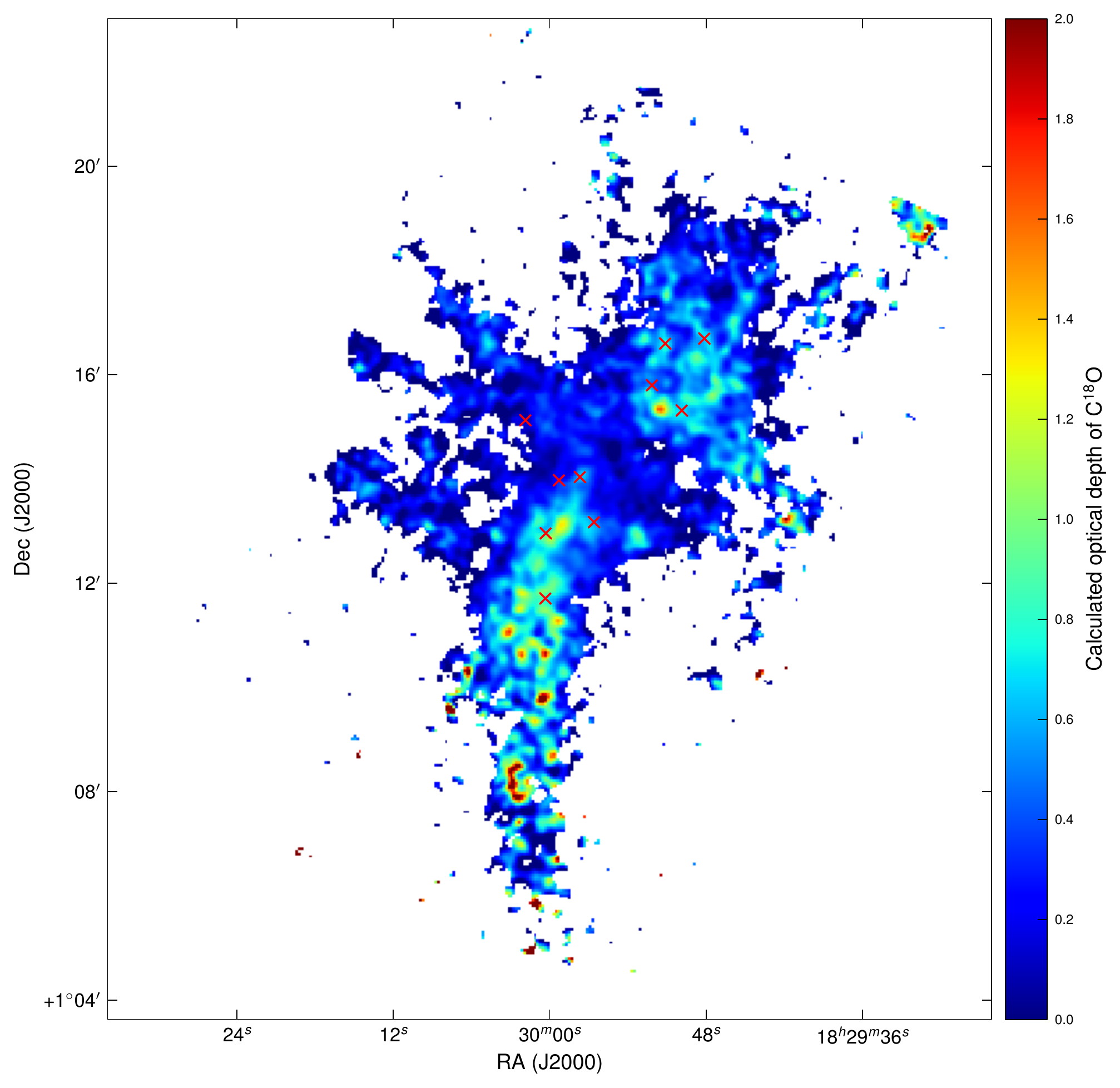}
   \caption{Calculated $\tau_{18}$ across the map \label{fig:tau18}}
 \end{figure}

Across the map the \thirtco/\ceighto\ ratio varies from around 3-8
across the central region to values from 2.6-7.0 towards the cores, with
most clustered around a value of 3. This gives $\tau_{18}$ values of
0.07 to 0.9 towards the cores, and $\tau_{13}$ values of 0.61 to 7.3.
This suggests that the \ceighto\ 3--2 emission is optically thin
across the map, and is even (marginally) optically thin towards the
position of the submillimetre cores, and that the \thirtco\ 3--2 emission is
transitional between optically thick and thin across much of the
cloud, and optically thick towards the dense
cores. Fig.~\ref{fig:tau18} shows the \ceighto\ optical depth
calculated across the map. It is very noticeable that the regions of
high optical depth do not correspond to the positions of the submillimetre
SMM cores. This is not unexpected, as the peaks of the \ceighto\
emission do not align exactly with that of the continuum
cores. However, the effect of the \ceighto\ ceasing to be optically
thin towards extremely dense regions, as well as the possibility of
depletion towards the centre of dense cores and the slight self
absorption present in the \thirtco\ emission prevent us from
investigating this further. 

\subsection{Excitation Temperature}

The excitation temperature of the molecular gas can be  calculated from the
peak temperature of emission from optically thick gas (via the assumption of
local thermodynamic equilibrium). In our observations however, the
most optically thick tracer -- \twelco\ -- is complicated by the strong
self absorption features seen in the \twelco\ line profiles across
much of the cloud. This would underestimate the excitation temperature
in these dense regions. However, as we have calculated that the
\thirtco\ emission is just optically thick across much of the cloud,
we can also calculate the excitation temperature from this emission, using:
\begin{equation}
  T_B =T_0 \left( \frac{1}{e^{T_o/T_{ex}} -1} - \frac{1}{e^{T_0/T_{bg}} -1} \right) (1 - e^{-\tau_{\nu}})
\end{equation}
 
where $T_B$, the main beam temperature, is assumed to be the same as
the peak temperature of the transition, $T_0$ is $h\nu/k_B = 16.6$\,K
for \twelco\ and $15.9$\,K for \thirtco, and $T_{bg} = 2.73$\,K is
the temperature of the cosmic background. $\tau$ is the optical depth
for the transition of interest and is assumed to be infinity
here. This equation assumes that the emission completely fills the
beam.
This gives:
\begin{equation}
  T_{ex}(\twelco) = \frac{16.6\,K}{\ln{\frac{16.6\,K}{T_{peak}+0.0377\,K} +1}}
\end{equation}
\begin{equation}
  T_{ex}(\thirtco)=\frac{15.9\,K}{\ln{\frac{15.9\,K}{T_{peak}+0.0472\,K} +1}}
\end{equation}

\begin{figure}
\begin{center}
  \includegraphics{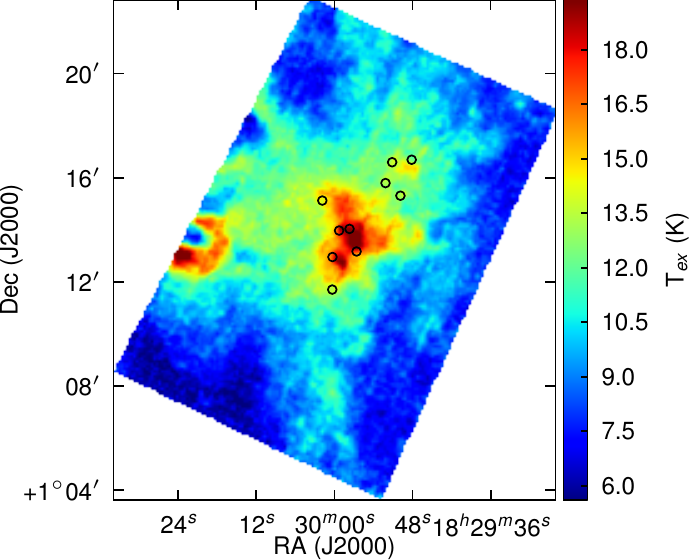}
  \caption{Excitation temperature from \thirtco\ emission in
    K. Positions of the submillimetre cores are shown as circles
    (i.e.~SMM1-11) from \citet{Davis1999}. \textrm{This was calculated assuming high optical depth.}}
  \label{fig:13co-ex}
\end{center}
\end{figure}

\begin{figure}
\begin{center}
  \includegraphics{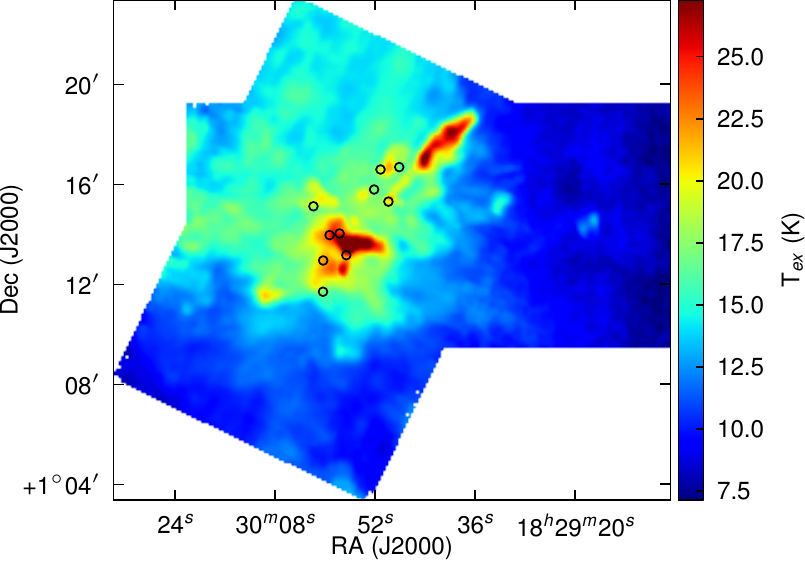}
  \caption{Excitation temperature from \twelco\ emission in
    K. Positions of the submillimetre cores are shown as circles
    (i.e.~SMM1-11) from \citet{Davis1999}}
  \label{fig:12co-ex}
\end{center}
\end{figure}

\noindent The results for both \thirtco\ and \twelco\ are shown in
Figs.~\ref{fig:13co-ex} \& \ref{fig:12co-ex}. \twelco\ gives us a mean
excitation temperature of 17\,K across the map, and temperatures
towards the positions of the cores of 21-33\,K. These positions will,
however, be looking at the excited gas from molecular outflows instead
of the cores themselves. These temperatures may be underestimated as
the \twelco\ emission is strongly self absorbed towards the bulk of
the cloud. However, by contrast the excitation temperature calculated
from the \thirtco\ emission is significantly lower---a mean
temperature across the cloud of 13\,K, and temperatures towards the
cores ranging from 16-22\,K.  \textrm{This \thirtco\ temperatures have
  been calculated assuming high optical depth. Although we have
  calculated the \thirtco\ opacities towards the centre of the map
  (where they are highest), we do not have \thirtco\ opacities for
  those regions towards the edges of the map where the \ceighto\
  emission is weak or not detected. Therefore we have not been able to
  correct for the \thirtco\ opacity in these regions, but we expect
  that the assumption of high optically depth will be least valid in
  these areas. Caution must used when drawing conclusions from the
  excitation temperatures in these regions. }

\textrm{The difference in temperatures derived from the \thirtco\ and
  \twelco\ maps is interesting, particularly given that we expect the
  \twelco\ temperature to be significantly underestimated due to the
  stronger self absorption. Fig.~\ref{fig:12co-ex} suggests that the
  excitation temperature estimated from the \twelco\ appears to trace
  some of the outflow structure in this cloud. This can be most clearly seen in the
  hot lobe-like structure visible in the NW of Fig.~\ref{fig:12co-ex}}
This structure corresponds to the position of HH 460, and there is
known bright \htwo\ emission in the region.  \textrm{This suggests
  that shocks are primarily responsible for producing the higher
  \twelco\ temperatures}

Also noticeable in this analysis is an arc of high temperatures
towards the east of the \thirtco\ map in Fig.~\ref{fig:13co-ex}. Near
to this position there is both a reflection nebula and the B type star
HIP 90707. \textrm{It is possible that these are influencing the gas
  temperatures in this region. The \thirtco\ map shows narrow but
  bright spectral line profiles in this region}
\\

\subsection{Cloud Mass} The mass of the cloud has been estimated from
the integrated \ceighto\ intensity. This calculation assumes that the
\ceighto\ emission is 
optically thin. This is generally
expected to be true everywhere in molecular clouds apart from the
centre of the very densest cores, and for Serpens this was confirmed
above in Sec.~\ref{sec-optdepth}. These extremely dense regions contain
only an insignificant proportion of the total mass in this region. The
ratio of \ceighto\ to \htwo\ is taken to be $10^{-7}$, and the
distance to the cloud was assumed to be 230~pc
\citep{Eiroa2008}. $T_{ex}$, the excitation temperature, was assumed
to be 12\,K. This seems reasonable both as a fairly standard value for
cold gas in molecular clouds, and from the excitation calculations
given above from the \twelco\ and \thirtco\ emission.

\begin{equation}
  M_{\ceighto} = 4.0\times 9.34\times 10^{-4} \left(\frac{\mathrm{Pixel size}}
    {3.6\times 10^{-3}\,\mathrm{pc}}\right)^2 \left(\frac{\eta_{MB}}{0.63}\right)^{-1}\int T_A^*(\nu)\ M_{\odot}
\end{equation}
This gives a mass for the cloud of
203\,$M_{\odot}$. \textrm{Quantifying the error on this number is not
  trivial given the large number of uncertainties in the input
  parameters and assumptions; however, the largest sources of
  systematic errors are probably contained in the estimate of the
  abundance and in the assumption of LTE. The relative abundance of CO
  to \htwo\ has not been measured in Serpens, therefore we have
  adopted the standard literature value.}

An estimate of the virial mass of the cloud can be calculated assuming
a spherical cloud of uniform density via \citep{Rohlfs2000}:
\begin{equation}
M_{virial} = 250\left( \frac{v_{fwhm}}{\mathrm{km\,s^{-1}}} \right)^2
\left( \frac{R}{\mathrm{pc}} \right)
\end{equation}
\noindent where $v_{fwhm}$ is the FWHM 1-D velocity in \kms\
and $R$ is the radius of the cloud in pc.  From the \ceighto\
emission, we see a full-width half-max line of between one and two
km\,s$^{-1}$, therefore we take a value of $v_{fwhm}$ of
1.5\,\kms. The bright \ceighto\ emission in the cloud is roughly
12\arcmin\ long and approximately 6\arcmin\ wide at the widest region,
so we assume a diameter of 9\arcmin\ and therefore a radius of
4.5\arcmin, which corresponds to 0.3\,pc at 230\,pc. This gives a
virial mass of 169\,$M_{\odot}$. 

This virial mass is approximately 80 percent of the measured \ceighto\
mass, suggesting that the cloud \textrm{is in a bound state. However,
  this result is of course subject to the same sort of systematic
  uncertainties as the mass estimate,. These large uncertainties
  prevent us from drawing a more firm or quantitative conclusion about
  the state of the cloud other than to estimate that the measured
  \ceighto\ mass is close the virial mass, even if we attempted to use
  a more complex model of the density.}

\subsection{Global Energetics}
\label{sec-global-energetics}

We can compare the different energetics present in the cloud by making
some simple approximations.  We can estimate the gravitational binding
energy and the turbulent kinetic energy using the following equations:

\begin{equation}
E_{grav} = -\frac{3GM^2}{5R}
\end{equation}
\begin{equation}
E_{kin}=\frac{3}{2}M\sigma^2
\end{equation}
where $G$ is the gravitational constant, $M$ is the mass of the gas in
the cloud, $R$ is the estimated size of the cloud, and $\sigma$ is the
estimated 1-D velocity dispersion in the cloud, equal to
$v_{fwhm}/2.35$. \textrm{This assumes a uniform density cloud. A more
  centrally condensed cloud would be more tightly bound (i.e.~would
  have a more negative gravitational binding energy). Similarly, due to
  the $M^2$ factor involved, the uncertainties in the mass of the
  cloud will produce a large uncertainty in the gravitational binding
  energy. The error in the assumption of a single velocity dispersion
  for the entire cloud will also produce large uncertainties in these
  values, due to the $\sigma^2$ factor.}

The virial mass calculated above corresponds to a gravitational
binding energy of 245\,\Msun\,km$^2$\,s$^{-2}$
The turbulent kinetic energy can be estimated by measuring the CO line
width in regions where the emission is not affected by outflow
activity, therefore we use the optically thin tracers. The
\thirtco\ and \ceighto\ maps give roughly similar results of
$\sim1.5-2$\,\vunits for the line width, 2\kms\ is used here, giving a
turbulent kinetic energy of 220\,\Msun\,km$^2$\,s$^{-2}$.
 
The energetics of the cloud are summarised in Table
\ref{energy_table}, along with the global outflow energy from
Sec.~\ref{sec-global-outflow}.

\begin{table}
\caption{Summary of the energetics of the Serpens Molecular
  cloud\label{energy_table}}\centering
\begin{tabular}{l r} \hline & Energy (\Msun\,km$^2$\,s$^{-2}$)\\ \hline
  Gravitational Binding  & 245 \\
  Turbulent kinetic      & 220 \\
  Global Outflow Kinetic Energy & \\ 
  \hspace{1cm}-along line of sight & 51 \\
  \hspace{1cm}-assuming random inclination & 153\\

\hline
\end{tabular}
\end{table}

This analysis suggests that the cloud is gravitationally
bound. \textrm{ However, it is important to reiterate that there are
  large uncertainties in the estimates of the values presented here.
  \\
  We can also examine the importance of the outflows in this
  region. The outflow momenta and energy have been derived by assuming
  that the measured speeds are along the line of sight; this will
  underestimate the values. To estimate the correction to this, we
  assume the direction of the outflows is distributed
  isotropically. As we have many flows contributing to the global
  outflow momentum and energetics, this is a good approximation.
  Table~\ref{energy_table} shows the corrected values. These suggest
  that the outflow energy is approximately 70 percent of the turbulent
  energy. This result would suggest that in this dense, clustered
  region the outflows are a very important effect and could be the
  dominant factor driving the local turbulence. However, it is
  important to note that there are significant uncertainties in the
  estimation of these variables (as mentioned in the discussion of the
  mass, and in further discussion in
  Sec.~\ref{sec-global-outflow}). Further work on disentangling the
  individual outflows and on analysing the detailed structure of the
  entire region will be required to quantify this more accurately,
  possibly using observations of other tracers.}

\section{Velocity Structure from C$^{18}$O}

The C$^{18}$O emission, which is not tracing outflows and is optically
thin across the whole extent of the Serpens cloud, allows us to study
the velocity structure of the cloud in detail, with no obvious
contamination from infall and outflow motions at the sensitivity of
these observations. In the following sections we will analyse the
velocity structure using velocity-coded 3-colour plots and
position-velocity diagrams through the region, to try to understand
the general dynamics of the region.

\subsection{Global velocity}

\begin{figure}
\includegraphics[width=0.4\textwidth]{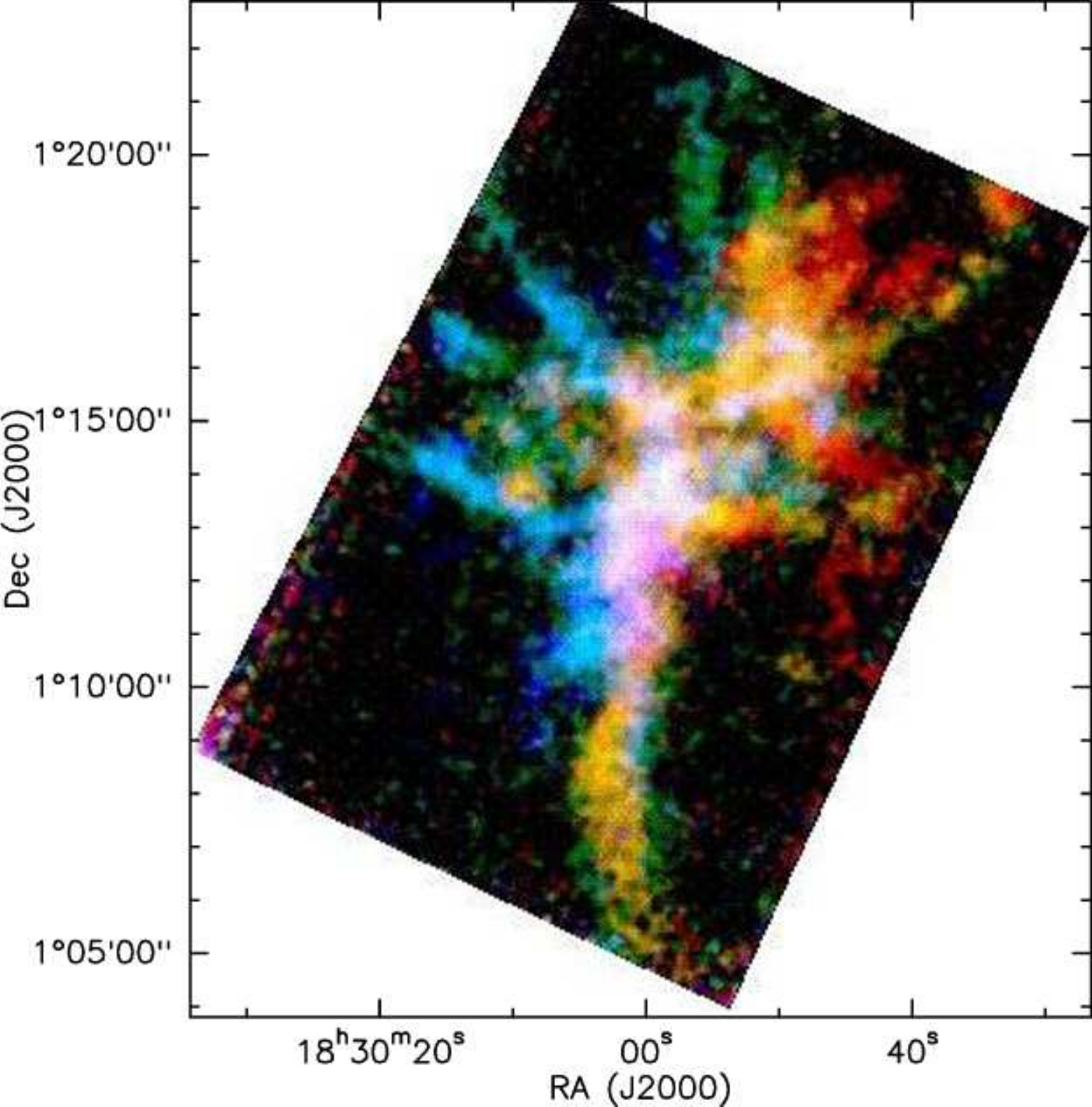}
\caption{Velocity coded 3 colour plot of the C$^{18}$O
  \jtt\ Each colour represents the maximum value in the
  following velocity intervals: blue: 5~-~7.7~kms$^{-1}$; green:
  7.7~~8.3~kms$^{-1}$; and red: 8.3~~11~kms$^{-1}$}
\label{fig:3c}
\end{figure}

The velocity structure of the dense gas of the Serpens cloud can be
seen on Fig.~\ref{fig:3c} as a velocity coded 3 colour plot. A first
glance at this figure shows that the blue shifted emission is east
of the Serpens filament, whereas the red emission is mainly west of
it. This gradient has been previously detected by other authors
\citep{Olmi2002}, who interpreted it as a global rotation of the
cloud. However, this conclusion was limited by the resolution of
their data, which could not resolve the velocity structure in as much
detail as we can.

Figure~\ref{fig:3c} demonstrates that this velocity structure is more complex
than simple rotation, by showing a 3 colour plot of the \ceighto\
red-shifted, blue-shifted and central-velocity emission. For instance,
we can see, in the SE end of the Serpens filament (the SE sub-cluster)
a sharp merger/overlap of the blue and the red emission, not expected
in a solid body rotation scenario. On the other hand, the NW end of
the filament (the NW sub-cluster) has very little blue-shifted
emission.  This velocity structure was also suggested to be possibly a
shear motion \citep{Olmi2002} or a collision of two clouds/flows where
the interface coincides with the SE end of the filament
\citep{DuarteCabral2010}.

Another noticeable feature visible in this view of the \ceighto\
velocity structure, are the
gas streams east of the filament. Note that these are not seen at the
west, nor do they correlate with any of the outflows seen in
$^{12}$CO. These small filaments roughly perpendicular to the main
filament are commonly seen in filamentary molecular clouds, but their
origin is not yet clear \citep{Myers2009}.

\subsection{Position-Velocity Diagrams}

\begin{figure}

\includegraphics[width=0.4\textwidth]{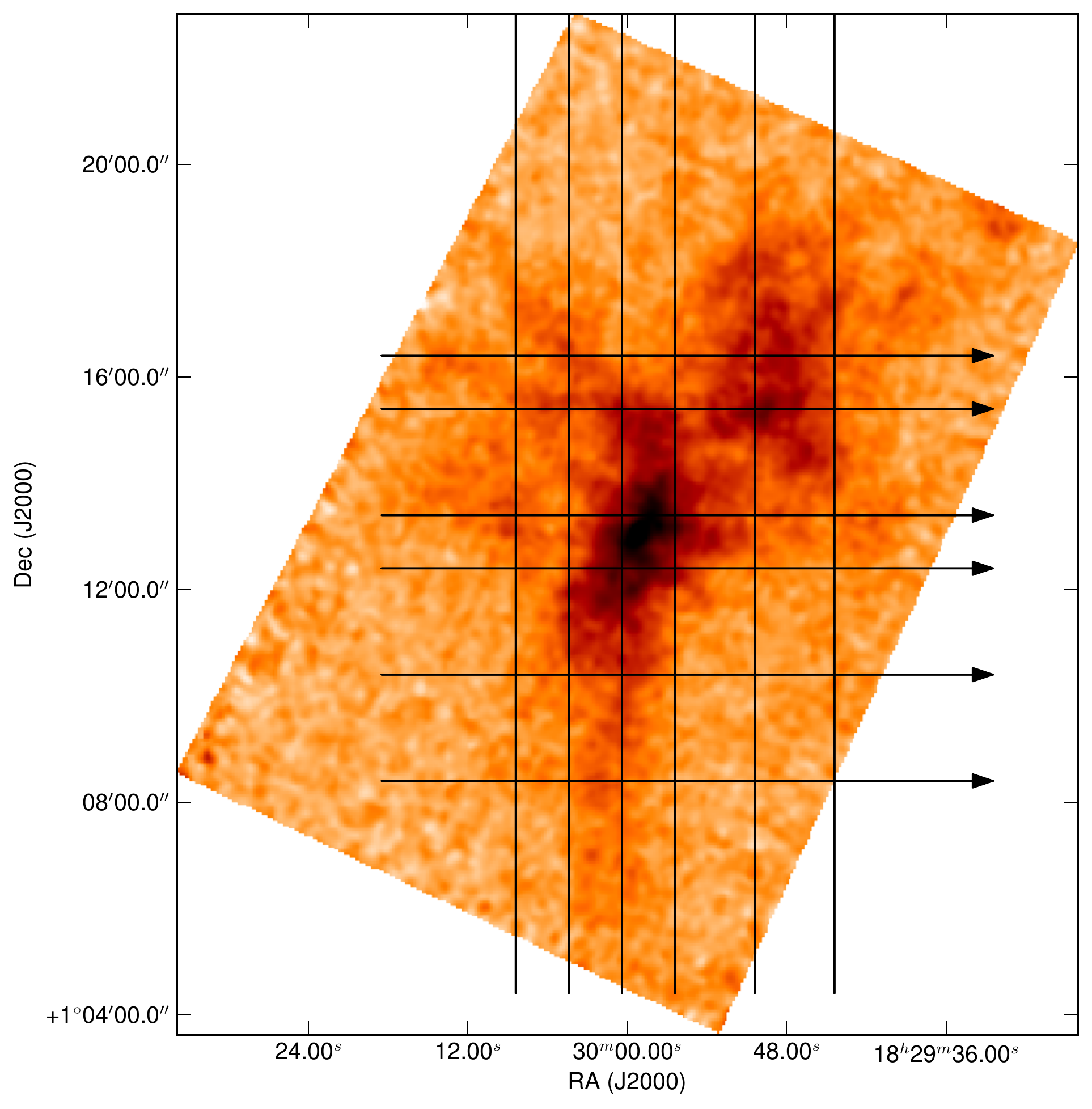}
\caption{Integrated \ceighto\ map showing the positions of the PV diagrams used in this section. \label{fig:pvfinder}}
\end{figure}

In order to look at the velocity structure at specific positions we
have created PV diagrams. By studying several PV
diagrams from our \ceighto\ maps, we can get a better sense of how the
velocity changes and evolves across the cloud. Our approach to such a
study involves several cuts both with constant Declination and with
constant RA. Fig.~\ref{fig:pvfinder} shows the positions of the PV cuts used here. 

\begin{figure}
\centering
\includegraphics[angle=0,width=0.3\textwidth]{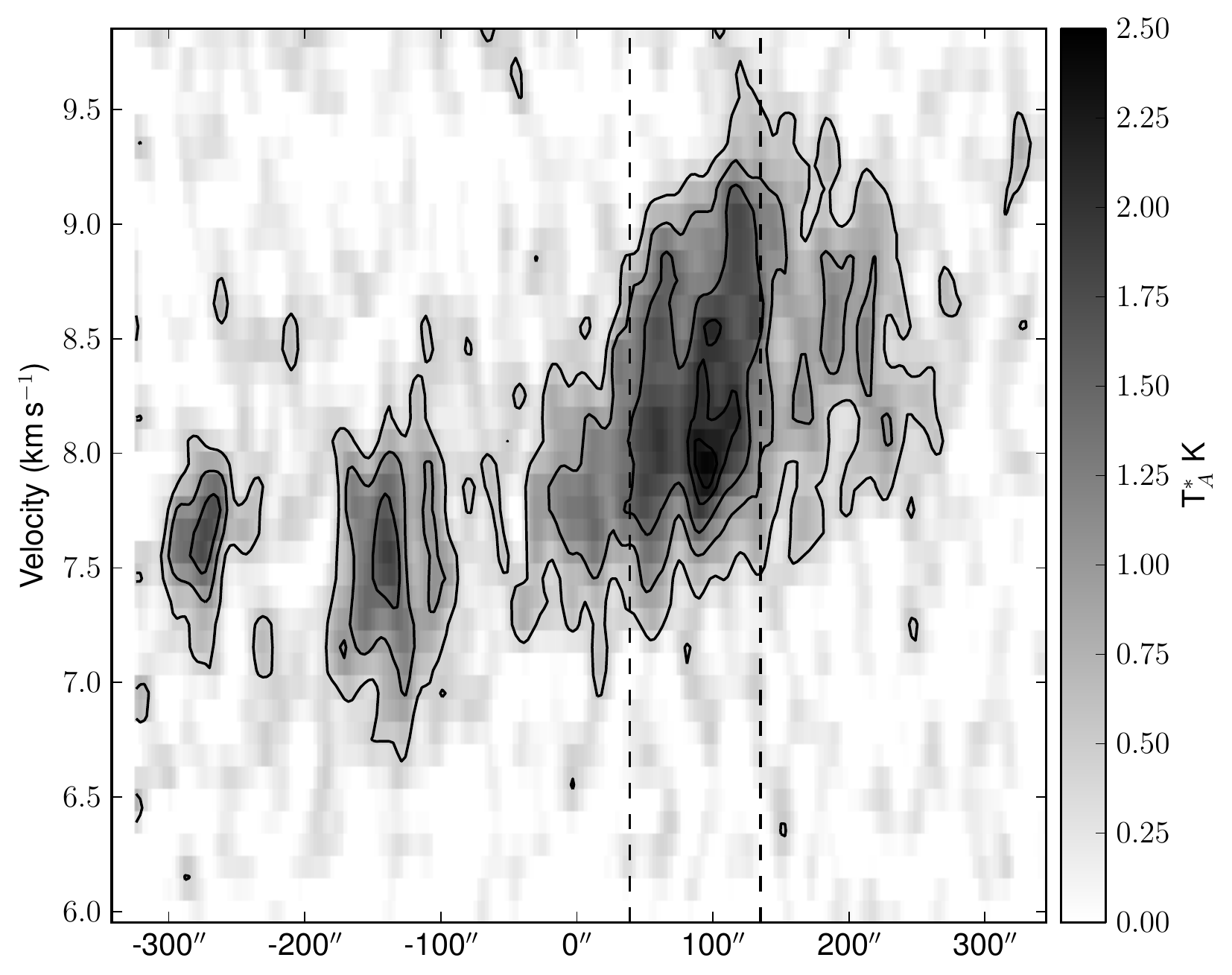}
\includegraphics[angle=0,width=0.3\textwidth]{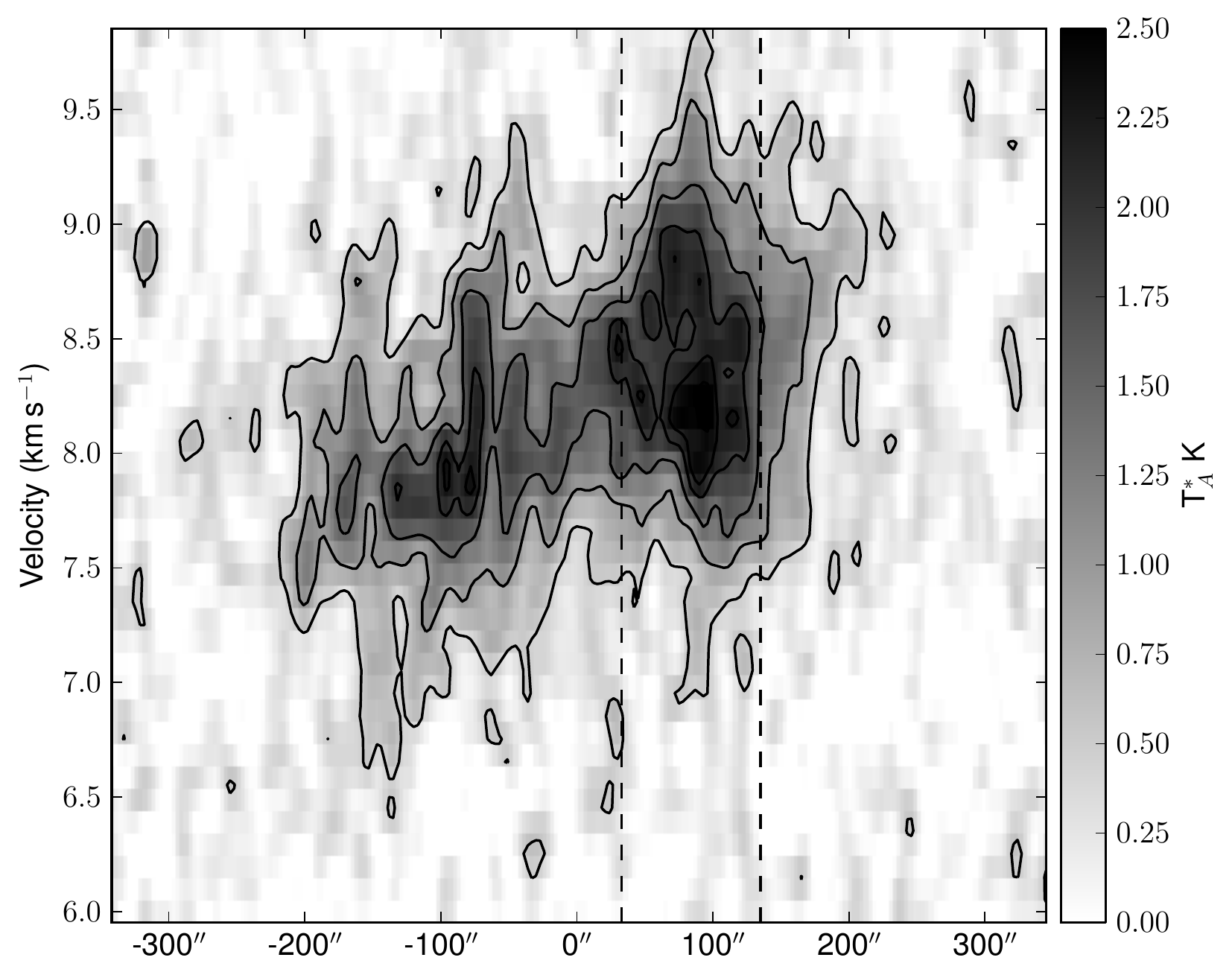}

\caption{Horizontal position-velocity diagrams in grey scale and
  contours of the C$^{18}$O \jtt\ emission. These
  represent the typical velocity structure in the NW sub-cluster. The
  diagrams are displayed with decreasing Declination, i.e. from North
  to South, and at constant Declinations of 1$^o$16'24'' (top) and
  1$^o$15'24'' (bottom). Right Ascension varies from
  18$^h$30$^m$18.5$^s$ to 18$^h$29$^m$32.5$^s$ (offset of 277 to -412
  respectively) on both diagrams. The dust emission seen in 850~$\mu$
  lies between the dotted lines.}
\label{fig:pvconstantDec1}
\end{figure}

\begin{figure}
\centering

\includegraphics[angle=0,width=0.3\textwidth]{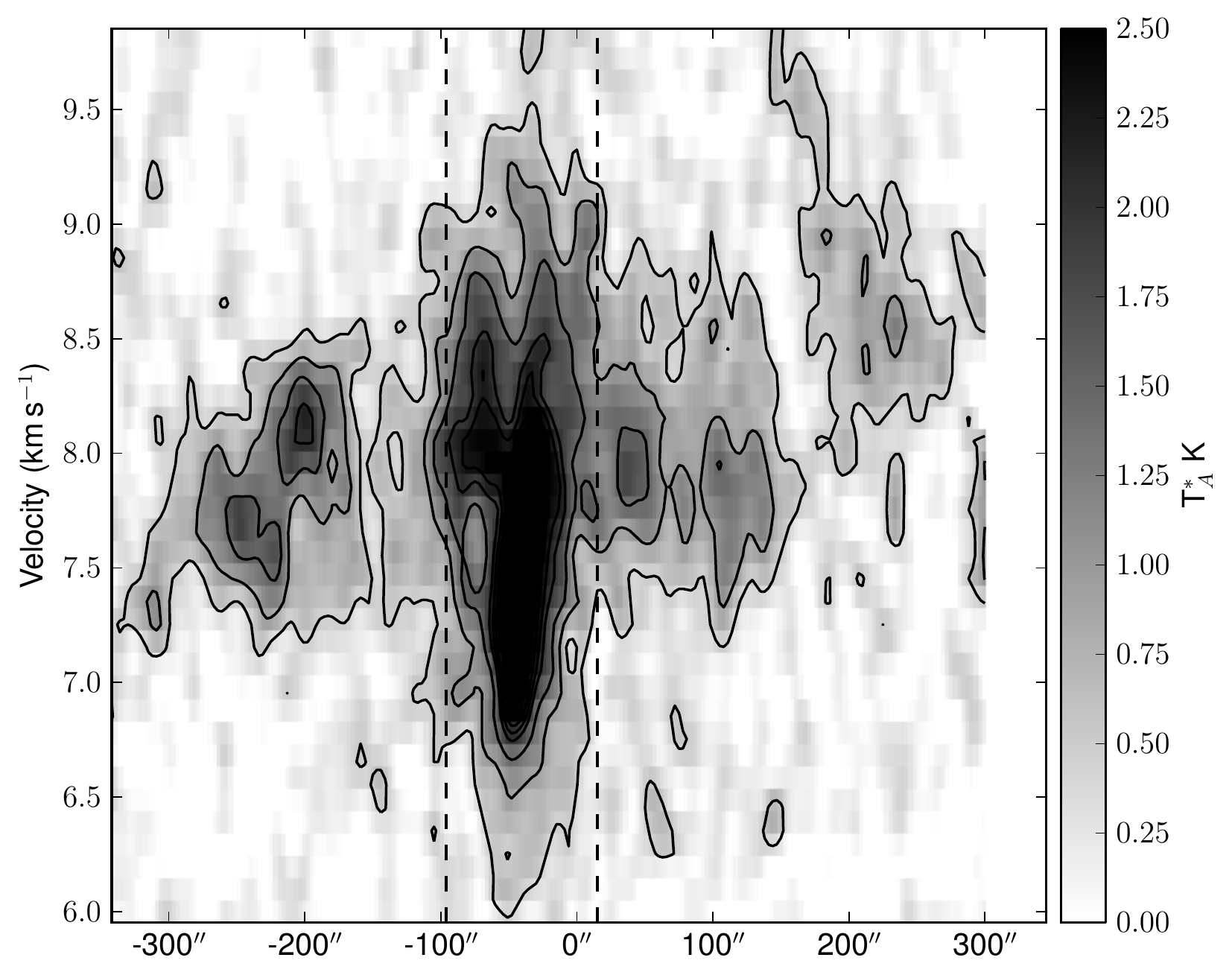}
\includegraphics[angle=0,width=0.3\textwidth]{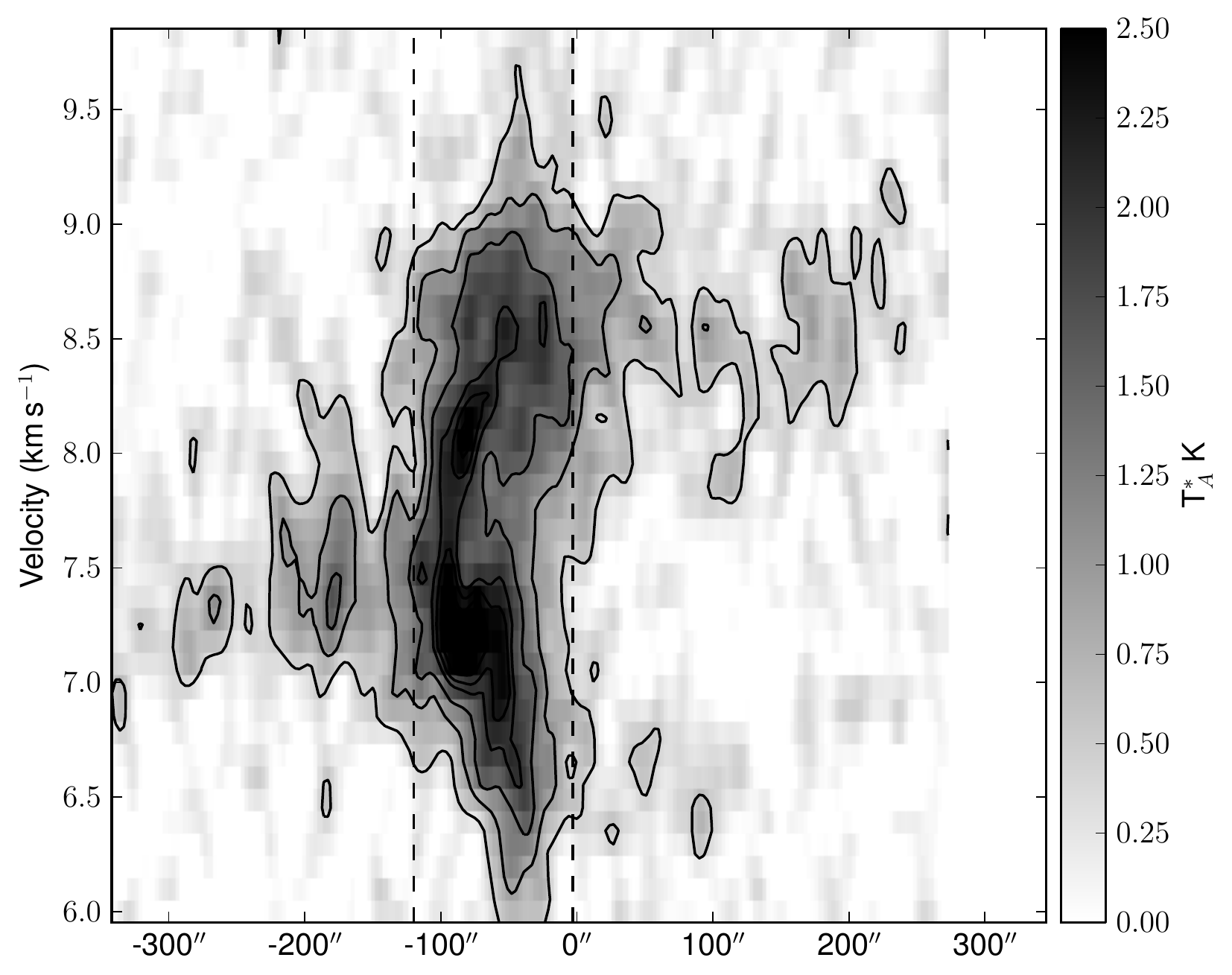}
\includegraphics[angle=0,width=0.3\textwidth]{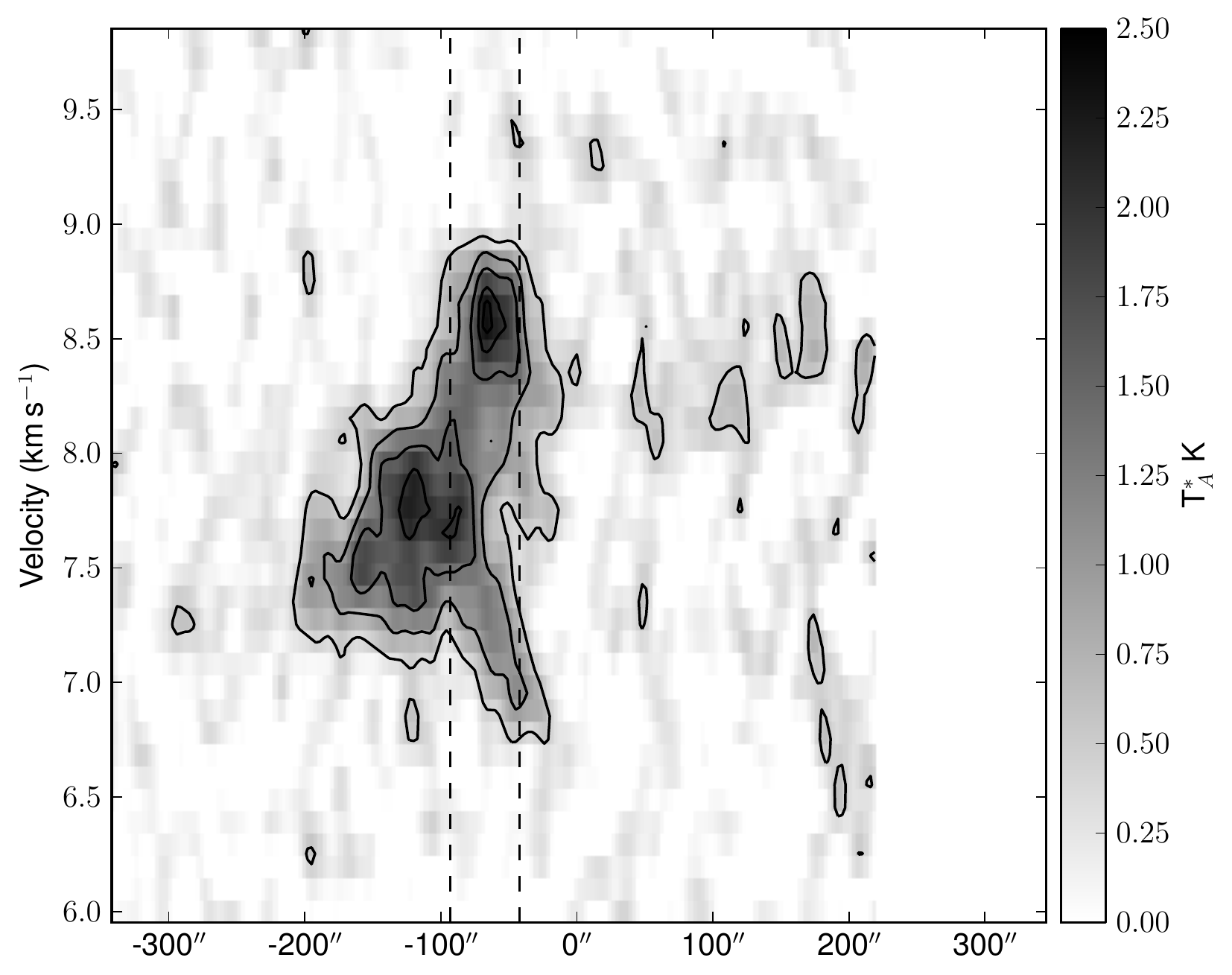}
\includegraphics[angle=0,width=0.3\textwidth]{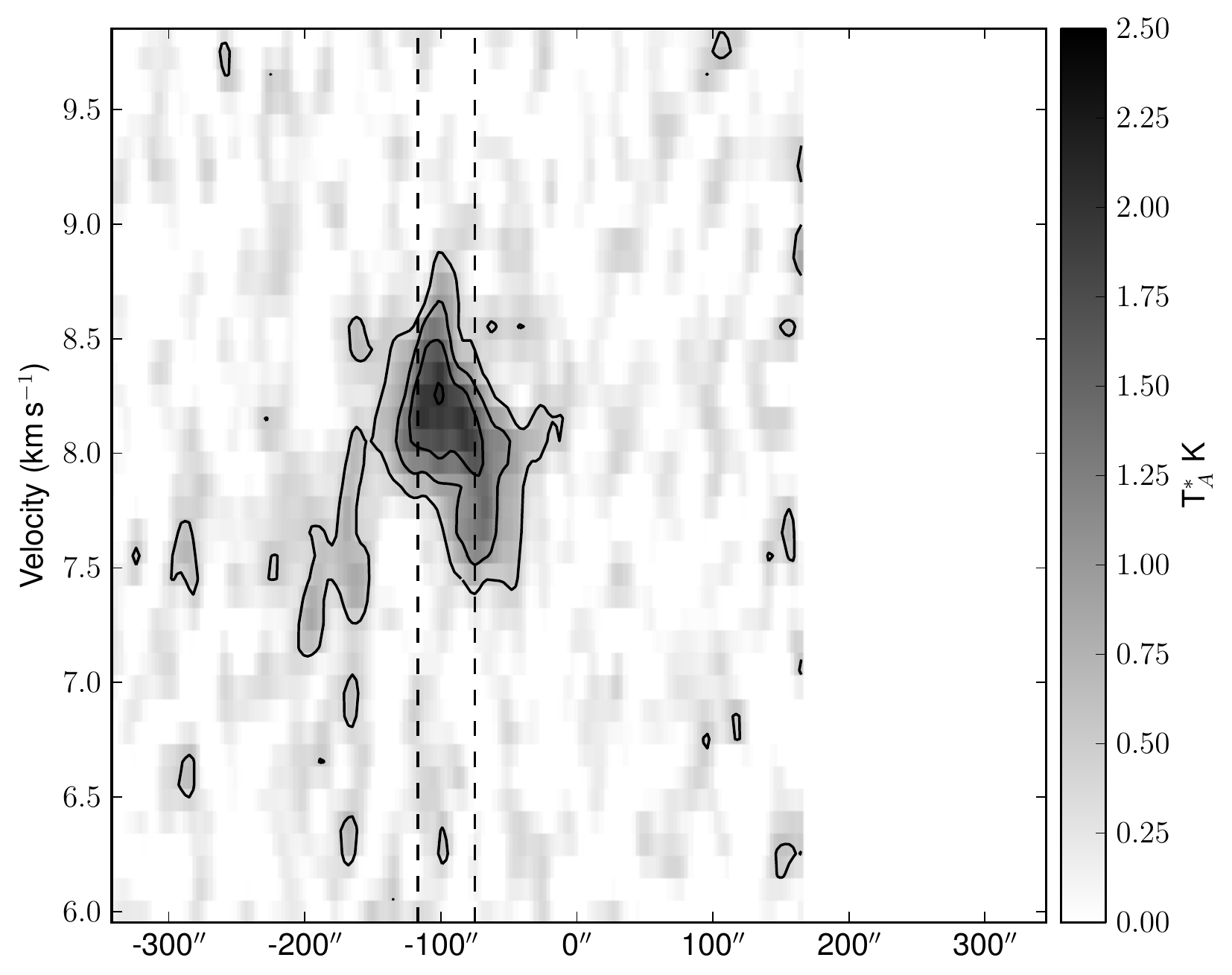}

\caption{Horizontal position-velocity diagrams in grey scale and
  contours of the C$^{18}$O \jtt\ emission. These
  represent the typical velocity structure in SE sub-clusters. The
  diagrams are at constant Declination (from top to bottom panel) of
  1$^o$13'24'', 1$^o$12'24'', 1$^o$10'24'' and 1$^o$08'24''. Right
  Ascension varies from 18$^h$30$^m$18.5$^s$ to 18$^h$29$^m$32.5$^s$
  (offset of +277'' to -412'' respectively) on all the diagrams. The
  dust filament seen in 850~$\mu$m lies between RA offsets of -120''
  and -180''.}
\label{fig:pvconstantDec2}
\end{figure}

Figs.~\ref{fig:pvconstantDec1} and \ref{fig:pvconstantDec2} shows PV
diagrams (horizontal slices of the data cube) at five fixed
Declinations, varying the Right
Ascension. Fig.~\ref{fig:pvconstantDec1} cuts through the NW
sub-cluster, where we can note that this sub-cluster is represented by
a velocity mainly between 8 and 8.5~kms$^{-1}$. However, there is some
lower velocity gas weakly emitting east of the sub-cluster, with
velocities around 7.5~kms$^{-1}$. Whether this gas is physically
connected with the higher velocity gas is not clear, as it could
either represent a separate cloud with a slightly different velocity,
or the same cloud undergoing rotation. Therefore, we must examine the
the evolution of these components as we look southwards through the
cloud in order to understand what these gas components are tracing.

Fig.~\ref{fig:pvconstantDec2} shows the velocity structure of the SE
sub-cluster. Here the two velocity components, which
in the north are very well separated, begin to merge. As we move
south the components  become more offset from each other in
velocity, and we see regions that are clearly reproduced by two
velocities (one around 7~kms$^{-1}$ and the other around
8.5~kms$^{-1}$). Note that the regions which also show dust emission
are coincident with regions where the two velocities coexist along the
line of sight. Finally, the lower velocity gas ceases to exist, and at
the very southern end of the filament  solely consists of the
higher velocity gas, at around 8.3~kms$^{-1}$.
\\

\begin{figure*}
\centering
\includegraphics[angle=270,width=0.3\textwidth]{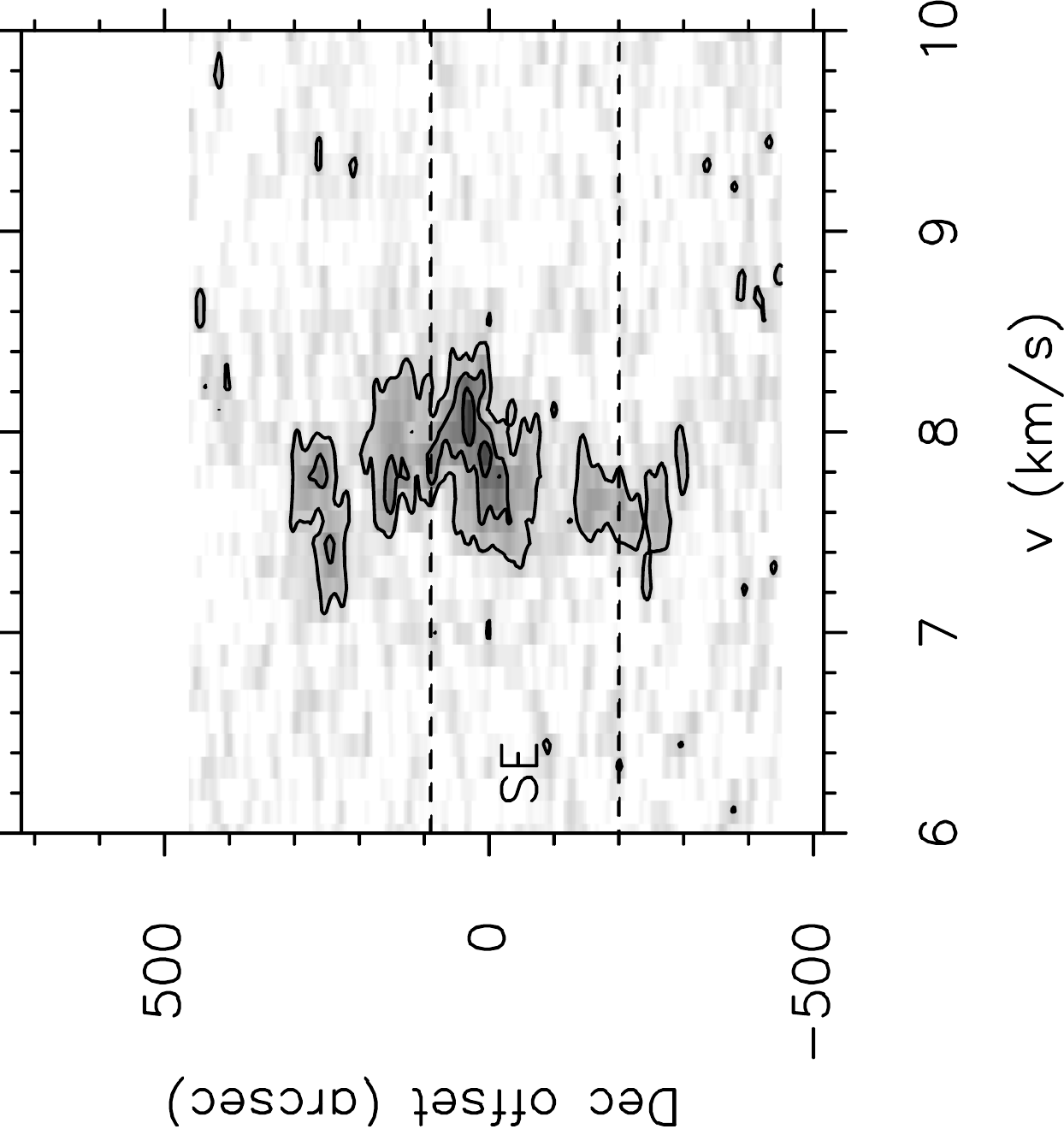}
\includegraphics[angle=270,width=0.25\textwidth]{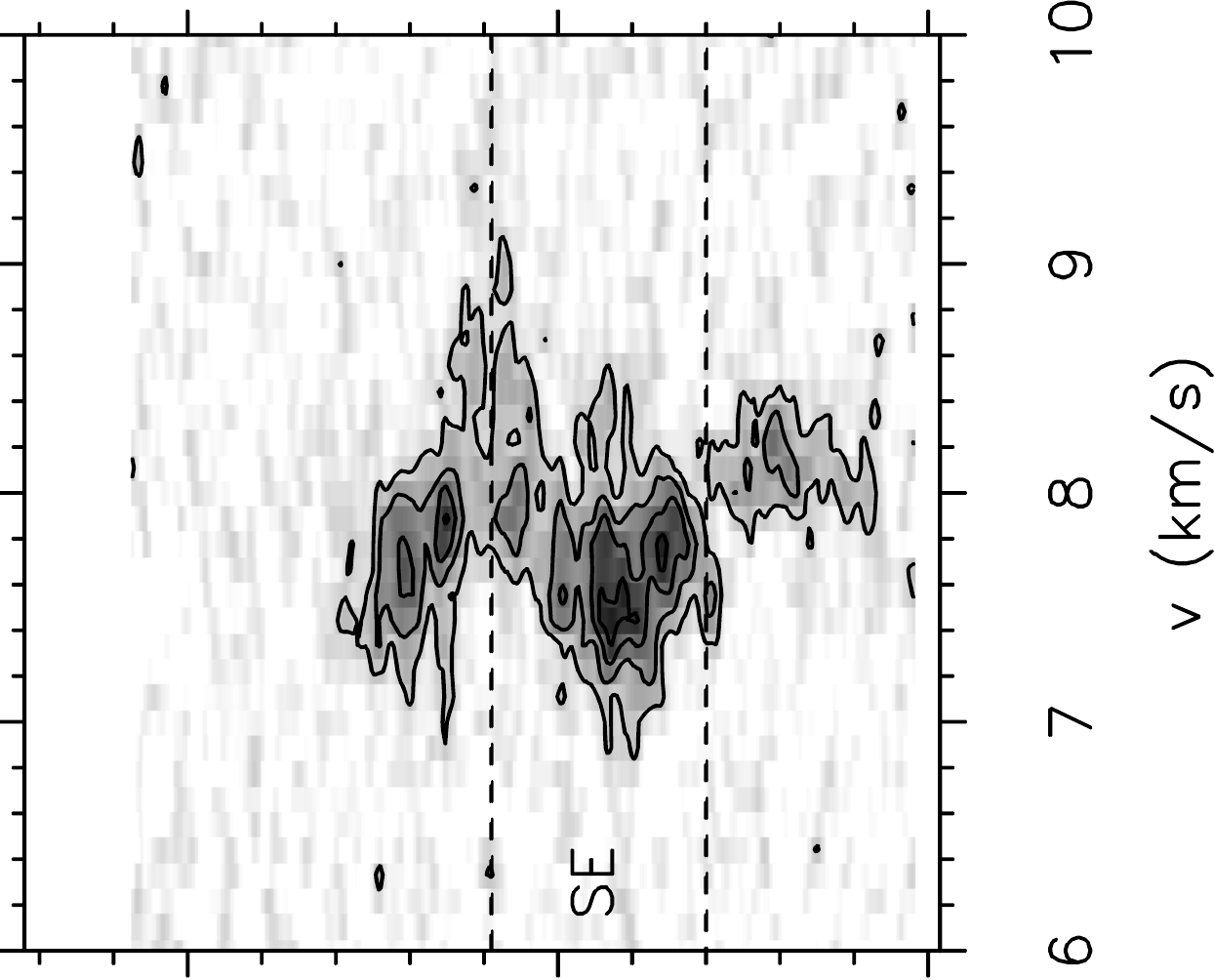}
\includegraphics[angle=270,width=0.29\textwidth]{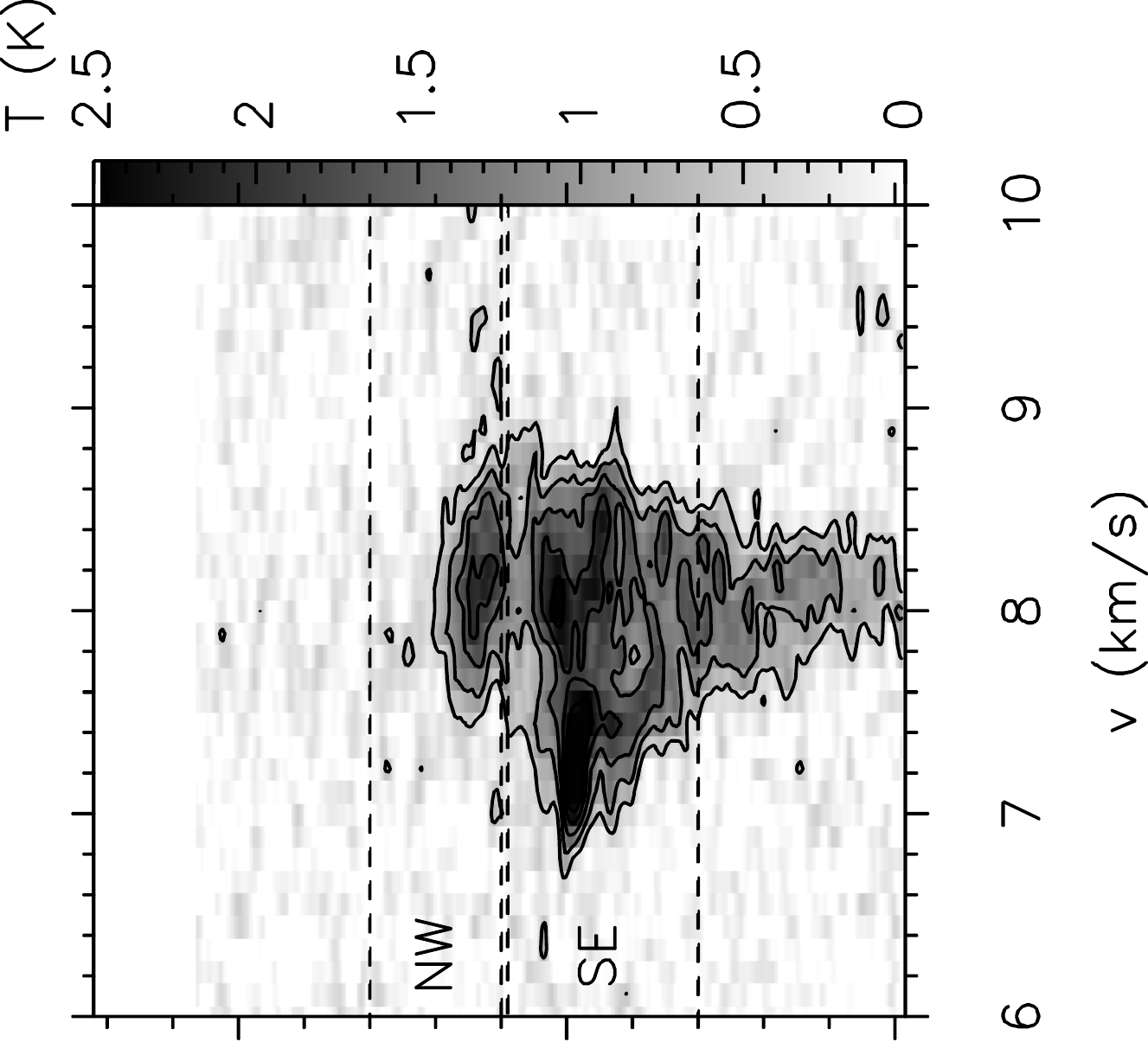}
\label{fig:pvconstantRA1}
\end{figure*}

\begin{figure*}
\centering
\includegraphics[angle=270,width=0.3\textwidth]{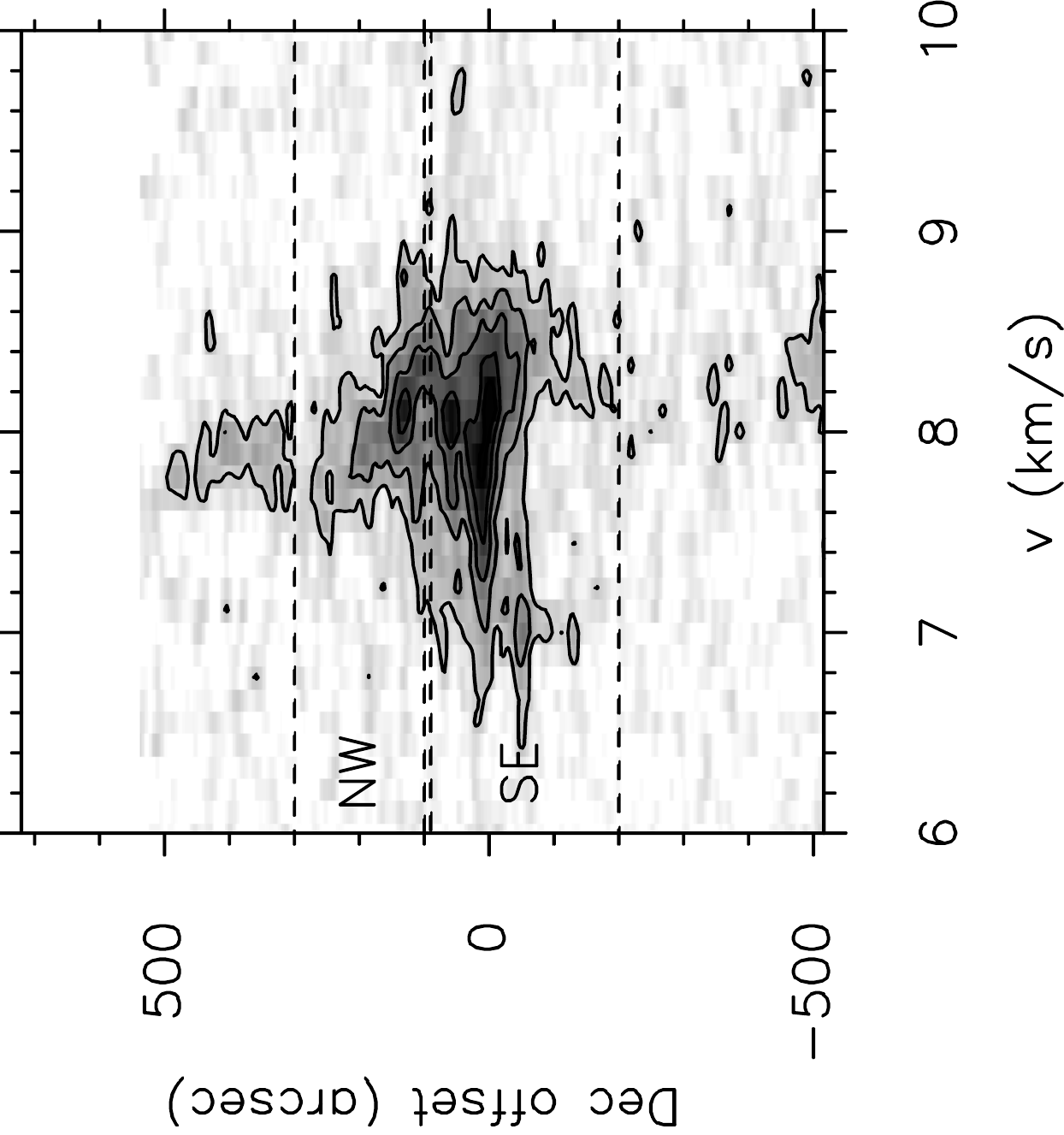}
\includegraphics[angle=270,width=0.255\textwidth]{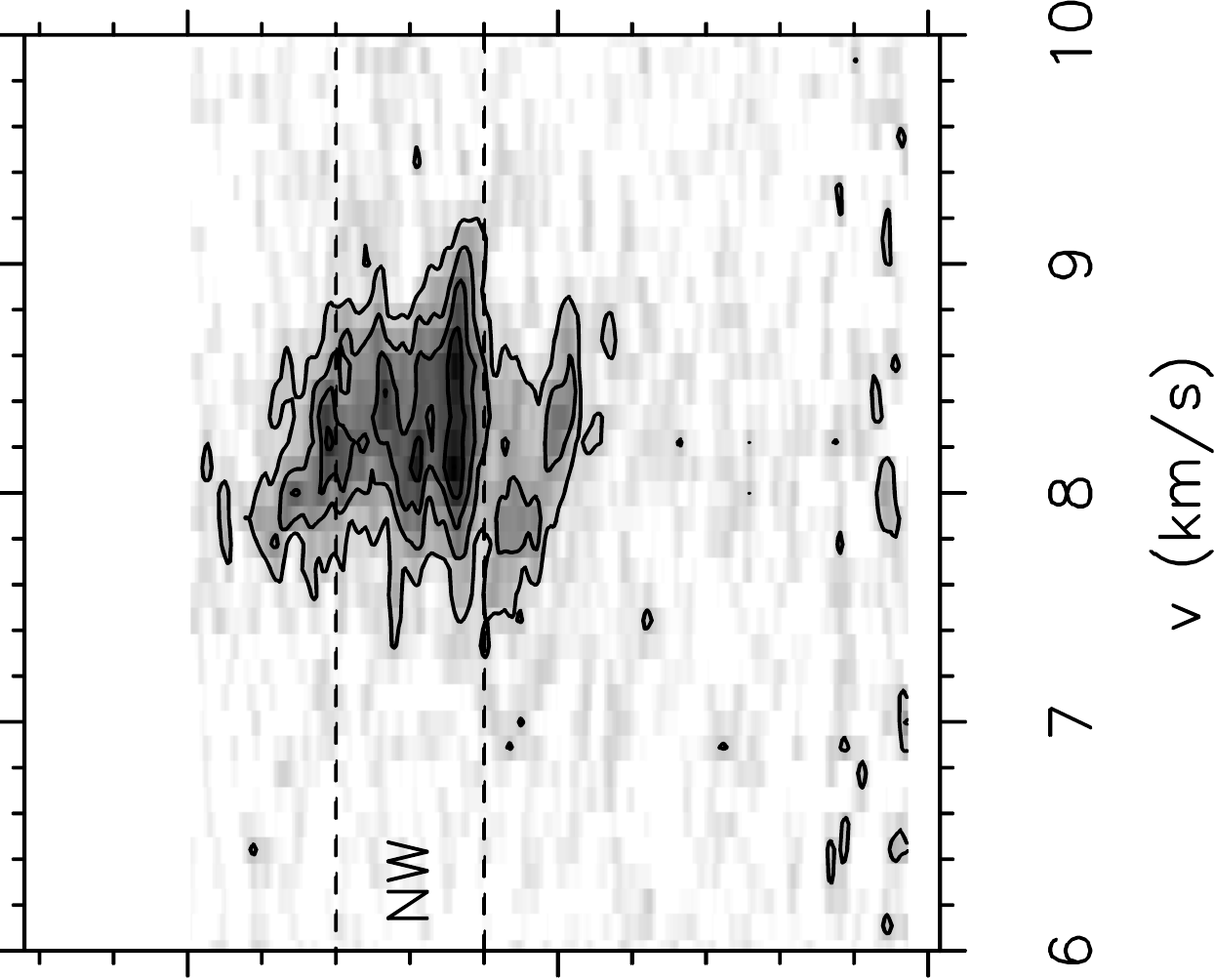}
\includegraphics[angle=270,width=0.29\textwidth]{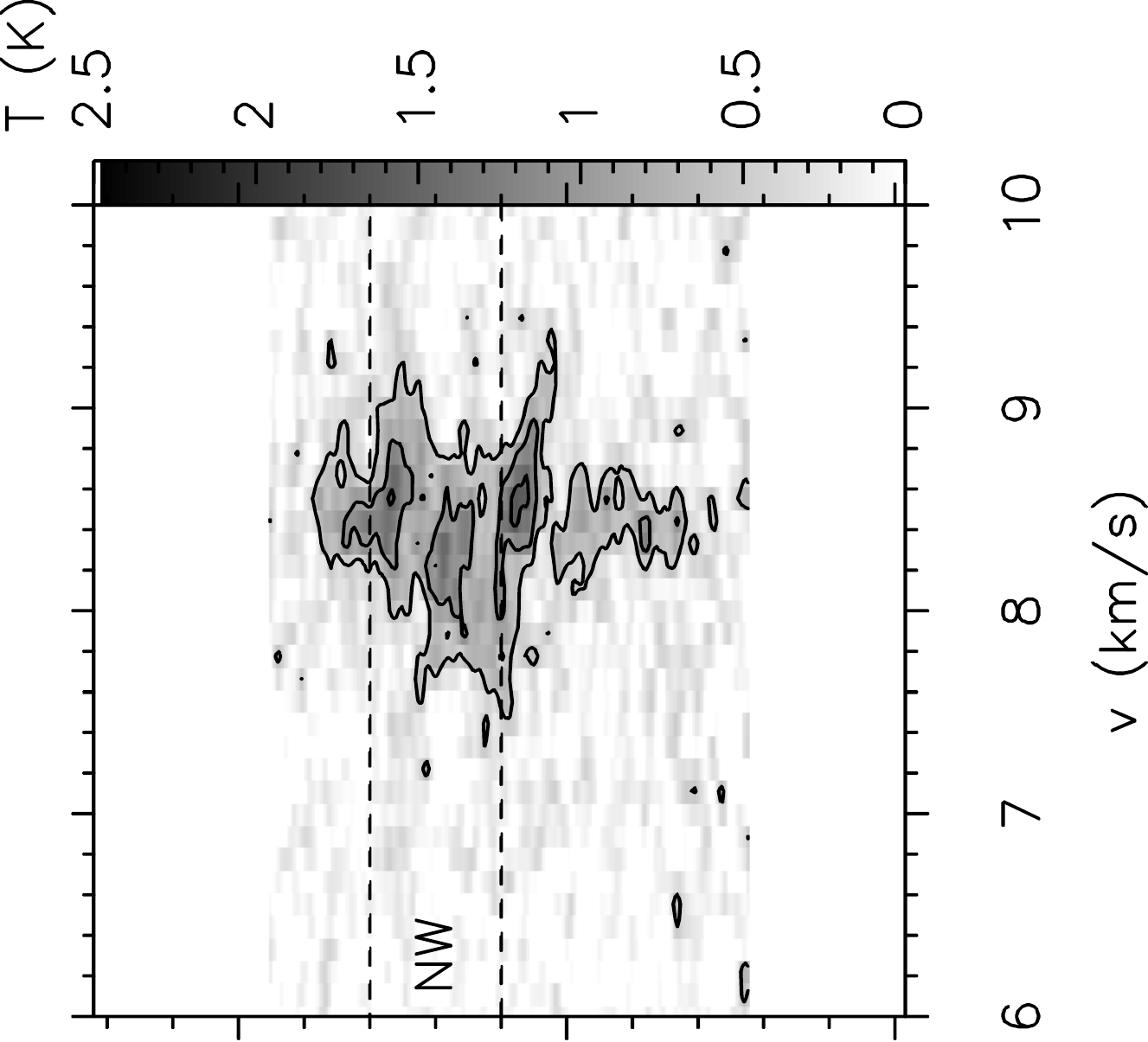}
\caption{Vertical position-velocity diagrams in grey scale and
  contours representing the C$^{18}$O \jtt. Declination
  ranging for 1$^o$4'24.1'' to 1$^o$25'0.8'' (offset from -515'' to
  720'' respectively) on all diagrams. The NW sub-cluster lies in Dec
  offsets of +90'' to +270'' and the SE sub-cluster between -180'' and
  +90''. Panels are displayed with decreasing Right Ascension
  (i.e. from East to West), with upper row being mostly representative
  of the SE sub-cluster whereas the lower row represents the NW
  sub-cluster. From top-left to lower-right diagrams, the cuts are at
  constant RA of 18$^h$30$^m$8.4$^s$, 18$^h$30$^m$4.4$^s$,
  18$^h$30$^m$0.4$^s$, 18$^h$29$^m$56.4$^s$, 18$^h$29$^m$50.4$^s$ and
  18$^h$29$^m$44.4$^s$.}
\label{fig:pvconstantRA2}
\end{figure*}

We also study the velocity structure using vertical PV diagrams
(constant RA) in 
Fig.~\ref{fig:pvconstantRA2}. The information we get from here adds
to the horizontal diagrams, in the sense that we can confirm the
complex velocity structure, including the double velocity component
towards the south sub-cluster (best seen on
Fig.~\ref{fig:pvconstantRA2}, top right panel). We can also see
velocities below 8~kms$^{-1}$ towards the northern region, in the
first two panels. Because this corresponds to higher RA positions,
these lower velocities in the north are only seen in regions offset to
east from the sub-cluster seen in 850~$\mu$m emission. The two last
panels of Fig.~\ref{fig:pvconstantRA2} (bottom row, centre and
right panels) show the bulk of emission towards the NW sub-cluster,
which traces well the 850~$\mu$m continuum emission, showing us the
velocities all concentrated between 8 and 8.5~kms$^{-1}$.
the outflow structure of the cloud is complex. Although there appear
to be many outflows in this region, many are overlapping. Instead of
attempting to assign spectra arbitrarily to one or another of an
overlapping outflow in order to calculate their properties, we have
decided to look first at the global outflow properties
(i.e.~energetics for the all the high velocity gas), and then to look
individually at those outflows that can be spatially located via their
very high velocity emission.

\section{Outflows}
\subsection{Global Outflow Properties}
\label{sec-global-outflow}

To calculate the global outflow properties, we wish to look at all the
gas that is moving at a significantly red- or blue-shifted velocity from
the line centre. However, it is not possible to completely exclude the
ambient emission, and some proportion of this will also be included in
these calculations.

We can calculate some basic kinematic properties for the global
properties of the outflowing gas.  Examination of the spectra in the
cloud shows that ambient \twelco\ spectra  away from the
outflows appear to cover a velocity range of 6 to 10\,\kms.  This is
assumed to include ambient emission for the entire cloud and is
excluded from the integration. Therefore the integration ranges used
are -15 to 6\,\kms for the blue-shifted emission and 10 to 30\kms\ for the red-shifted emission.

The masses were calculated under the assumption of LTE, a distance of
230\,pc and assuming optically thin emission in the line wings with an
excitation temperature of 50\,K \citep{Davis1999,White1995}. Although
this is higher than the peak excitation value we calculated earlier,
it is expected that the gas in the outflows will be considerably
warmer than that in the bulk of the gas.

\begin{eqnarray}
M_{^{12}\mathrm{CO}}= &2.2\times 10^{-6} \left(\frac{X_{CO}}{10^{-4}}\right)^{-1}\left(\frac{\eta_{mb}}{0.63}\right)^{-1}\left(\frac{A_{pixel}}{9\,\mathrm{square arcseconds}}\right) \nonumber \\
&\int T_A^*(v)dv\ M_{\odot}
\label{eqn-outflow}
\end{eqnarray}

The momentum and energy are estimated similarly, by replacing $\int
T_A^* dv$ by $\int T_A^* |v-v_0| dv$ and $0.5\times\int T_A^*(v-v_0)^2
dv$, where $v_0$ is the central velocity. Here, an average central velocity value is
used for the whole cloud.

The global results for the Serpens cloud are summarised in Table
\ref{tabglobal}. 
In Table \ref{energy_table} we give the total energy from both global
outflow values: 50 \,\Msun\,km$^{2}$\,s$^{-2}$.  \textrm{It is of
  course again important to consider the sources of systematic
  uncertainty affecting these figures. Our assumption of the
  excitation temperature of 50\,K, our chosen value for the CO
  abundance, and our assumption of LTE are sources of uncertainty in
  these values. We also expect some proportion of the shocked gas to
  be disassociated from its molecular form, and this will be a larger
  effect at the high velocities. We have also assumed that the entire
  line wing is optically thin; in fact this may well not be true,
  especially at the velocities close to the ambient cloud
  emission. This effect may mean that our values for the outflow
  masses are significantly under estimated. There may also be an
  effect on the momentum and energy estimated, but the optical depth
  is likely to be a less important factor and the disassociation a
  larger factor as these values are dominated by the high-velocity
  emission. These effects would tend to increase the values shown
  here. It is also important to reiterate that there will be some
  contamination of the values shown here by the ambient gas. }

Eqn.~\ref{eqn-outflow} only provides us with the energy and momentum
along the line-of-sight, and we must correct for the inclination of
the outflows if we wish to calculate 3-D values. \textrm{The outflow
  velocity should be corrected by $1/cos(i)$ from the line-of-sight
  velocity, where $i$ is the inclination angle. If we assume that the
  outflows are in an isotropic random distribution, this will increase
  the momentum by a factor of $2$ and the kinetic energy by $3$.  }

\begin{table}
  \caption{Global outflow properties, both along the line-of-sight only (l.o.s.) and corrected for a random distribution of inclinations (corr.).}

\begin{tabular}{c c c c} \hline 
&Mass/$M_{\odot}$&Momentum/$M_{\odot}$ \kms&Energy/\Msun\,km$^{2}$\,s$^{-2}$\\ 
\hline 
\textbf{l.o.s.} & & & \\ 
Blue & 1.3 & 6.1 & 28\\ 
Red  & 1.8 & 6.3 & 23 \\ 
\textbf{corr.} & & &\\
Blue  & - & 12.2 & 84\\
Red  & - & 12.6 & 69\\
\hline

\label{tabglobal}
\end{tabular}
\end{table}

We have also made a rough estimate of the dynamical age of the
outflows in the region. An estimate of the average flow lobe length
was found from Fig.~\ref{fig:co_outflows} of 210\,arcseconds (0.23\,pc
at $D=230$\,pc) in  the plane of the sky. The average maximum
velocity in the line of sight is $\sim$14\,\kms, thus obtaining a
value of 1.6$\times10^4$\,yr.  This very young age agrees with the
conclusions in the literature that the Serpens sources are Class 0/I
YSOs \citep{Casali1993,Testi1998}. The values for these energetic
properties correspond well to those found in \citet{Davis1999}.

\subsection{Effect of outflows on global energetics}
The results in Tables \ref{energy_table} and \ref{tabglobal} allow us to
discuss the validity of the theory of supersonic turbulence sustained
by molecular outflows. 

\textrm{If we apply a correction for random inclination, we estimate
  the total kinetic energy contained in the outflows to be of order 70
  per cent of the total turbulent kinetic energy (see
  sec.~\ref{sec-global-energetics}). This suggests that the outflows
  are a strong influence on the structure and energetics of the
  Serpens cloud core, and studies of this region must take care to
  account for them (particularly with respect to small scale
  structure).
  \\
  Our finding that the turbulent kinetic energy and the global outflow
  energy are of comparable size fits well with theories and
  simulations that involve the driving of supersonic turbulence by
  outflows in regions of active star formation (see
  e.g. \citet{Li2006,Matzner2007}). However, it is still necessary
  to identify a mechanism that can convert the outflow momentum (directed
  outwards from the cloud) into the turbulent energy. An analytical model for this process was developed by \citet{Matzner2007}, who proposed that molecular cloud turbulence can be driven by collimated protostellar outflows, resulting in a cascade of momentum rather than energy. Numerical simulations by \citet{Li2006} and \citet{Carroll2009} appear to be consistent with this picture.}

 A further interesting comparison can be made between the turbulent
 kinetic energy and the gravitational binding energy. Since the
 binding energy is very similar to the turbulent kinetic energy this
 suggests that the entire cloud is only weakly bound and that some
 small, local, 
 regions may well be in a state of collapse where as others are in a
 state of expansion. This is one of the expected observations of the
 theory of star formation control by supersonic turbulence
 \citep[see~][and references therein,
 e.g.~\citet{MacLow2004}]{Mckee2007}. 
 Further insight can be gained by considering the ratio of kinetic
 outflow energy to gravitational binding energy. \textrm{A value of
   $\sim0.7$ is indicative that outflows are extremely important to
   understanding the cloud support mechanism.  However, we expect the
   uncertainties in the estimation of the various parameters are large
   enough to prevent a more firm conclusion}

\subsection{Individual Outflows}

\begin{figure*}
\begin{center}
\hspace{0.27in}
\includegraphics[width=6.20in]{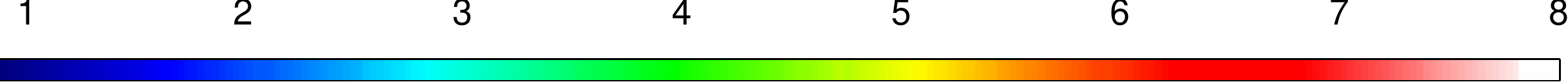}

\includegraphics[width=6.5in]{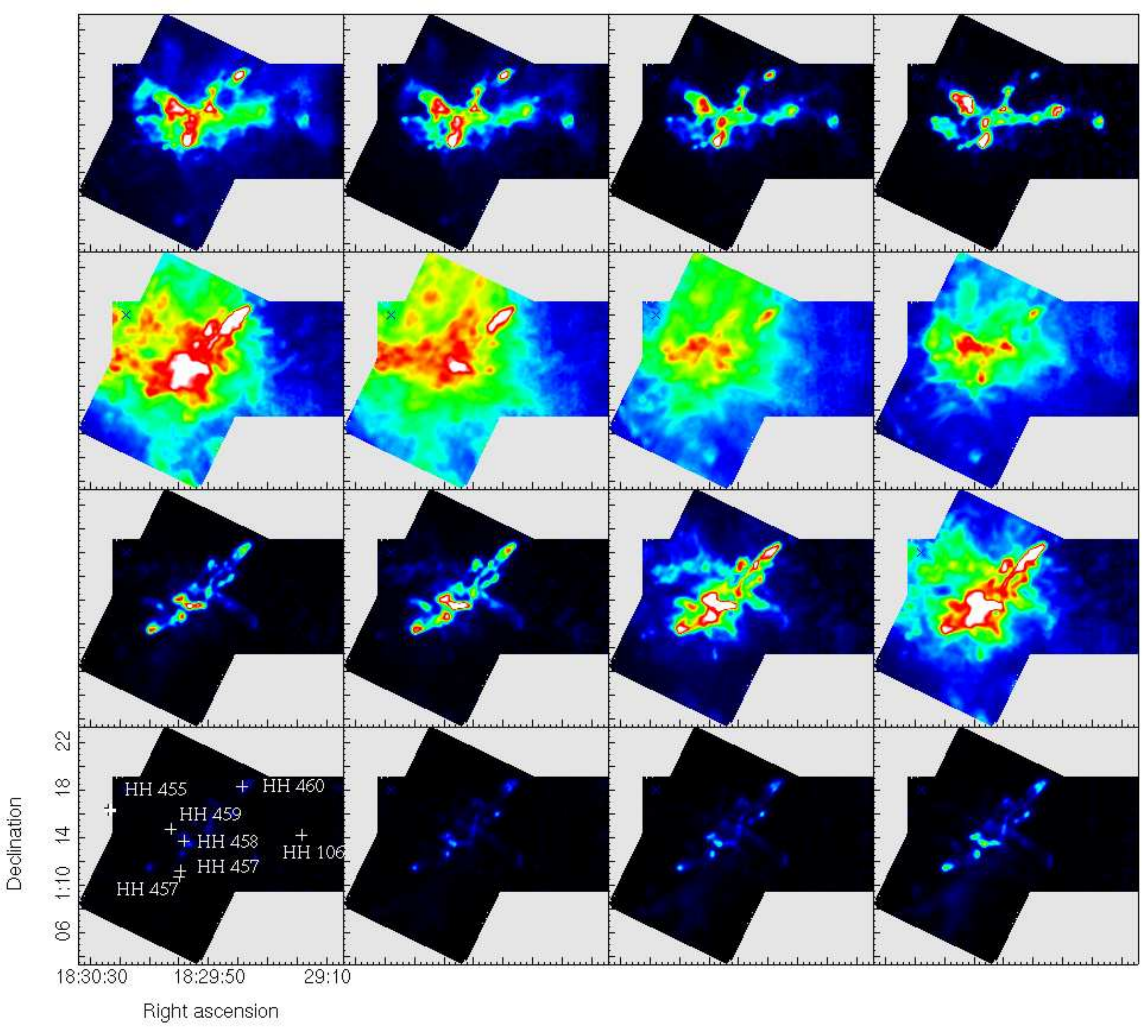}

\caption{\twelco\ channel maps of \Tstara. Beginning at -1.1\,\kms\ in
  lower left and increasing up to 14.9\,\kms\ in 1\,\kms\ steps. The
  colour scale used is shown above, in Kelvin. The HH objects known in
  this region are shown and labelled on the first
  channel. \label{fig-chanmaps}}
\end{center}
\end{figure*}

As can be seen from the \twelco\ channel maps presented in
Fig.~\ref{fig-chanmaps}, the \twelco\ high velocity structure is
extremely complex -- identifying outflows from the \twelco\ emission
alone is extremely difficult. We see considerable `finger-like'
structures in this region, that correspond to large scale outflow
structure across the cloud.  Although areas containing high velocity
red- and blue-shifted gas can be seen clearly in the channel maps,
categorising these into neat bipolar lobes is not so simple. However,
there are reasonable amounts of complementary data available. The
presence of known HH objects are indicated on
Fig.~\ref{fig:co_outflows}, and a comparison with the positions of known cores
allows a few more clear-cut outflows to be
identified. Figs.~\ref{fig:pv} and ~\ref{fig:pv-cont} show PV diagrams
along some of these outflows.

\subsubsection{Features from the channel maps}
\begin{figure}
  \centering
  \includegraphics[width=3.5in]{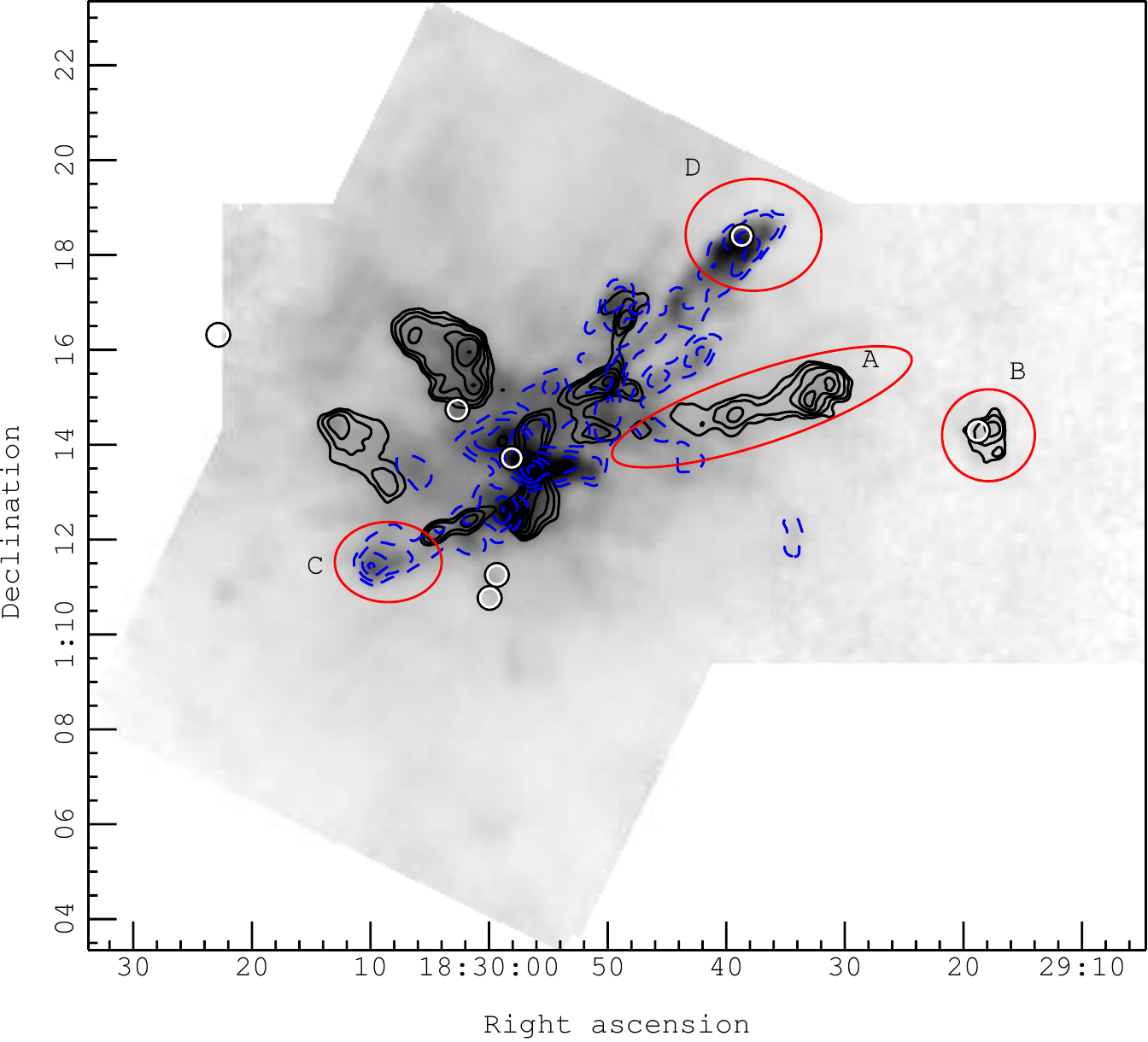}
  \caption{Finder chart for discussion of outflow regions identified
    from the \twelco\ channel map in Fig.~\ref{fig-chanmaps}. The
    regions discussed are shown in the red ellipses labelled A-D. The
    image shows an integrated \twelco\ map with red- and blue-shifted
    contours shown as black solid lines and blue dashed lines
    respectively. The HH objects are shown as circles.}
  \label{fig:finder-chan}
\end{figure}
Fig.~\ref{fig:finder-chan} displays some of the clearer potential
outflow structures from the channel maps, circled and labelled.

\begin{description}
\item[A:]
Examining the velocity structure from the \twelco\ channel maps
(Fig.~\ref{fig-chanmaps}) and the red and blue shifted contours allows
us to immediately see one strong feature -- the long extended
red-shifted emission westwards from the centre of the map, leading to
HH 106. 
There is no blue-shifted emission obviously lining up with this large
feature, but examination of the spectra in this region clearly show
the strong red-shifted line-wing. The spectra also show that the
maximum velocity of the emission increases along the flow as it heads
away from the main cloud, reaching up to 25\,\kms.

\item[B:]
 We also see a clump of emission towards the position of HH 106 in the
 extended region of the \twelco\ map; the spectral line profile for
 this region does indicate some red-shifted line-wing excess, although
 there is no corresponding \ceighto\ observations towards this
 position to estimate the central velocity from (and indeed there may
 not be any \ceighto\ emission towards this
 position). Fig.~\ref{fig:pv-cont} shows a position-velocity cut along
 this feature. The bow shock like feature at the end of the outflow
 can be clearly seen in the bright lobe of red-shifted emission in the
 PV diagram at $\sim1.5$\arcmin\ offset.

\item[C:] East and south of SMM 11 by 2.25 arcminutes we see a strong
  blue lobe. It has a minimum velocity of 9 \kms, with the fastest
  part of the feature being present at the farthest spatial distance
  from the main cloud. It does not clearly trace back to any source
  within the cloud. There is some red-shifted emission slightly north
  and west of this feature, but this appears to connect back further
  within the cloud.
 
\item[D:]
  Towards the position of HH 460 we see the end of a strong blue-shifted
  lobe. There is also some much weaker red-shifted emission. The maximum
  blue-shifted velocity in this region is at around -11\kms.
\end{description}

\subsubsection{Extremely high velocity emission}

\begin{figure}
\includegraphics[width=3.1in]{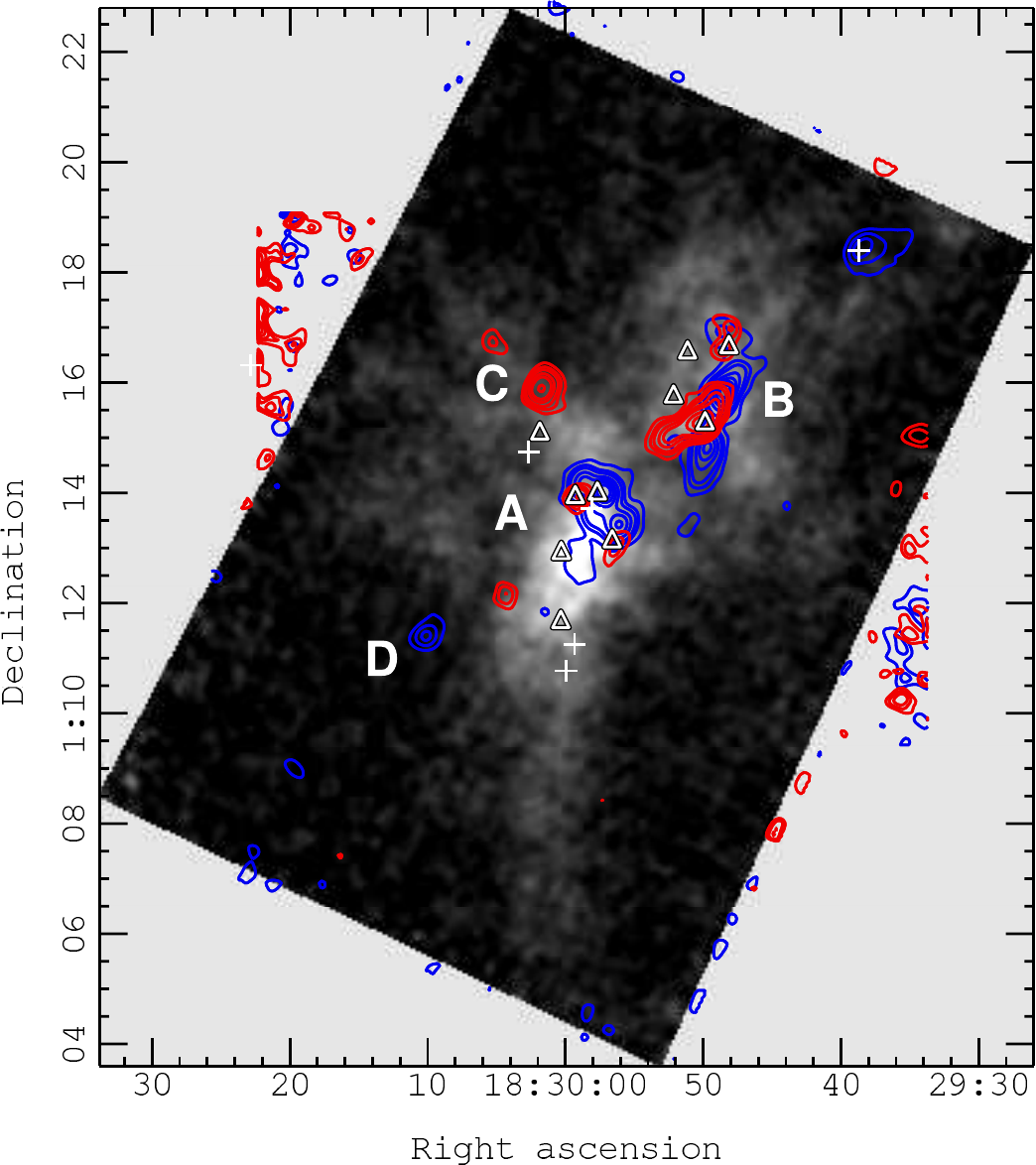}
\caption{Integrated \ceighto\ emission with contours of the highest
  velocity \twelco\ emission overlaid. Red contours represent emission
  from 24 to 34\,\kms, and blue contours represent emission from -15
  to -5\,\kms. Positions of the submillimetre cores are shown as white triangles
  and positions of HH objects are shown as white
  crosses. Letters label positions of interested discussed below. \label{fig:c18o-highv}}
\end{figure}

In order to attempt to separate some of this confused outflow
structure in this region, we can look at those outflows that have
emission at very high velocities -- of order 10\,\kms\ away from the
line centre. Fig.~\ref{fig:c18o-highv} shows contours
of intensity maps of the \twelco\ data integrated from 24-34\,\kms\
and from -15 to -5\,\kms. These are overlaid on the \ceighto\
integrated intensity map. By only looking at the outflows with this
extreme velocity range, we are missing out much of the outflowing
material in the cloud. However, this reduces the complexity of the velocity
structure, and allows us to be fairly
confident that we are not including a large amount of ambient
emission. This method also allows us to spatially separate otherwise confused
and overlapping outflows, and thus calculate properties of specific outflows.

As can be seen in Fig.~\ref{fig:c18o-highv}, several regions seem to
show compact features around cores, suggesting the presence of
outflowing material. 

\begin{description}
\item[A:] 
Specifically around SMM 6, 3 and 4, in the centre
of the map, what appears to be at least two overlapping outflows can be
seen. The blue shifted contours show two peaks, one at SMM4 and one
around SMM 6 and 3. HH 458 is also centred on this region.

\item[B:]
 SMM 1 and 9 also both show overlapping red and
blue lobes, which again appear to belong to two separate
outflows. Here the emission from SMM 1 appears to be much larger than
that around SMM 9.

\item[C:]
 There is a strong region of compact red-shifted emission to the north
of SMM 8, however visual inspection of the \twelco\ data cube does not
reveal any adjacent very-high-velocity blue-shifted emission that can
be separated from the main line centre emission. The red shifted
emission is however also near HH 459 so may be connected with that. In
the absence of any blue-shifted emission it is not possible to
identify this feature as indicating a molecular outflow, therefore we
have identified this as tentatively connected with the  HH object 459,
but have not attempted to link it directly to a submillimetre source. 

\item[D:]
 South-east of SMM 11 there is some evidence of adjacent red and blue
compact regions of emission (in Fig.~\ref{fig:c18o-highv} the blue
contoured region is more clearly seen). Inspection of the data cube
suggests that the red and blue structure exists more clearly at less
extreme velocities. However, given the lack of an obvious source it is
not clear whether or not this represent a bipolar
outflow. Additionally, inspection of spectra in the red shifted region
suggests it is possible that the high velocity red-shifted component
may be due to an additional velocity component along the line of sight
instead of excess line wing emission, so the spatial
connection of the red and blue emission may be coincidental. It has
not been possible to associate this red/blue region with a
submillimetre source.
\end{description}

\begin{table}
  \caption{Outflow masses and line of sight energetics for \twelco\ emission in range -30 to 1.8 \kms\ (blue) and 17.8 to 36.8 \kms\ (red) for each outflow. \textrm{Note that these outflows are spatially defined on the basis of their extreme velocity emission -- closer to the ambient velocity they are spatially substantially larger (and more confused), and the values presented here are very much lower limits for these specific outflows}}
    \begin{tabular}{c|c c c c}\hline
      & SMM 4 & SMM 6\&3 & SMM 1 & SMM 8\\\hline\hline
      Mass (\Msun) & &  & &\\
      Red & 0.0061 & 0.0085 & 0.012 & 0.003\\
      Blue & 0.0021 & 0.0016 & 0.0052 & 0.0015\\\hline
      Mom (\Msun\,\kms ) & & & &\\
      Red & 0.0014 & 0.0012 & 0.0043 & 0.0010\\
      Blue & 0.0018 & 0.0028 & 0.0042 & 0.00084\\\hline
      Energy (\Msun\,km$^2\,$s$^{-2}$) & & & &\\
      Red & 0.0095 & 0.0097 & 0.040 & 0.0080\\
      Blue & 0.012 & 0.0021 & 0.034 & 0.0058\\\hline\hline

    \end{tabular}

    \label{tab:outflowmasses}
  \end{table}

  For the outflows in the SMM 3, 4 and 6 region and in the SMM 1 and 8
  region we have calculated their properties. As these appear to
  slightly overlap spatially, it was not possible to perfectly
  separate these into the four separate outflows we believe
  exist. Instead, a simple best guess has been taken 
 (on the
  basis of the contours of the extremely high velocity gas), and
  spatial regions chosen that appear to contain most of the outflow
  emission. These spatial regions have then been used to calculate red
  and blue shifted masses, momentum and kinetic energy for each
  outflow, using the same assumptions as for the global outflow
  properties. No attempt has been made to correct for the angle of
  inclination.

The velocity ranges used in the calculations were -30 to 1.88\,\kms\
and 17.8 to 36.8\,\kms. The spatial regions were defined separately
for the red- and blue-shifted emission for each outflow. The masses
calculated are shown in Table \ref{tab:outflowmasses}. Again, no
attempt was made to correct for inclination.

The total mass for these outflows comes to merely 0.03 \Msun\ for the
blue-shifted emission and 0.01 \Msun\ for the red-shifted emission;
this seems low considering the global values we have previously
calculated. However, we have been very conservative with the velocity
range used here, attempting to be sure that we are not being
contaminated by the line centre emission or by overlapping outflows
that do not reach such high velocities. If we use the same velocity
ranges as used for the global properties than these values
approximately double, making these outflows responsible for most of
the high velocity mass in this cloud.

\begin{figure}
 
  \begin{minipage}[t]{1.0\linewidth}
    \includegraphics[width=3.4in]{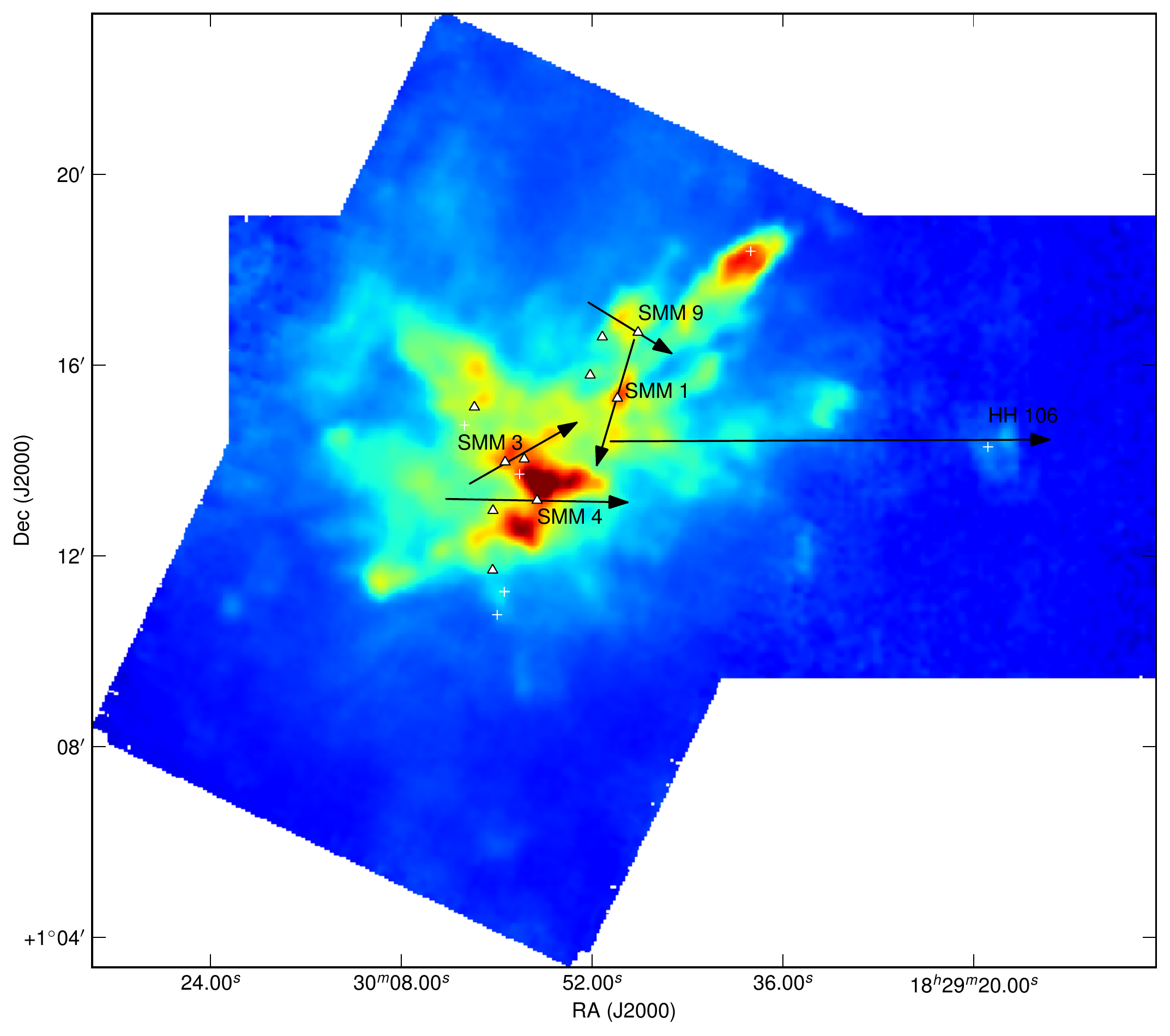}
  \end{minipage}

\centering
\includegraphics[height=2.0in]{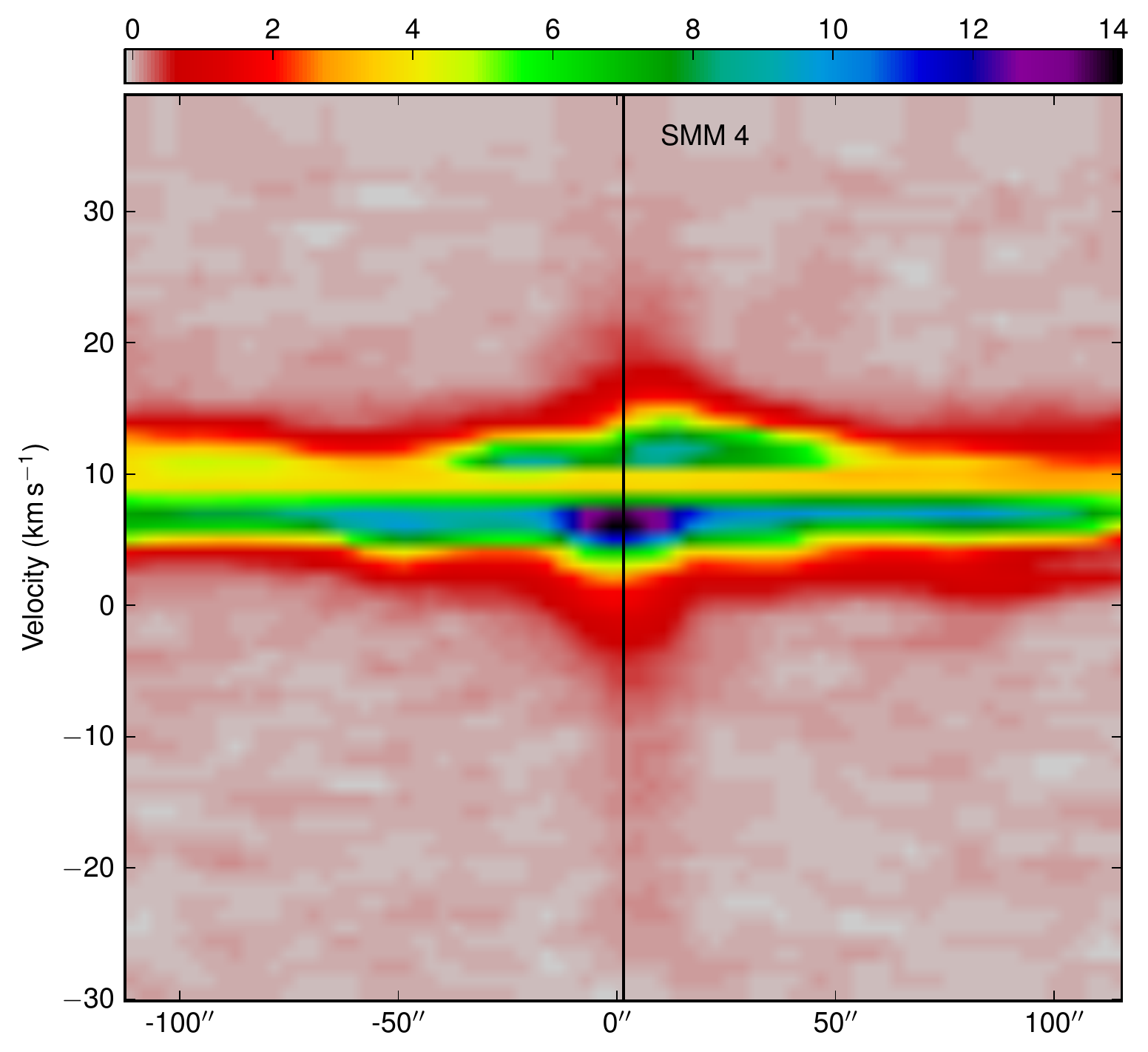}%
\includegraphics[height=2.0in]{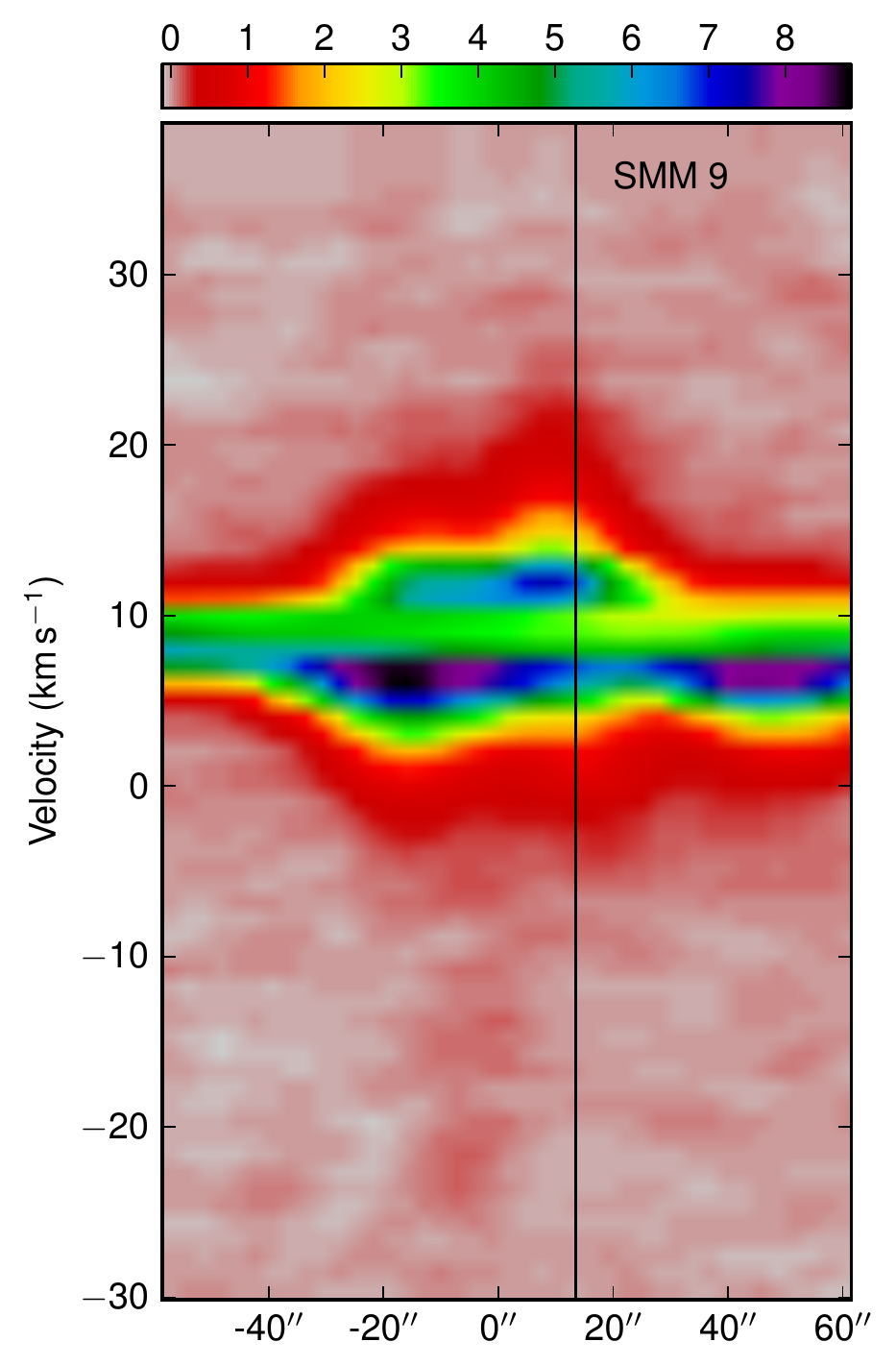}
     \includegraphics[height=2.0in]{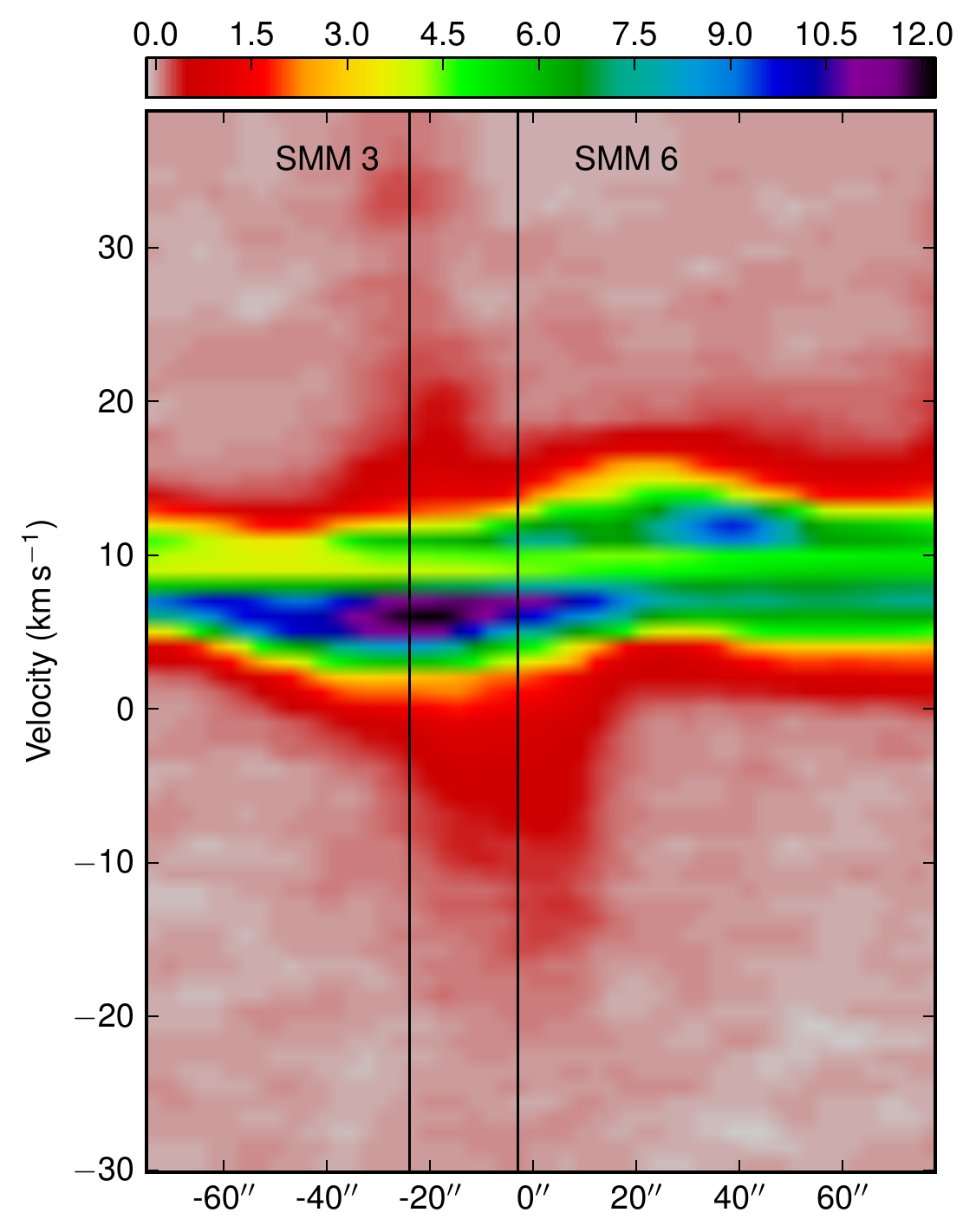}
     \includegraphics[height=2.0in]{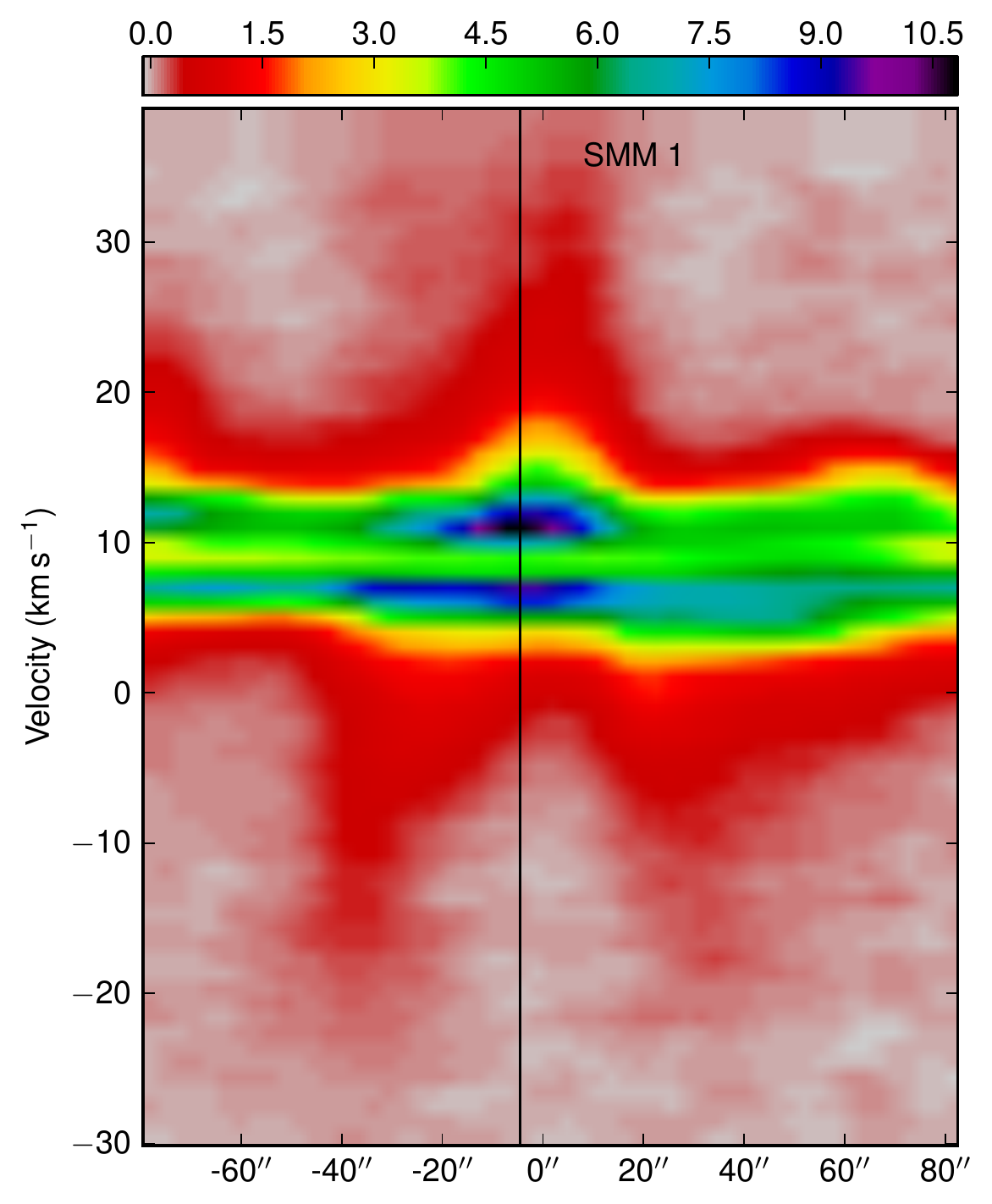}%

     \caption{\twelco\ position-velocity diagrams along the positions of the
       identified compact outflows, as shown on the integrated image
       intensity image (top). PV
       diagrams shown are from SMM 4 (middle left), SMM 9 (middle right),
       SMM 3 (bottom left) and SMM 1 (bottom right). The position of
       the appropriate cores are shown as black lines on the PV
       diagrams. The top image shows the locations of the slices. All
       intensity values are of \Tstara. \label{fig:pv}}
\end{figure}

\begin{figure}
  \includegraphics[angle=0,width=3.4in]{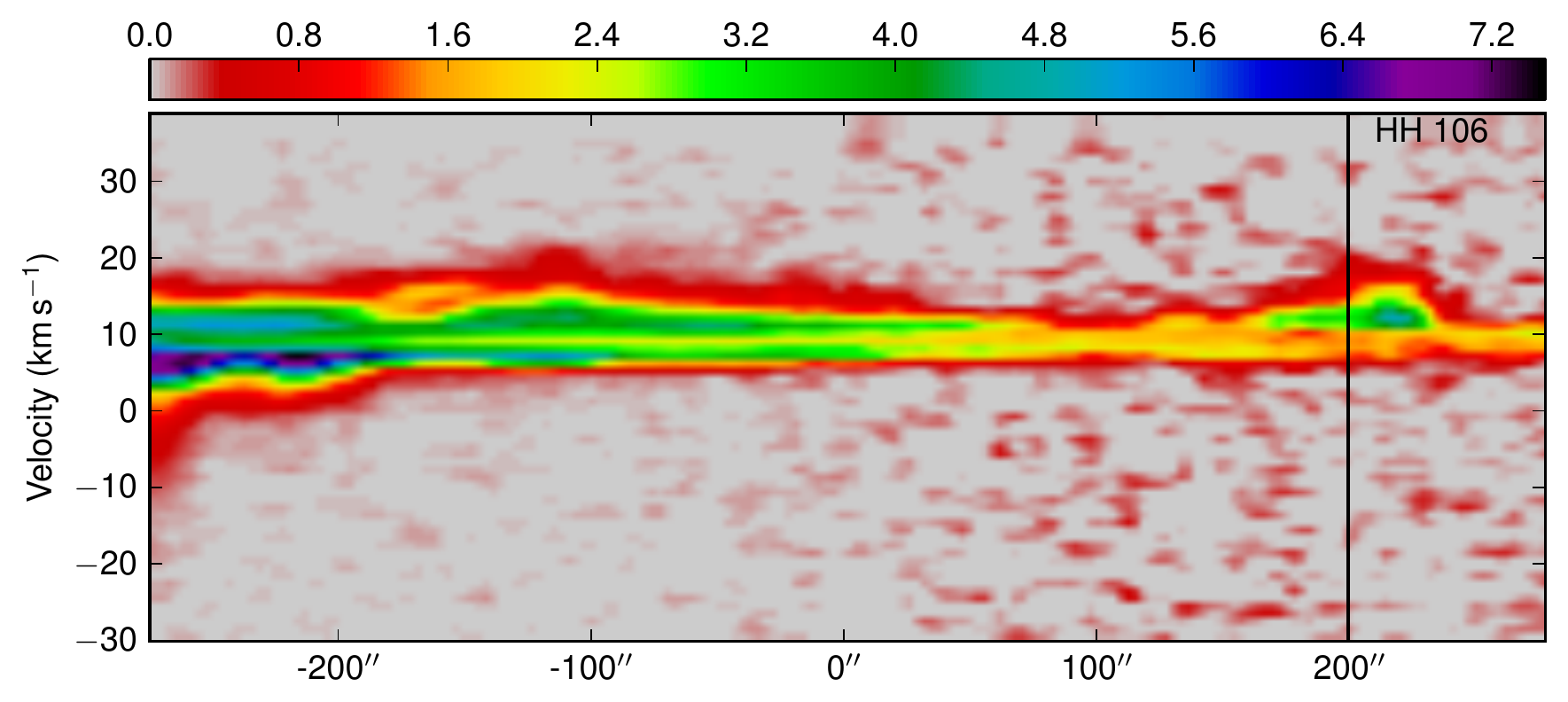}
  \caption{Position-velocity diagrams continued from
    Fig.~\ref{fig:pv}. HH 106:a PV diagram along the long east-west
    red-shifted lobe towards HH 106 is shown.}
  \label{fig:pv-cont}
\end{figure}

To further show that these outflows are indeed connected with the
cores, Fig.~\ref{fig:pv} shows position-velocity diagrams along
the outflows identified. Evidence of the outflow emission can be
seen in both the red and blue shifted regions, although it is the
central line emission that dominates. The dip in the centre of the
line shows the self absorption in the \twelco\ emission.

\section{Summary}

High resolution mapping of the Serpens molecular cloud in the $J=3-2$
transition of \twelco\, \thirtco\ and \ceighto\ was performed with 
HARP at the JCMT. The \twelco\ map traces the numerous
outflows in the region, with line wings detected out to $-30$ and
$+37$\,\vunits. The \ceighto\ map was used to examine the global
velocity structure and to estimate the mass of the region, which was
found to be $\sim$ 203 M$_\odot$. Intensity ratios were also used to
estimate an abundance ratio for \thirtco\ to \ceighto\ and in turn an
estimate of the optical depth. Additionally, the kinetic temperature was examined.

The global velocity structure of Serpens was examined in the \ceighto\
emission, and evidence was found to support the conclusion of
\citet{DuarteCabral2010} that the cloud consists of two sub clouds at
different velocities.

Although across the bulk of the cloud it would not be straightforward
to identify the individual outflows present as they overlap spatially,
the global properties of the outflowing gas were examined, and
\textrm{we found that the total outflow energy (corrected for random
  inclination distribution) was approximately 70 percent of the total
  turbulent energy in the region. This suggests that the outflows may
  be a significant factor in driving turbulence in this star-forming
  region, in broad agreement with many recent simulations and theory.
}

Four compact outflows towards five of the submillimetre sources were identified
on the basis of extremely high velocity emission and their properties
examined, finding that this extreme gas does not contribute a large
portion of the global outflow energy.

\section{Acknowledgements}
The JCMT is supported by the Science and Technology Facilities
Council, the National Research Council Canada, and the Netherlands
Organisation for Scientific Research.

\label{lastpage}

\end{document}